\newcommand\submitms{n}		

\if\submitms y
    \documentclass[12pt,preprint]{aastex}
\else
    \documentclass[apj]{emulateapj}
\fi

\if\submitms y
\else
\usepackage{apjfonts}
\fi

\usepackage{amssymb, amsmath}
\usepackage{natbib}
\usepackage{ifthen}
\usepackage{appendix}
\newcommand\degree{\degr}
\newcommand\degrees\degree
\newcommand\vs{{\em vs.}\ }
\newcounter{fignum}

\DeclareSymbolFont{UPM}{U}{eur}{m}{n}
\DeclareMathSymbol{\umu}{0}{UPM}{"16}
\let\oldumu=\umu
\renewcommand\umu{\ifmmode\oldumu\else\math{\oldumu}\fi}
\newcommand\micro{\umu}
\renewcommand\micron{\micro m}
\newcommand\microns \micron

\let\oldsim=\sim
\renewcommand\sim{\ifmmode\oldsim\else\math{\oldsim}\fi}
\let\oldpm=\pm
\renewcommand\pm{\ifmmode\oldpm\else\math{\oldpm}\fi}
\newcommand\by{\ifmmode\times\else\math{\times}\fi}

\newbox{\wdbox}
\renewcommand\c{\setbox\wdbox=\hbox{,}\hspace{\wd\wdbox}}
\renewcommand\i{\setbox\wdbox=\hbox{i}\hspace{\wd\wdbox}}

\newcount\timect
\newcount\hourct
\newcount\minct
\newcommand\now{\timect=\time \divide\timect by 60
         \hourct=\timect \multiply\hourct by 60
         \minct=\time \advance\minct by -\hourct
         \number\timect:\ifnum \minct < 10 0\fi\number\minct}

\newcommand\mctc{\multicolumn{2}{c}}

\catcode`@=11

\newcommand\comment[1]{}
\newcommand\commenton{\catcode`\%=14}
\newcommand\commentoff{\catcode`\%=12}

\renewcommand\math[1]{$#1$}
\newcommand\mathshifton{\catcode`\$=3}
\newcommand\mathshiftoff{\catcode`\$=12}

\comment{the backslash is necessary}

\comment{alignment tab}

\let\atab=&
\newcommand\atabon{\catcode`\&=4}
\newcommand\ataboff{\catcode`\&=12}

\let\oldmsp=\sp
\let\oldmsb=\sb
\def\sp#1{\ifmmode
           \oldmsp{#1}%
         \else\strut\raise.85ex\hbox{\scriptsize #1}\fi}
\def\sb#1{\ifmmode
           \oldmsb{#1}%
         \else\strut\raise-.54ex\hbox{\scriptsize #1}\fi}
\newbox\@sp
\newbox\@sb
\def\sbp#1#2{\ifmmode%
           \oldmsb{#1}\oldmsp{#2}%
         \else
           \setbox\@sb=\hbox{\sb{#1}}%
           \setbox\@sp=\hbox{\sp{#2}}%
           \rlap{\copy\@sb}\copy\@sp
           \ifdim \wd\@sb >\wd\@sp
             \hskip -\wd\@sp \hskip \wd\@sb
           \fi
        \fi}
\def\msp#1{\ifmmode
           \oldmsp{#1}
         \else \math{\oldmsp{#1}}\fi}
\def\msb#1{\ifmmode
           \oldmsb{#1}
         \else \math{\oldmsb{#1}}\fi}
\def\supon{\catcode`\^=7}
\def\supoff{\catcode`\^=12}
\def\subon{\catcode`\_=8}
\def\suboff{\catcode`\_=12}
\def\supsubon{\supon \subon}
\def\supsuboff{\supoff \suboff}

\newcommand\actcharon{\catcode`\~=13}
\newcommand\actcharoff{\catcode`\~=12}

\newcommand\paramon{\catcode`\#=6}
\newcommand\paramoff{\catcode`\#=12}

\comment{And now to turn us totally on and off...}

\newcommand\reservedcharson{\commenton \mathshifton \atabon \supsubon \actcharon
	\paramon}

\newcommand\reservedcharsoff{\commentoff \mathshiftoff \ataboff
	\supsuboff \actcharoff \paramoff}

\catcode`@=12
\reservedcharsoff

\reservedcharson

\comment{ Must have ONLY ONE of these... trust these macros, they work

}

\newcommand{\squishlist}{
 \begin{list}{$\bullet$}
  { \setlength{\itemsep}{1pt}
     \setlength{\parsep}{0pt}
     \setlength{\topsep}{3pt}
     \setlength{\partopsep}{0pt}
     \setlength{\leftmargin}{2.0em}
     \setlength{\labelwidth}{1.5em}
     \setlength{\labelsep}{0.5em} } }

\newcommand{\squishend}{
  \end{list}  }

\reservedcharsoff

\actcharon

\if\submitms y
\bibpunct[, ]{(}{)}{,}{a}{}{,}
\else
\fi

\actcharon

\shorttitle{Analysis of Exoplanet HD 149026b Using BLISS Mapping}
\shortauthors{Stevenson {\em et al.}}

\reservedcharson

\begin{document}

\title{Transit and Eclipse Analyses of the Exoplanet HD 149026\MakeLowercase{b} Using BLISS Mapping}

\if\submitms n
    \author{Kevin B.\ Stevenson and Joseph Harrington}
    \affil{Planetary Sciences Group, Department of Physics, University of Central Florida\\
    Orlando, FL 32816-2385}

    \author{Jonathan J. Fortney}
    \affil{Department of Astronomy \& Astrophysics, University of California Santa Cruz\\ Santa Cruz, CA 95064}

    \author{Thomas J. Loredo}
    \affil{Center for Radiophysics and Space Research, Space Sciences Building, Cornell University\\
    Ithaca, NY 14853-6801}
    
    \author{Ryan A.\ Hardy, Sarah Nymeyer, William C.\ Bowman, Patricio Cubillos, M. Oliver Bowman and Matthew Hardin}
    \affil{Planetary Sciences Group, Department of Physics, University of Central Florida\\
    Orlando, FL 32816-2385}

    \email{kevin218@knights.ucf.edu}
\else
    \author{Kevin B.\ Stevenson\altaffilmark{1}}
    \author{Joseph Harrington\altaffilmark{1}}
    \author{Jonathan Fortney\altaffilmark{2}}
    \author{Thomas J. Loredo\altaffilmark{3}}
    \author{Ryan A.\ Hardy\altaffilmark{1}}
    \author{Sarah Nymeyer\altaffilmark{1}}
    \author{William C.\ Bowman\altaffilmark{1}}
    \author{Patricio Cubillos\altaffilmark{1}}
    \author{M. Oliver Bowman\altaffilmark{1}}
    \author{Matthew Hardin\altaffilmark{1}}
    \affil{\sp{1}Planetary Sciences Group, Department of Physics, University of Central Florida, Orlando, FL 32816-2385, USA}
    \affil{\sp{2}Department of Astronomy \& Astrophysics, University of California Santa Cruz, Santa Cruz, CA 95064, USA}
    \affil{\sp{3}Astronomy Department, Cornell University, Ithaca, NY 14853-6801, USA}
\fi

\begin{abstract}

The dayside of HD 149026b is near the edge of detectability by the {\em Spitzer Space Telescope}.  We report on eleven secondary-eclipse events at 3.6, 4.5, $3\times 5.8$, $4\times 8.0$, and $2\times 16$ {\microns} plus three primary-transit events at 8.0 {\microns}.  The eclipse depths from jointly-fit models at each wavelength are 0.040 {\pm} 0.003\% at 3.6 {\microns}, 0.034 {\pm} 0.006\% at 4.5 {\microns}, 0.044 {\pm} 0.010\% at 5.8 {\microns}, 0.052 {\pm} 0.006\% at 8.0 {\microns}, and 0.085 {\pm} 0.032\% at 16 {\microns}.
Multiple observations at the longer wavelengths improved eclipse-depth signal-to-noise ratios by up to a factor of two and improved estimates of the planet-to-star radius ratio ($R\sb{p}/R\sb{\star} = 0.0518 \pm 0.0006$).  We also identify no significant deviations from a circular orbit and, using this model, report an improved period of 2.8758916 \pm\ 0.0000014 days.
Chemical-equilibrium models find no indication of a temperature inversion in the dayside atmosphere of HD 149026b.  Our best-fit model favors large amounts of CO and CO\sb{2}, moderate heat redistribution ($f=0.5$), and a strongly enhanced metallicity.
%
These analyses use BiLinearly-Interpolated Subpixel Sensitivity (BLISS) mapping, a new technique to model two position-dependent systematics (intrapixel variability and pixelation) by mapping the pixel surface at high resolution.  BLISS mapping outperforms previous methods in both speed and goodness of fit.
We also present an orthogonalization technique for linearly-correlated parameters that accelerates the convergence of Markov chains that employ the Metropolis random walk sampler.
The electronic supplement contains light-curve files and supplementary figures.

\end{abstract}
\keywords{planetary systems
--- stars: individual: HD 149026
--- techniques: photometric
}

\section{INTRODUCTION}
\label{intro}

Discovered in 2005 using Doppler measurements, the Saturn-sized extrasolar planet HD 149026b orbits (in 2.876 days) a G0IV star that is larger (1.45 solar radii), and hotter (6150 \pm\ 50 K) than most stars known to host transiting exoplanets.  The planet's small radius and high average density suggest that between 50\% and 90\% of the planet's mass must be in its rocky or icy core \citep[hereafter K09]{Knutson2009b}.  Invoking current theories, it is difficult to form this exoplanet by gravitational instability \citep{Sato2005}.


Shortly after detection, 
\citet{Fortney2006} computed models of the atmospheric temperature 
structure and spectra of HD 149026b.  They suggested that the planet was 
a strong candidate for having a day-side atmospheric temperature 
inversion.  The highly irradiated planet is hot enough to have gaseous 
TiO and VO molecules in the dayside atmosphere.  These molecules are 
strong optical absorbers and had been previously shown to cause 
temperature inversions in model atmospheres \citep{Hubeny03}.

Beginning in 2005, we used the photometric channels of the {\em Spitzer Space Telescope} \citep{SPITZER} to observe HD 149026b during secondary eclipse, when the planet passes behind its parent star, to characterize the planet's dayside atmosphere. \citet[hereafter H07]{Harrington2007} found an 8.0 {\micron} eclipse depth of $0.084\%\sp{+0.012}\sb{-0.009}$, indicating the hottest brightness temperature (2300 \pm\ 200 K) observed at that time.  This temperature matches an instantaneous re-emission model (zero albedo) from \citet{Fortney2006} that exhibits a temperature inversion (which tends to enhance the planet/star contrast at 8.0 {\microns}), thus suggesting the presence of absorbers, such as TiO and VO gas molecules, in the atmosphere.

\citet{Charbonneau2006} observed two primary transits, when the planet passes in front of its parent star, using the Fred Lawrence Whipple Observatory telescope through the Sloan $g$ and $r$ filters.  \citet{Winn2008} reported on five ground-based transits through Str\"{o}mgren $b$ and $y$ filters at the Fairborn Observatory.  In August of 2007, \citet{Nutzman2009} used {\em Spitzer} to monitor a transit of HD 149026b at 8.0 {\microns}.  \citet{Carter2009} used the NICMOS detector on board the {\em Hubble Space Telescope} to observe four transits of HD 149026b at 1.4 {\microns}.  Their data have the best photometric precision to date and, after combining their data with all previous transit measurements, provide improved estimates of orbital parameters, mass, and radius.

In 2008, K09 monitored the system for just over half an orbit to characterize the planet's phase variation at 8.0 {\microns}.  Their observations began slightly before the primary transit and finished slightly after the secondary eclipse.  Using the final 7.2 hours of data, K09 report an eclipse depth of 0.0411 \pm\ 0.0076\%, half that of H07.  As part of their paper, K09 reanalyzed the 2005 secondary-eclipse data and found eclipse depths ranging from 0.05 - 0.09\%, though the lower values are preferred in most of their models.  This large range of eclipse depths depends on the choice of systematic error model, fitting routines, and bad-pixel trimming methods.

HD 149026b is an interesting planet given its extremely unusual bulk abundances; the majority of the 
planet's mass must be in heavy elements, making the planet perhaps more 
akin to Uranus and Neptune than Jupiter and Saturn.  In the solar 
system, a bulk composition that is enhanced in metals goes hand in hand 
with an \emph{atmospheric} composition enhanced in metals \citep{Marley07}.  This suggests that HD 149026b could have an atmospheric metallicity far greater than that of most transiting 
exoplanets.  Verifying this by measurement would let us understand the makeup of this planet and the role of atmospheric composition in determining temperature structure.

In this paper we present {\em Spitzer Space Telescope} secondary-eclipse observations of HD 149026b that resolve the disagreement in eclipse depths at 8.0 {\microns}, characterize the planet's dayside atmosphere, and further constrain its orbital and physical parameters.  We give detailed descriptions of our techniques and results because how one handles {\em Spitzer's} systematics can lead to best-fit parameters that disagree by more than 1\math{\sigma}, as demonstrated in Section \ref{sec:eclresults}.

Below, we describe the observations and data analysis, present a new method for modeling one of {\em Spitzer's} systematics, explain how we arrived at the final fits and compare the results to previously published work, discuss implications for the planetary emission spectrum and planetary composition, give improved constraints on the orbital parameters, and state our conclusions.

\section{OBSERVATIONS AND DATA ANALYSIS}
\label{sec:obs}

\subsection{Observations}

We observed secondary eclipses of HD 149026b at 3.6, 4.5, 5.8, and 8.0 {\microns} with the Infrared Array Camera \citep[IRAC]{IRAC} and at 16 {\microns} using the Infrared Spectrograph's \citep[IRS]{IRS} photometric blue peak-up array.  The program also observed a primary transit at 8.0 {\microns}.  Including the four previously analyzed data sets labeled in Table \ref{table:ObsDates}, we present fourteen observations spanning more than 3.5 years.

\if\submitms y
\clearpage
\fi
\begin{table*}[ht]
\centering
\caption{\label{table:ObsDates} 
Observation Information}
\begin{tabular}{crccclcc}
    \hline
    \hline
    Label\tablenotemark{a}       
                & Observation Date      & Duration  & Frame Time    & Total Frames  & {\em Spitzer} & Wavelength    & Previous      \\
                &                       & [minutes] & [seconds]     &               & Pipeline      & [\microns]    & Publications\tablenotemark{b}  \\
    \hline
    HD149bs41   &    August 24, 2005    & 330       & 0.4           & 44352         & S18.7.0       & 8.0           & H07           \\
    HD149bs51   &     August 4, 2007    & 386       & 14            & 1050          & S18.18.0      & 16            & -             \\
    HD149bs31   &    August 13, 2007    & 386       & 0.4           & 54080         & S18.7.0       & 5.8           & -             \\
    HD149bp41   &    August 14, 2007    & 478       & 0.4           & 67008         & S18.7.0       & 8.0           & N09 \& C09    \\
    HD149bs52   &    August 30, 2007    & 386       & 14            & 1050          & S18.18.0      & 16            & -             \\
    HD149bp42   &     Sept. 12, 2007    & 386       & 0.4           & 54080         & S18.7.0       & 8.0           & -             \\
    HD149bs11   &     March 10, 2008    & 386       & 0.4           & 54080         & S18.7.0       & 3.6           & -             \\
    HD149bs42   &     April 11, 2008    & 386       & 0.4           & 54080         & S18.7.0       & 8.0           & -             \\
    HD149bs21   &        May 9, 2008    & 386       & 0.4           & 54080         & S18.7.0       & 4.5           & -             \\
    HD149bp43   &       May 11, 2008    & 499       & 0.4           & 70000         & S18.7.0       & 8.0           & K09           \\
    HD149bs43   &       May 12, 2008    & 432       & 0.4           & 60500         & S18.7.0       & 8.0           & K09           \\
    HD149bs32   &      June 16, 2008    & 386       & 0.4           & 54080         & S18.18.0      & 5.8           & -             \\
    HD149bs33   &     March 13, 2009    & 386       & 0.4           & 54080         & S18.18.0      & 5.8           & -             \\
    HD149bs44   &     March 22, 2009    & 386       & 0.4           & 54080         & S18.18.0      & 8.0           & -             \\
    \hline
\end{tabular}
\begin{minipage}[t]{0.75\linewidth}
\tablenotetext{1}{HD149b designates the planet, p/s specifies primary transit or secondary eclipse, and \#\# identifies the wavelength and observation number.}
\tablenotetext{2}{H07 = \citet{Harrington2007}, N09 = \citet{Nutzman2009}, C09 = \citet{Carter2009}, and K09 = \citet{Knutson2009b}.}
\end{minipage}
\end{table*}
\if\submitms y
\clearpage
\fi

\subsection{POET Pipeline}

Our Photometry for Orbits, Eclipses and Transits (POET) pipeline produces systematics-corrected light curves using {\em Spitzer}-supplied Basic Calibrated Data, fits a multitude of models with a wide range of analytic forms for systematic effects, chooses the best-fit model, and assesses the uncertainty of each free parameter.  Below, we describe each of these steps in detail.

We calculate the Julian date of each image at mid-exposure using the UTCS\_OBS and FRAMTIME keywords in the {\em Spitzer}-supplied headers.  Following \citet{Eastman2010}, we convert dates to Barycentric Julian Dates in the Coordinated Universal Time standard (BJD\sb{UTC}) using the JPL Horizons system\footnote{http://ssd.jpl.nasa.gov/?horizons} to interpolate {\em Spitzer's} position relative to our solar system's barycenter.  Additionally, converting from UTC to the Barycentric Dynamical Time (TDB) standard addresses any discontinuities due to leap seconds.  This paper reports both time standards to facilitate comparisons with previous work, which mostly does not apply the leap-second correction, and to ease the transition to the more-accurate standard.  

The POET pipeline flags bad pixels (from energetic particle hits and other causes) by grouping sets of 64 frames and doing a two-iteration, 4$\sigma$ rejection at each pixel location in the set.  Stellar centers for photometry come from a Gaussian fit and 5$\times$ interpolated aperture photometry (H07 Supplementary Information, SI) produces the light curves.  We test a broad range of aperture sizes in 0.25-pixel increments and omit frames with bad pixels in the photometry aperture.  The background, subtracted before photometry, is an average of the good pixels within the specified annulus centered on the star in each frame.

\subsection{{\em Spitzer} Systematics}

Exoplanet characterization requires photometric stability well beyond {\em Spitzer's} design criteria \citep{IRAC}.  Detector sensitivity models vary by channel and can have both temporal (detector ramp) and spatial (intrapixel variability) components.  The main systematic effect at 3.6 and 4.5 {\microns} is intrapixel sensitivity variations \citep{CharbonneauEtal2005apjTrES1}, in which the photometry depends on the precise location of the stellar center within its pixel.  We fit this systematic using the new BiLinearly-Interpolated Subpixel Sensitivity (BLISS) mapping technique described in Section \ref{sec:bilinint}.  This technique maps the spatial sensitivity variations at high resolution.

The 5.8, 8.0, and 16 {\micron} arrays primarily suffer from temporal variability, attributed to charge trapping (K09) in the 8.0 {\micron} case.  Weak spatial dependencies can also occur at these wavelengths \citep{Stevenson2010}, so we consider both systematics when determining our best-fit model.  Typically, we omit the initial portion of each light curve from the model fit to avoid the worst of the ramp effect and to allow the telescope pointing and detector to stabilize.  The clipping parameter, $q$, defines the number of unmodeled points from the start of a data set.

\subsection{Pixelation}
\label{subsec:pixelation}

\begin{figure}[ht]
\if\submitms y
  \setcounter{fignum}{\value{figure}}
  \addtocounter{fignum}{1}
  \newcommand\fignam{f\arabic{fignum}.ps}
\else
  \newcommand\fignam{./figs/pixelation2.eps}
\fi
\centering
\includegraphics[width=\linewidth, clip]{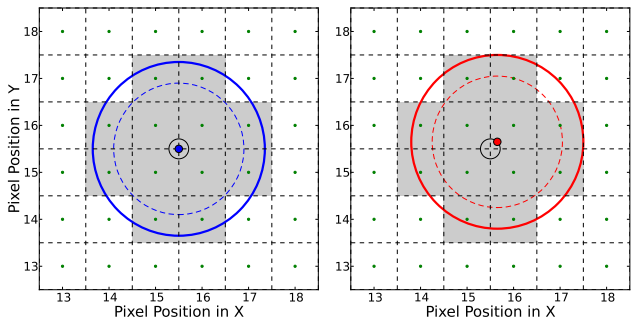}
\caption{Illustrative example of the pixelation effect, which arises when small changes in stellar centers do not add or subtract pixels within the aperture but do change the collected flux.
The left panel uses a solid blue circle to depict a photometry aperture centered at (15.5, 15.5), which is on the corner of four pixels (defined by dashed black lines).  The 12 shaded pixels have centers (green dots) that fall within the aperture and are summed to determine the stellar flux in this image.  All of the incoming photons that land within the dashed blue circle (where there is a greater density of incoming photons) count towards the flux.
So long as the stellar center falls within the small black circle, the photometry aperture will encompass the same green dots; thus, the specific pixels that contribute to the total flux will not change.  One such example is depicted in the right panel, where we apply an offset of (0.15, 0.15) pixels.  In this case, not all of the photons that fall within the dashed red circle count towards the flux.  Instead, the shaded pixels count additional flux from photons that land outside the aperture (solid red circle) where the density of photons is less.
The net effect is that, due to a change in centering and a non-uniform photon density, the shaded pixels in the right panel will record fewer incoming photons.  The result is a position-dependent systematic called pixelation.
}
\label{fig:pixelation}
\end{figure}

\begin{figure}[ht]
\if\submitms y
  \setcounter{fignum}{\value{figure}}
  \addtocounter{fignum}{1}
  \newcommand\fignam{f\arabic{fignum}.ps}
\else
  \newcommand\fignam{./figs/hd149bs31-fig906.ps}
\fi
\centering
\includegraphics[width=\linewidth, clip]{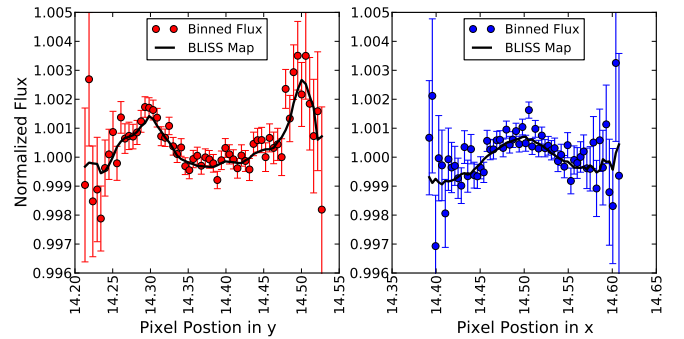}
\caption{Projected flux from HD149bs31 integrated along the $x$ (left) and $y$ (right) axes.  The non-uniform flux in both panels is clear evidence of pixelation.  We use 5$\times$-interpolated aperture photometry, which results in 0.2-pixel spacing between peaks.  In the left panel, low-order polynomial models would fit the pixelation effect poorly at the peaks; the BLISS map (see Section \ref{sec:bilinint}) has no such limitation.  The systematic is weakly constrained near the edges due to low sampling, as indicated by the large error bars.  Whether a specific point on a pixel is a local maximum or minimum (due to pixelation) is a function of aperture size, which defines which subpixels to include in the aperture at any given point on the detector.
}
\label{fig:hd149bs31-ip}
\end{figure}

Pixelation is an infrequently discussed systematic error inherent to all array detectors.  Sufficiently small stellar center motions between frames will not add or subtract pixels from an aperture, but these motions will cause the total flux within the aperture to vary. This means there is a (potentially large) range of stellar centers that utilizes the same set of aperture pixels, introducing a position dependence to the photometric sensitivity.
We provide an illustrative example in Figure \ref{fig:pixelation} and display the magnitude of this effect in Figure \ref{fig:hd149bs31-ip}. 
A flux-conserving, subpixel image interpolation, combined with precise centering and applied before photometry, mitigates the pixelation effect by decreasing its range and amplitude.  As demonstrated by our BLISS maps in Section \ref{sec:bilinint}, uninterpolated photometry exhibits strong sensitivity peaks at one-pixel increments, while 2$\times$- and 5$\times$-interpolated pixels exhibit progressively weaker peaks at 0.5 and 0.2 pixel increments, respectively, in each spatial direction.  

\if\submitms y
\clearpage
\fi
\if\submitms y
\clearpage
\fi

\if\submitms y
\clearpage
\fi
\if\submitms y
\clearpage
\fi

Pixelation is most apparent with small apertures placed on under-resolved point-response functions (PRFs) such as {\em Spitzer's}.  It may not be apparent in other situations, such as when the aperture contains almost all of the integrated PRF, when the centroid wander is small relative to the subpixel size, and when other noise sources dominate (e.g., systematics and variable PRFs).
Increasing the aperture size lessens the pixelation effect by decreasing the fraction of uncaptured light outside the aperture, but doing so may decrease the signal-to-noise ratio (S/N) by increasing the amount of background noise included in the aperture.  Thus, choosing the best aperture size may introduce this position-dependent systematic.  We have determined the intrapixel variability at 5.8, 8.0, and 16 {\microns} to be a pixelation effect, previously reported at 5.8 {\microns} by \citet{Stevenson2010} and at 8.0 {\microns} by \citet{Anderson2011}.  

There are several ways to correct pixelation.  First, one could shift model PRFs to match each frame's precisely determined stellar center.  Dividing the stellar flux by the PRF flux in the aperture should remove the effect, but requires a highly accurate model PRF.  Second, using smaller subpixels could decrease the amplitude of the pixelation effect until it is insignificant relative to the noise, but there is a limit: interpolation can only approximate the information destroyed when photons fall into the detector's finite-sized pixels.  Third, if the pointing is sufficiently consistent and compact, one could choose an interpolation factor that happens to place the flat portion of the pixelation response on the stellar centers (since the peaks in Figure \ref{fig:hd149bs31-ip} move with different interpolation factors).  Fourth, with high-precision centering, one could use a series of images taken at slightly different positions to model the position sensitivity analytically or with pixel-mapping techniques such as BLISS, but one would first have to remove any time-dependent components from the light curve model (see Section \ref{subsec:lcmodel}).
The accuracy of these models depends on the centering and photometric precisions, the former being \sim0.01 pixels for 0.4 second IRAC subarray exposures of bright sources (see below and \citealp{Stevenson2010}).  We test for pixelation in all data sets using BLISS mapping, which corrects the effect when it is significant.  

\subsection{Light-curve Modeling}
\label{subsec:lcmodel}

The full light-curve model is:

\begin{eqnarray}
\label{eqn:full}
F(x, y, t) = F\sb{\rm s}E(t)R(t)M(x,y)V(\upsilon)P(p),
\end{eqnarray}

\noindent where \math{F(x, y, t)} is the measured flux centered at position $(x,y)$ on the detector at time $t$, \math{F\sb{\rm s}} is the (constant) system flux outside of secondary eclipse or primary transit, \math{E(t)} is the primary-transit or secondary-eclipse model component, \math{R(t)} is the time-dependent ramp model component, \math{M(x,y)} is the position-dependent intrapixel model component or sensitivity map, \math{V(\upsilon)} is the visit sensitivity as a function of visit frame number $\upsilon$, and \math{P(p)} is the flat-field correction at position $p$.  Below, we discuss some of these components in more detail.

\subsubsection{Eclipse and Transit Models}
\label{subsec:eclmodels}

The uniform-source and small-planet equations from \citet{MandelAgol2002ApJtransits} describe the secondary-eclipse and primary-transit model components, \math{E(t)}, respectively.  Transit light curves at 8.0 {\microns} exhibit weak limb darkening that is not well constrained by fitting limb-darkening models to the data.  We follow the method of \citet{Beaulieu2008} in deriving limb-darkening coefficients for HD 149026.  {\em Spitzer's} 8.0 {\micron} spectral response curve weights the intensities of a Kurucz ATLAS stellar atmosphere model \citep[\math{T\sb{\rm eff} = 6250} K, \math{\log(g) = 4.5} cgs, and \math{[M/H] = 0.3}]{Kurucz2004}, given as a function of wavelength and angle from the star's center.  A least-squares minimization of the resulting curve determines the non-linear limb-darkening coefficients \citep[$a_1$-$a_4$ = 0.51477, -0.80525, 0.75683, -2.6168]{Claret2000}, which are then fixed for the three transit light-curve fits.

\subsubsection{Ramp Models}
\label{subsec:rampmodels}

We consider a multitude of ramp equations, \math{R(t)}, all of which stem from three basic forms: exponential, logarithmic, and/or polynomial.


\begin{equation}
\label{eqnse}
R(t) = 1 {\pm} e\sp{-r\sb{0}t + r\sb{1}}
\end{equation}
\begin{equation}
\label{eqnsel}
R(t) = 1 {\pm} e\sp{-r\sb{0}t + r\sb{1}} + r\sb{2}(t-0.5)
\end{equation}
\begin{equation}
\label{eqnseq}
R(t) = 1 {\pm} e\sp{-r\sb{0}t + r\sb{1}} + r\sb{2}(t-0.5) + r\sb{3}(t-0.5)\sp{2}
\end{equation}
\begin{equation}
\label{eqnse2}
R(t) = 1 {\pm} e\sp{-r\sb{0}t + r\sb{1}} {\pm} e\sp{-r\sb{4}t + r\sb{5}}
\end{equation}

\begin{equation}
\label{eqnll}
R(t) = 1 + r\sb{1}(t-0.5) + r\sb{6} \ln(t-t\sb{0})
\end{equation}
\begin{equation}
\label{eqnlq}
R(t) = 1 + r\sb{1}(t-0.5) + r\sb{2}(t-0.5)\sp{2} + r\sb{6} \ln(t-t\sb{0})
\end{equation}
\begin{equation}
\label{eqnlog}
R(t) = 1 + r\sb{6} \ln(t-t\sb{0}) + r\sb{7} [\ln(t-t\sb{0})]\sp{2}
\end{equation}
\begin{equation}
\label{eqnl4q}
R(t) = 1 + r\sb{1}(t-0.5) + r\sb{2}(t-0.5)\sp{2} + r\sb{6} \ln(t-t\sb{0}) + r\sb{7} [\ln(t-t\sb{0})]\sp{2}
\end{equation}

\begin{equation}
\label{eqnlin}
R(t) = 1 + r\sb{1}(t-0.5)
\end{equation}
\begin{equation}
\label{eqnquad}
R(t) = 1 + r\sb{1}(t-0.5) + r\sb{2}(t-0.5)\sp{2}
\end{equation}

\noindent The time, \math{t}, is in units of phase (days) for secondary-eclipse (primary-transit) events.  We use ``+'' and ``-'' subscripts in Eqs.\ \ref{eqnse} - \ref{eqnse2} to denote the corresponding functional form.  For example, Eq.\ \ref{eqnse}\sb{+} describes an exponentially decreasing, asymptotically constant ramp while Eq.\ \ref{eqnse}\sb{--} describes an exponentially increasing, asymptotically constant ramp.  

There is a physical interpretation applicable to the rising
exponential ramps \citep{Agol2010}.  Consider a population of charge traps due to an
impurity in the detector's infrared material and a flux of
photoelectrons through the material.  The traps collect some fraction
of electrons, releasing them randomly with some characteristic time
scale that depends on the impurity.  
Bright sources saturate the traps, decreasing the fraction of
captured electrons and raising the detected signal of a steady source
according to the asymptotic rising exponential function in Eq.\ \ref{eqnse}\sb{--}.  
A double rising exponential (Eq.\ \ref{eqnse2}\sb{--}) approximates a rapidly saturating PSF core and slowly saturating wings \citep{Knutson2008}.  
It could also represent two impurities;
the very short ramps of the HD149bs41 data set's visit sensitivity suggests the presence of an impurity that has a very short time scale and releases many of its electrons over the course of one cycle ($\sim$30 minutes, see H07 Supplementary Figure 6).  We tested for a common set of characteristic time scales in all of the rising exponential free parameters; however, even for the same planet in the same array, we did not achieve reasonable fits with all data sets.  This may be due to inadvertent and inconsistent pre-flashing by the objects observed prior to our own observations.
Despite its potential for providing a physical
explanation for the ramps, the rising exponential does not always
provide the best model according to the criteria in Section \ref{subsec:modelselect}.  This
may be due to our fitting the final photometry and not the individual
pixels' data and/or to the pointing instability producing unsteady
illumination at the precision levels relevant here.

\citet{Agol2010} advocate using a double rising exponential (Eq.\ \ref{eqnse2}\sb{--}) for 8.0 {\micron} data due to its improved fit and smaller residuals, weaker dependence on aperture size, and less sensitivity in the clipping parameter.  Similarly, \citet{Knutson2011} find that Eq.\ \ref{eqnse}\sb{--} is sufficient for their 8.0 {\micron} preflashed data sets, while Eq.\ \ref{eqnse2}\sb{--} is necessary for their non-preflashed data.  
We test all relevant ramp equations on data sets that exhibit time-dependent systematics and orthogonalize any correlated parameters that inhibit convergence (see Sections \ref{subsec:ortho} and \ref{subsec:errest}).  Upon doing so, we find that we cannot corroborate the claims by \citet{Agol2010}.  Equation \ref{eqnse2}\sb{--} should not be used for all 8.0 {\micron} data sets because we typically find correlations between the eclipse depth and its ramps parameters that can double the latter's uncertainty relative to other models (see HD149bp42).  Rather, we recommend a comprehensive examination of all relevant ramp equations before selecting the final model.  See Sections \ref{sec:tranresults} and \ref{sec:eclresults} for discussion relevant to particular events.

\subsubsection{Orthogonalization}
\label{subsec:ortho}

The exponential model components have one difficulty: the Markov-Chain method used to assess the uncertainties does not converge to the posterior probability distribution, according to the criteria discussed in Section \ref{subsec:errest}, even after tens of millions of iterations.  The problem is a correlation between the exponential ramp parameters.  Several recent analyses \citep[e.g., K09,][]{Stevenson2010, Campo2011} have ``solved'' the problem by fixing one of the exponential parameters.  
This is effectively profiling (slicing) rather than marginalizing (integrating) over the posterior parameter distribution, an approach that ignores correlated errors and may reduce the calculated error bar (see
\citealp{NumRecipes}, Figure 15.6.3 and related text).  In one case discussed below (HD149bp42), fixing parameters incorrectly decreased the calculated eclipse depth uncertainty by 50\%.
There are two legitimate escapes from the problem.  The first is to write a Monte Carlo sampler that explores the phase space more intelligently than the Metropolis random walk we and many others use.  The second is to re-cast the equations in a form that eliminates the non-linear correlation of the ramp parameters.  

For this paper, we have expediently chosen the second approach.  The version of Eq.\ \ref{eqnse}\sb{--} presented here produces a more linear correlation between the \math{r\sb{0}} and \math{r\sb{1}} parameters than the original version in H07, whose exponent was \math{-r(t-t\sb{0})}.  In some cases this converges quickly and the job is done.  
In others it still does not converge, so we run at least $10^5$
iterations to sample the posterior distribution and then rewrite the
model with a change of variables that orthogonalizes the most
correlated parameters.  
This method does not modify the number of free parameters, nor involve any interpolations or approximations.
For the selected correlated parameters, the
orthogonalization shifts the origin to the center of the distribution,
divides each parameter by its standard deviation to give a uniform
scale in all directions, and rotates the subspace to minimize
correlations  (see Figure \ref{fig:hd149bs21-ortho}).  
A routine that prepares for principal components
analysis (PCA) finds the transformation matrix from the distribution sample (we do not actually perform the PCA).
We rerun the Markov chain using the new equations until it converges
according to the criteria given below.  Then we transform the points
back to the familiar equations to assess parameter uncertainties.  A simple
example of a two-parameter orthogonalization of Eq.\ \ref{eqnse} is:

\begin{equation}
\label{eqnse+theta}
R(t) = 1 {\pm} e\sp{-c\sb{0}[t\cos(\theta)-\sin(\theta)]}e\sp{c\sb{1}[t\sin(\theta)+\cos(\theta)]},
\end{equation}

\noindent where $c\sb{0}$ and $c\sb{1}$ are the ramp parameters in the rotated frame and \math{\theta} is the rotation angle between coordinate systems.  The arctangent of the slope of the best-fit line through the initial sample, projected into the \math{r\sb{0}-r\sb{1}} plane, determines $\theta$.  
Such approximate orthogonalization of the posterior distribution has long been standard practice for improving MCMC performance \citep[e.g., ][]{Hills1992, Brooks1998}.  
Similar discussion can be found in \citet{Connolly1995}, \citet{Pal2008}, and \citet{Cowan2009}.

In effect, this method is the same as the first, accomplished by rotating the data and using the original sampler rather than writing a new one.  This method works best for linearly-correlated parameters and achieves moderate success with more exotic correlations.  Converting to curvilinear coordinates or implementing manifold learning algorithms from nonlinear dimensionality reduction may offer further improvements in convergence time for the extreme cases \citep{Lee2007}.


\comment{
\if\submitms y
\clearpage
\fi
\begin{figure}[ht]
\if\submitms y
  \setcounter{fignum}{\value{figure}}
  \newcommand\fignam{f\arabic{fignum}.ps}
\else
\fi
\centering
\includegraphics[width=\linewidth, clip]{\fignam}
\label{fig:hd149bs21-ortho}
\end{figure}
\if\submitms y
\clearpage
\fi
}

\if\submitms y
\clearpage
\fi
\begin{figure}[ht]
\if\submitms y
  \setcounter{fignum}{\value{figure}}
  \addtocounter{fignum}{1}
  \newcommand\fignama{f\arabic{fignum}a.ps}
  \newcommand\fignamb{f\arabic{fignum}b.ps}
\else
  \newcommand\fignama{./figs/hd149bs21-OrthoHist1.ps}
  \newcommand\fignamb{./figs/hd149bs21-OrthoHist2.ps}
\fi
\centering
\includegraphics[width=\linewidth, clip]{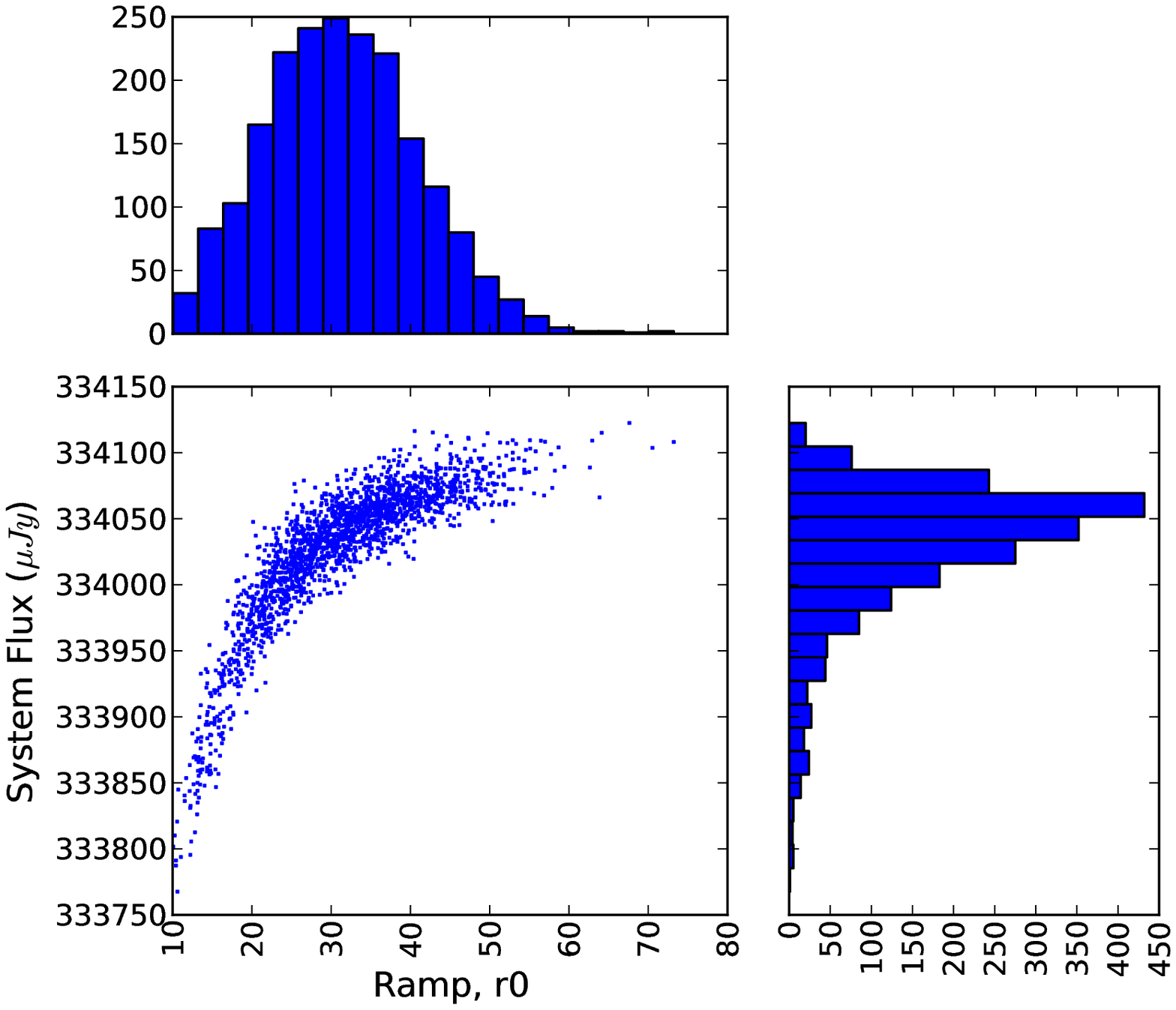}
\includegraphics[width=\linewidth, clip]{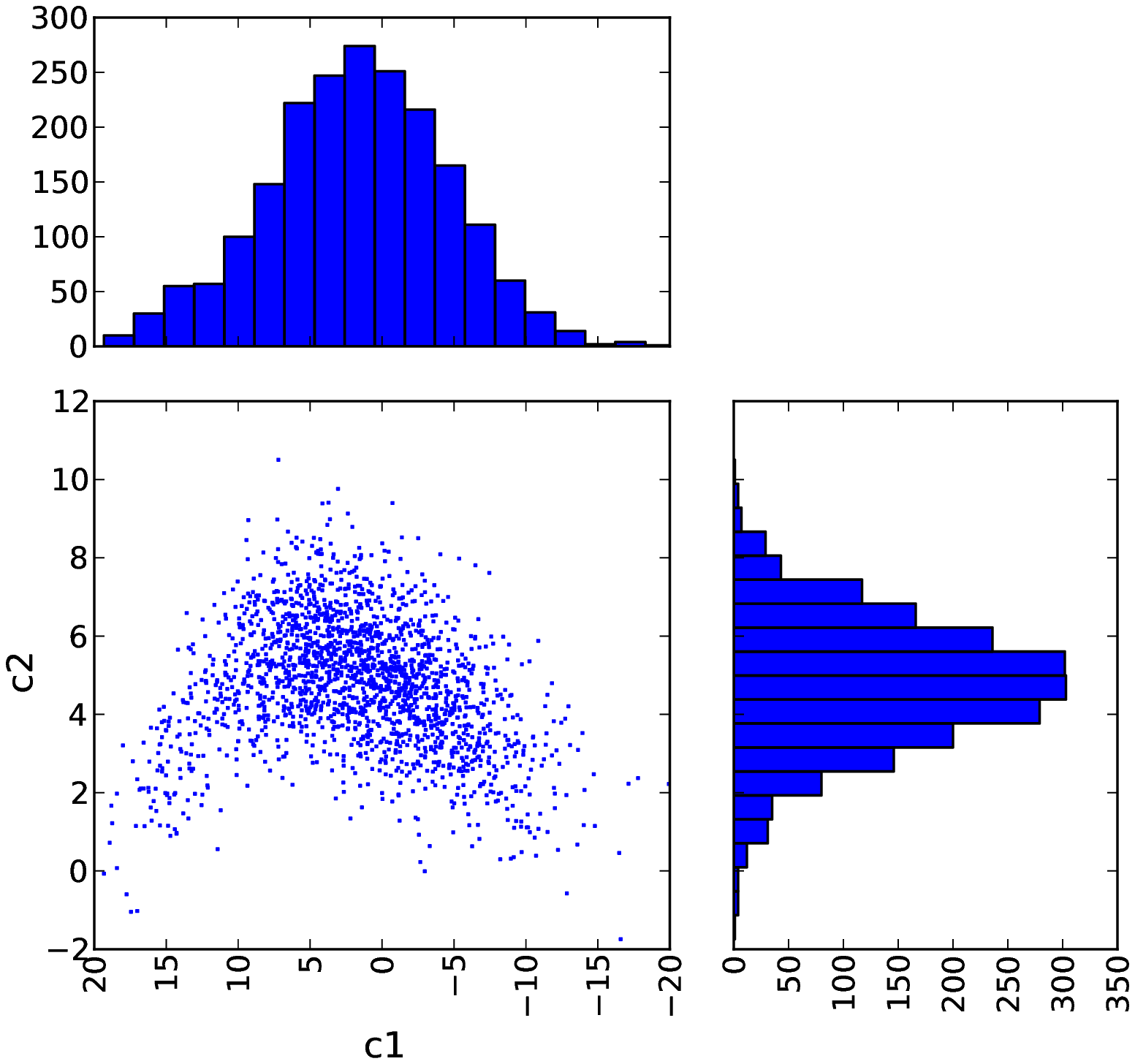}
\caption{Two-parameter orthogonalization example for HD149bs21 with histograms.  The physical parameters (top panels) show a strong, non-linear correlation and asymmetric histograms; however, the orthogonalized parameters (bottom panels) are nearly uncorrelated and have symmetric histograms.  Running a Markov-chain method with the orthogonalized parameters reduces the convergence time.
}
\label{fig:hd149bs21-ortho}
\end{figure}
\if\submitms y
\clearpage
\fi

\subsubsection{Flat Field (Position) Sensitivity Models}
\label{subsec:posmodels}

Most of the data sets presented here follow the standard time-series observing practice of keeping the object fixed to one location on the array (staring) to minimize position-dependent sensitivity effects.  However, the H07 observation (HD149bs41) cycled through nine different nod positions ($p=0-8$) in an attempt to use the unobserved positions in each frame to make a high-quality flat field that would correct the entire data set.  This approach was unsuccessful and not repeated.  Each position in the HD149bs41 data requires a flat-field correction, \math{P(p)}, to account for the difference in pixel sensitivity. To eliminate correlations with the system flux (i.e., keep the
correction from floating), we require the mean of all of the corrections to equal unity.  We do this by freely varying \math{p=1-8} and equating $p=0$ to the number of positions minus the sum of the other corrections.  

For both 16 {\micron} data sets, the telescope reacquires the target at one third and two thirds of the way into the observing runs.  This action is similar to a nod because after reacquisition, the three sets of measured stellar centers are non-overlapping.  We apply the same model component to these data as with HD149bs41, but only use three flat-field correction parameters (\math{p=0,1,2}).

\subsubsection{Visit Sensitivity Model}
\label{subsec:vsmodels}

With HD149bs41, {\em Spitzer} completed twelve cycles through the nine nod positions mentioned above for a total of 108 visits.  Each visit has a briefer and steeper ramp compared to the overall ramp, \math{R(t)}.  As discussed in their SI, H07 use a 12-knot spline to model the visit-sensitivity effect, \math{V(\upsilon)}.  They fix the final three knots to unity and allow the remaining nine parameters to vary.  We model the visit sensitivity ramp using:
\begin{equation}
\label{eqnvs}
V(\upsilon) = \upsilon\sb{1}\cdot \ln(\upsilon-\upsilon\sb{0}) + 1,
\end{equation}
\noindent which is identical in form to the model component used by K09.  The only difference is that K09 use time as the independent variable while we use visit frame number, \math{\upsilon}.  In either case, the independent variable resets to zero upon moving to a new nod position.

\subsection{Model Selection}
\label{subsec:modelselect}

For each data set, we test different photometry aperture sizes, detector ramp model components, and intrapixel sensitivity model components looking for the best combination. One must be careful in assessing which model is the ``best'' because \math{\chi^2}-like comparisons must derive from the same data set and different photometric apertures produce different data sets for this purpose.  We use the standard deviation of the normalized residuals (SDNR) when comparing models of the same analytic form to different data sets. Once we have identified the best aperture size, we use the Bayesian Information Criterion \citep[BIC,][]{Schwarz1978, Liddle2008, Stevenson2010, Campo2011} to compare models with different numbers of free parameters:

\begin{equation}
\label{eqnbic}
{\rm BIC} = \sum\limits\sb{1<i<n} \frac{\epsilon\sb{i}\sp{2}}{\sigma\sb{i}\sp{2}} + k\ln n.
\end{equation}

\noindent Here, \math{\epsilon\sb{i}} and \math{\sigma\sb{i}} are the residual and uncertainty of the $i$\sp{th} data point, \math{k} is the total number of free parameters, and \math{n} is the number of data points used in the model fit.  
The {\em Spitzer}-supplied uncertainty frames are typically overestimated, so we scale \math{\sigma\sb{i}} such that the reduced \math{\chi^2}, \math{\chi_{\nu}^{2}}, equals unity for our best-fit model.  Using the unscaled \math{\sigma\sb{i}} values in a least-squares minimizer improperly weights the data and results in a sub-optimal fit.
When selecting between competing models for an observed dataset, what matters is how the models compare in predicting the actually observed data, not how variable individual model fits may be among hypothetical realizations of the data. The BIC is defined so that the marginal likelihood ratio---the ratio of predictive probabilities for the observed data---is approximately $e^{0.5 \Delta BIC}$. Note that what matters is the {\em difference} in BIC values (and thus the difference of maximum likelihood or minimum $\chi^2$ values), not the absolute values.
Information criteria such as BIC may not apply when comparing fits with intrapixel maps to those with polynomial model components.  See Appendix \ref{sec:tom} for more discussion on the matter.


\subsection{Error Estimation}
\label{subsec:errest}

We explore phase space and estimate uncertainties using a Markov-chain Monte Carlo (MCMC) routine following the Metropolis random walk algorithm.  See \citet{Campo2011} for more discussion on MCMC.  The POET pipeline can test any combination of systematic model components, computing the SDNR and BIC for each fit, and can model multiple events simultaneously while sharing parameters such as the eclipse midpoint and depth.  Each MCMC run begins with a least-squares minimization, a rescaling of the {\em Spitzer}-supplied uncertainties, a second least-squares minimization using the new uncertainties and, finally, at least $10^5$ MCMC iterations to populate the posterior parameter distributions, from which we derive parameter uncertainties.  We study parameter correlation plots and the posterior distribution to ensure that we have a reliable result, then publish them so that others may evaluate our work and compare to their own.  We test for convergence every $10^5$ steps, terminating only when the \citet{Gelman1992} diagnostic for all free parameters has dropped to within 1\% of unity using all four quarters of the chain.
We also examine trace and autocorrelation plots of each parameter to confirm convergence visually.  We estimate the effective sample size \citep[ESS,][]{Kass1998} and autocorrelation time, $\tau$, for the $i$\sp{th} free parameter as follows:

\begin{equation}
\label{eqness}
{\rm ESS\sb{i}} = \frac{m}{\tau\sb{i}} = \frac{m}{1 + \sum\limits\sb{k=1}\sp{k\sp{\prime}} \rho\sb{k}(\theta\sb{i})},
\end{equation}

\noindent where $m$ is the length of the MCMC chain, $\rho\sb{k}(\theta\sb{i})$ is the autocorrelation of lag $k$ for the free parameter $\theta\sb{i}$, and $k\sp{\prime}$ is the cutoff point such that $\rho\sb{k} < 0.01$.  
We use the longest autocorrelation time from each event to determine the number of steps between effectively independent samples for thinning each MCMC chain.  The distribution of thinned samples quantifies parameter uncertainties.

\section{BLISS MAPPING TECHNIQUE}
\label{sec:bilinint}

\subsection{Background}

The change in pixel sensitivity with respect to stellar position on the detector is a well known systematic with the {\em Spitzer Space Telescope} \citep{CharbonneauEtal2005apjTrES1, Knutson2008}.  This effect is particularly strong in IRAC's 3.6 and 4.5 {\micron} channels but has also been seen at 5.8, 8.0, and 16 {\microns} \citep{Stevenson2010, Anderson2011}.  The position sensitivity in the latter channels is due to a pixelation effect (see Section \ref{sec:obs} for a description) rather than an actual change in sensitivity over the pixel surface.  In this section, we present a new technique for modeling these position-dependent systematics, called BiLinearly-Interpolated Subpixel Sensitivity (BLISS) mapping.

The most common method for removing the intrapixel variability is to fit a polynomial in both spatial directions \citep{Knutson2008}.  The polynomial order typically ranges from quadratic to sextic and may include cross terms.  Other variations of this modeling method have applied multiple polynomials, one for each pixel quadrant, in order to find the best fit.  Polynomial methods work reasonably well for data sets with small stellar position wander, resulting in a smoothly varying intrapixel sensitivity.
An analysis becomes exceedingly complicated if the variation is not smooth or if strong correlations arise between parameters.  These complications can increase uncertainty estimates in the best case scenario and, in the worst case, lead to incorrect results.

A new approach, pioneered by \citet[hereafter B10]{Ballard2010b}, attempts to map the intrapixel variability on a subpixel-scale grid without assuming a specific functional form.  For their particular light curve, they bin their flux and stellar positions into 20-second bins (\sim145 points/bin) before computing a sensitivity correction for each binned point.  Each correction considers a set of flux values that does not include in-transit frames or frames from the current binned position.  This set of flux values is Gaussian-weighted in both spatial directions relative to the position of the binned point being corrected, summed to a single value, then normalized by dividing by the summed Gaussian weighting function.  This method effectively models not only the large-scale features in their data, but also a smaller-scale ``corrugation'' effect that a low-order polynomial cannot remove.

\begin{figure}[ht]
\if\submitms y
  \setcounter{fignum}{\value{figure}}
  \addtocounter{fignum}{1}
  \newcommand\fignam{f\arabic{fignum}.ps}
\else
  \newcommand\fignam{./figs/ypos.ps}
\fi
\centering
\includegraphics[width=\linewidth, clip]{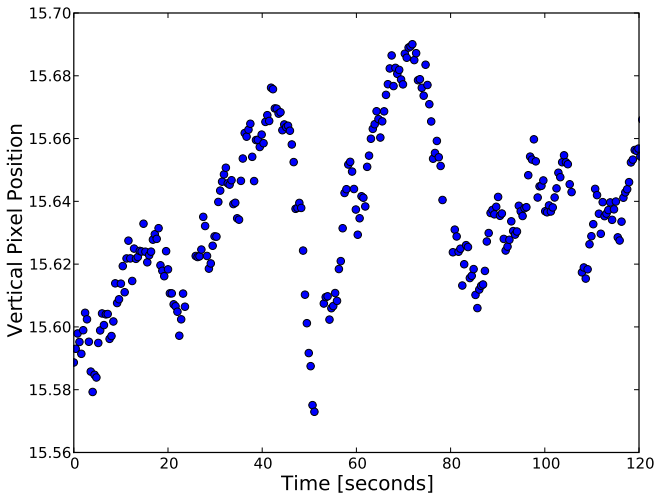}
\caption{Vertical pixel positions from HD149bs11.  We can track the motion of the stellar center to high precision ($\sim$0.01 pixels) over a duration of only a few seconds.  These oscillations are much faster than {\em Spitzer's} 1/2--1 hour oscillations, initially reported by \citet{CharbonneauEtal2005apjTrES1}.  In a span of 20 seconds, the stellar centers can vary by > 0.1 pixels.  All positions are zero-based.}
\label{fig:yPos}
\end{figure}

Moving away from a polynomial model is an excellent concept; however, this particular implementation has some drawbacks that limit its scope.  For instance, ignoring the points during secondary eclipse requires that the out-of-eclipse portion of the data set be significantly longer than the in-eclipse portion, which is atypical in primary-transit and secondary-eclipse observations.  Also, the weighting function computes too slowly to be used in a MCMC routine and is even slow when using a minimizer.  The calculation would be even slower if the data were not binned into relatively long, 20-second time intervals.  Figure \ref{fig:yPos} shows 120 seconds of vertical pixel positions from HD149bs11, which has a 0.4-second exposure duration \vs 0.1 seconds for B10.  The stellar center can vary significantly over a 20-second interval, indicating that positional information is lost when binning over such time scales.

\if\submitms y
\clearpage
\fi
\if\submitms y
\clearpage
\fi

\begin{figure}[tb]
\if\submitms y
  \setcounter{fignum}{\value{figure}}
  \addtocounter{fignum}{1}
  \newcommand\fignama{f\arabic{fignum}a.ps}
  \newcommand\fignamb{f\arabic{fignum}b.ps}
  \newcommand\fignamc{f\arabic{fignum}c.ps}
\else
  \newcommand\fignama{./figs/gj436bo41-BLISS-ap2500715-1x.ps}
  \newcommand\fignamb{./figs/gj436bo41-BLISS-ap2500715-2x.ps}
  \newcommand\fignamc{./figs/gj436bo41-BLISS-ap2500715-5x.ps}
\fi
\centering
\includegraphics[height=5.8cm, clip]{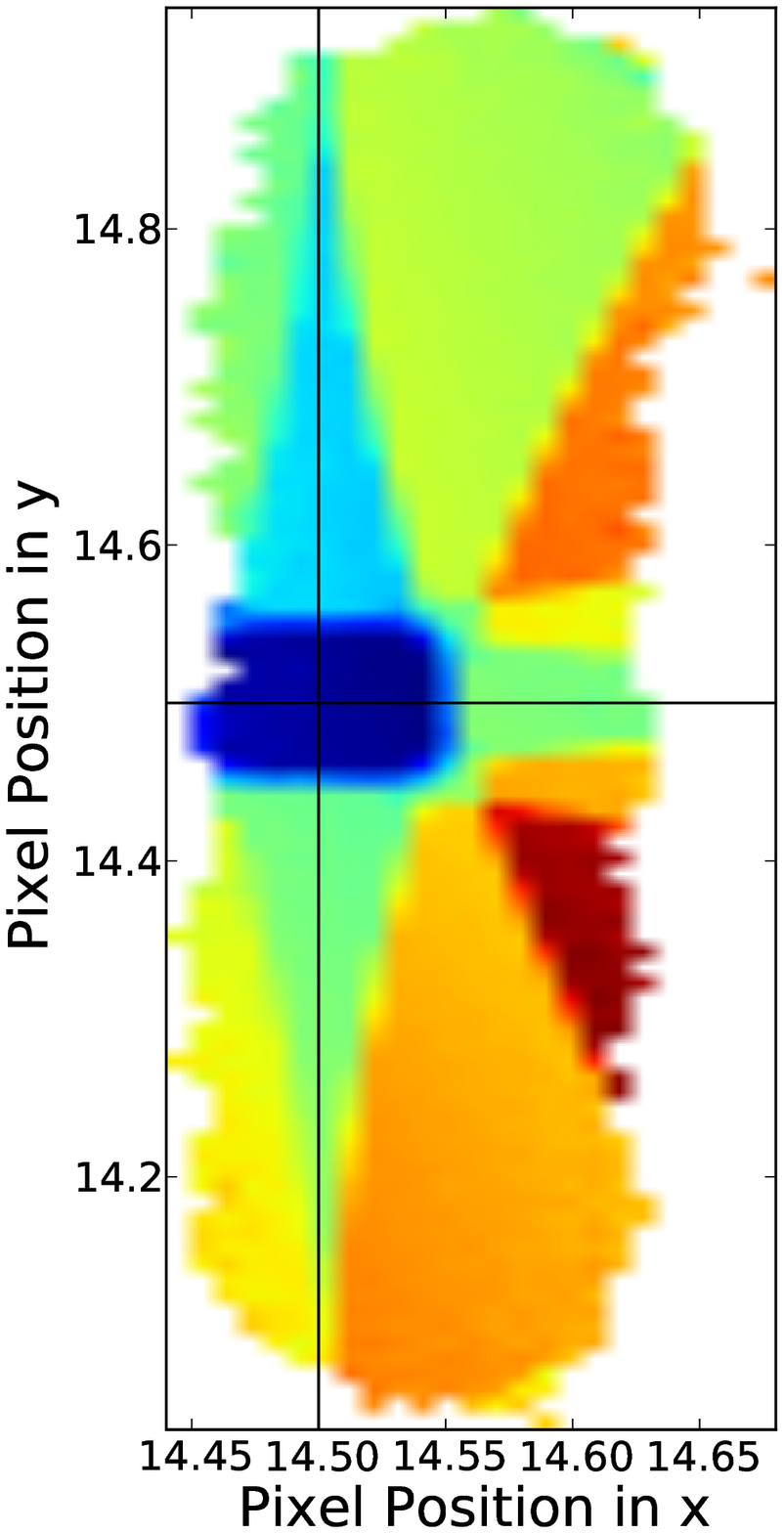}
\includegraphics[height=5.8cm, clip]{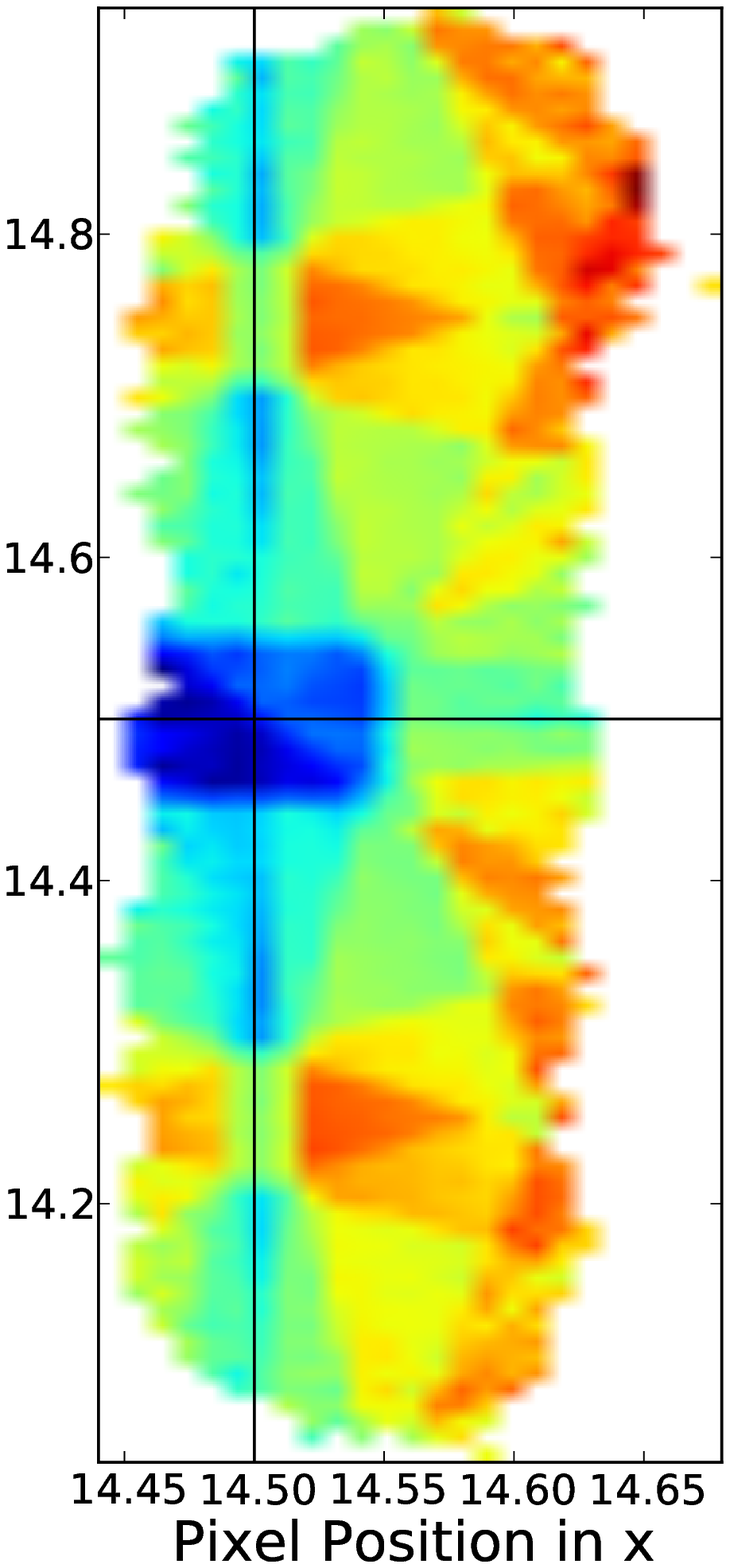}
\includegraphics[height=5.8cm, clip]{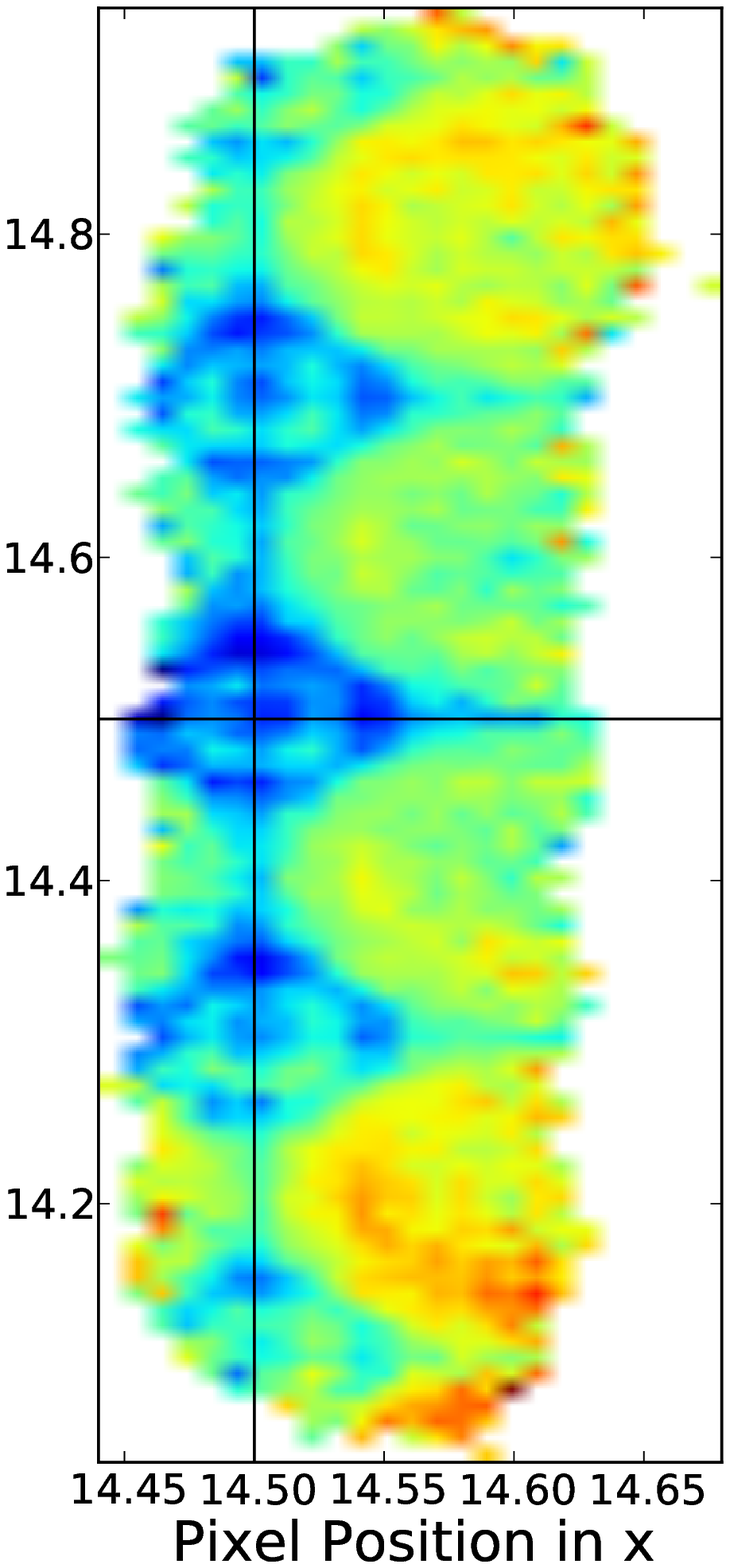}
\caption{BLISS maps illustrating the position-dependent pixelation effect.  The maps use 1{\by}- (left), 2{\by}- (center), and 5{\by}- (right) interpolated, 2.5 pixel-aperture photometry.  Redder (bluer) colors indicate more (less) flux within the aperture.  The bin size is 0.01 pixels for all maps.  The horizontal and vertical black lines depict pixel boundaries.  Without subpixel interpolation, the pixelation effect is significant, but it is progressively reduced with 2{\by}- and 5{\by}-interpolated photometry.  For time-series data such as these, one can calculate a BLISS map to correct for pixelation.}
\label{fig:gj436bo41-BLISS}
\end{figure}

The idea of using a spline to interpolate position-dependent systematics stems from observed fine-scale sensitivity variations in some of our data sets that cannot be modeled by a low-order polynomial (see the pixelation effect in Figures \ref{fig:hd149bs31-ip} and \ref{fig:gj436bo41-BLISS} for examples).  We initially attempted to map the pixel surface using a bicubic spline because we wanted a smoothly-varying model; however, this type of interpolation is prohibitively computationally expensive.  A typical secondary-eclipse observation spans a 0.3 {\by} 0.3 pixel region.  Placing knots at 0.05-pixel intervals requires 49 free parameters.  Polynomial models typically require less than 10 parameters.  Adequately describing the fine-scale sensitivity variations requires a large number of knots, but varying all of these knot parameters at each step of an MCMC routine leads to extremely slow convergence.  Our new BLISS mapping technique circumvents the problem of slow convergence by directly computing the knot parameters, rather then allowing them to vary freely.  Thus, we can use $>$1000 knots to map the pixel surface at high resolution (see Figure \ref{fig:gj436bo41-BLISS}). 

\if\submitms y
\clearpage
\fi
\if\submitms y
\clearpage
\fi


\subsection{Implementation}



In seeking methods that compute faster than bicubic interpolation, we give up the constraint of differentiability.  Nearest-neighbor interpolation (NNI) is simple, assigning each point the value of its nearest knot.  Bilinear interpolation (BLI) is a straightforward calculation (see Eq.\ \ref{eqnBLI}) that maintains continuity over the pixel surface, unlike NNI, and, given sufficiently precise centering, should be more accurate.  Using BLI, the flux at a point (\math{x,y}) is:

\begin{equation}
\label{eqnBLI}
\begin{split}
    M(x,y) = F\sb{\rm IP}(x\sb{1},y\sb{1}) \frac{x\sb{2}-x}{x\sb{2}-x\sb{1}} \frac{y\sb{2}-y}{y\sb{2}-y\sb{1}} \\
         +\; F\sb{\rm IP}(x\sb{2},y\sb{1}) \frac{x-x\sb{1}}{x\sb{2}-x\sb{1}} \frac{y\sb{2}-y}{y\sb{2}-y\sb{1}} \\
         +\; F\sb{\rm IP}(x\sb{1},y\sb{2}) \frac{x\sb{2}-x}{x\sb{2}-x\sb{1}} \frac{y-y\sb{1}}{y\sb{2}-y\sb{1}} \\
         +\; F\sb{\rm IP}(x\sb{2},y\sb{2}) \frac{x-x\sb{1}}{x\sb{2}-x\sb{1}} \frac{y-y\sb{1}}{y\sb{2}-y\sb{1}}.
\end{split}
\end{equation}

\noindent 
This is a distance-weighted average of the flux of the four nearest knots, \math{F\sb{\rm IP}(x\sb{i},y\sb{j})}, where $i$ and $j$ are horizontal and vertical indices for a rectangular grid of knots.  This method computes faster than bicubic interpolation and may achieve comparable smoothness within the errors with less computing time simply by increasing the number of knots (see Figure \ref{fig:hd149bs31-ip}).

We create a rectangular grid of knots that spans the range of centers in $x$ and $y$.  Each point in the data set associates with its nearest knot.   For BLI, we compute the distances from each point to its four nearest knots, for Eq.\ \ref{eqnBLI}.  If one or more knots in Eq.\ \ref{eqnBLI} does not have any assigned points, we use NNI there instead, or the calculation would fail.  This usually only occurs near the boundary of the grid of knots.  We precompute the knot associations and distances prior to initiating the MCMC as they remain constant from iteration to iteration.

\begin{figure}[ht]
\if\submitms y
  \setcounter{fignum}{\value{figure}}
  \addtocounter{fignum}{1}
  \newcommand\fignama{f\arabic{fignum}a.ps}
  \newcommand\fignamb{f\arabic{fignum}b.ps}
  \centering
  \includegraphics[width=0.5\linewidth, clip]{\fignama}
  \includegraphics[width=0.5\linewidth, clip]{\fignamb}
\else
  \newcommand\fignama{./figs/hd149bs11-BLISS.ps}
  \newcommand\fignamb{./figs/hd149bs11-PointingHist.ps}
  \centering
  \includegraphics[width=\linewidth, clip]{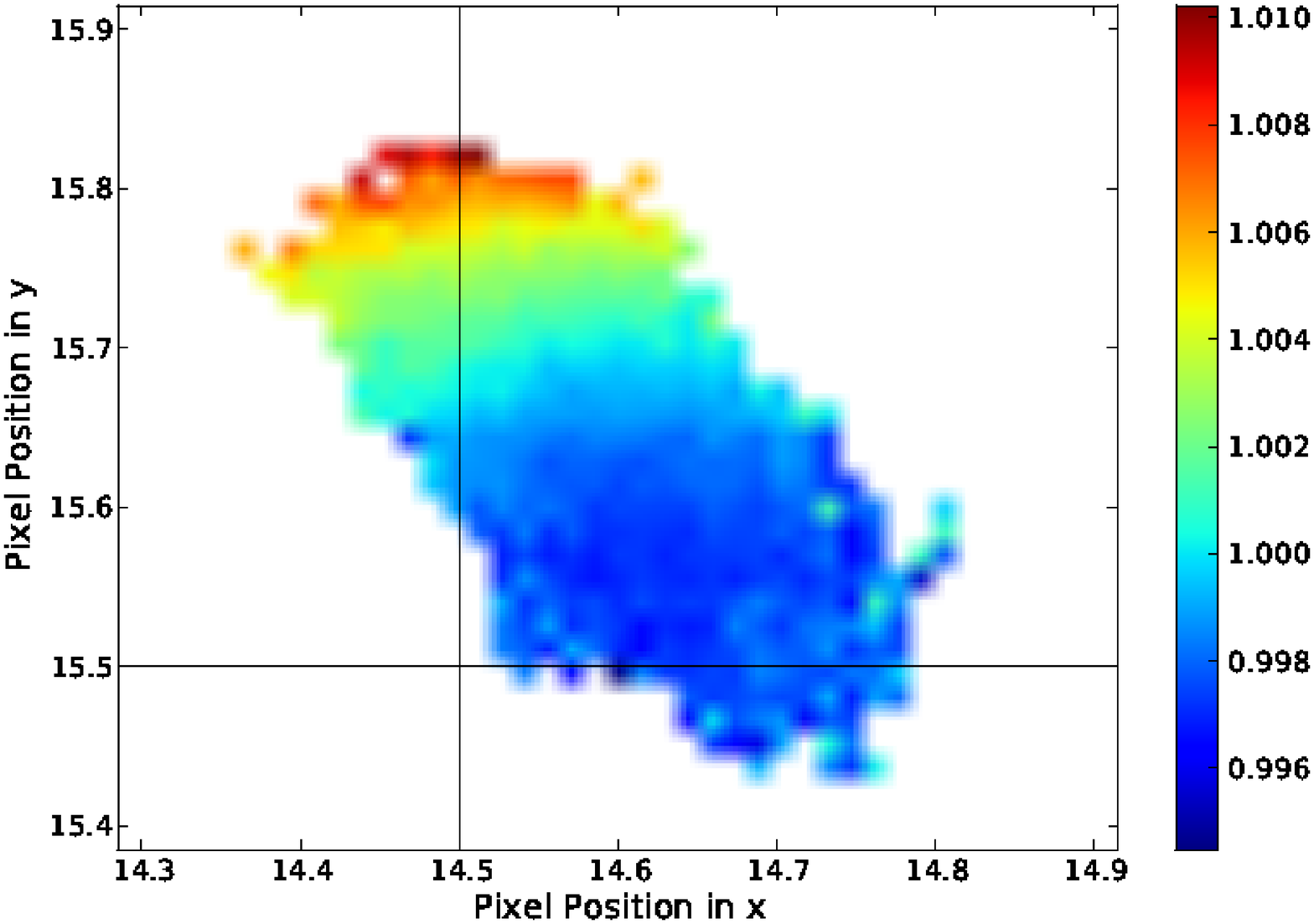}
  \includegraphics[width=\linewidth, clip]{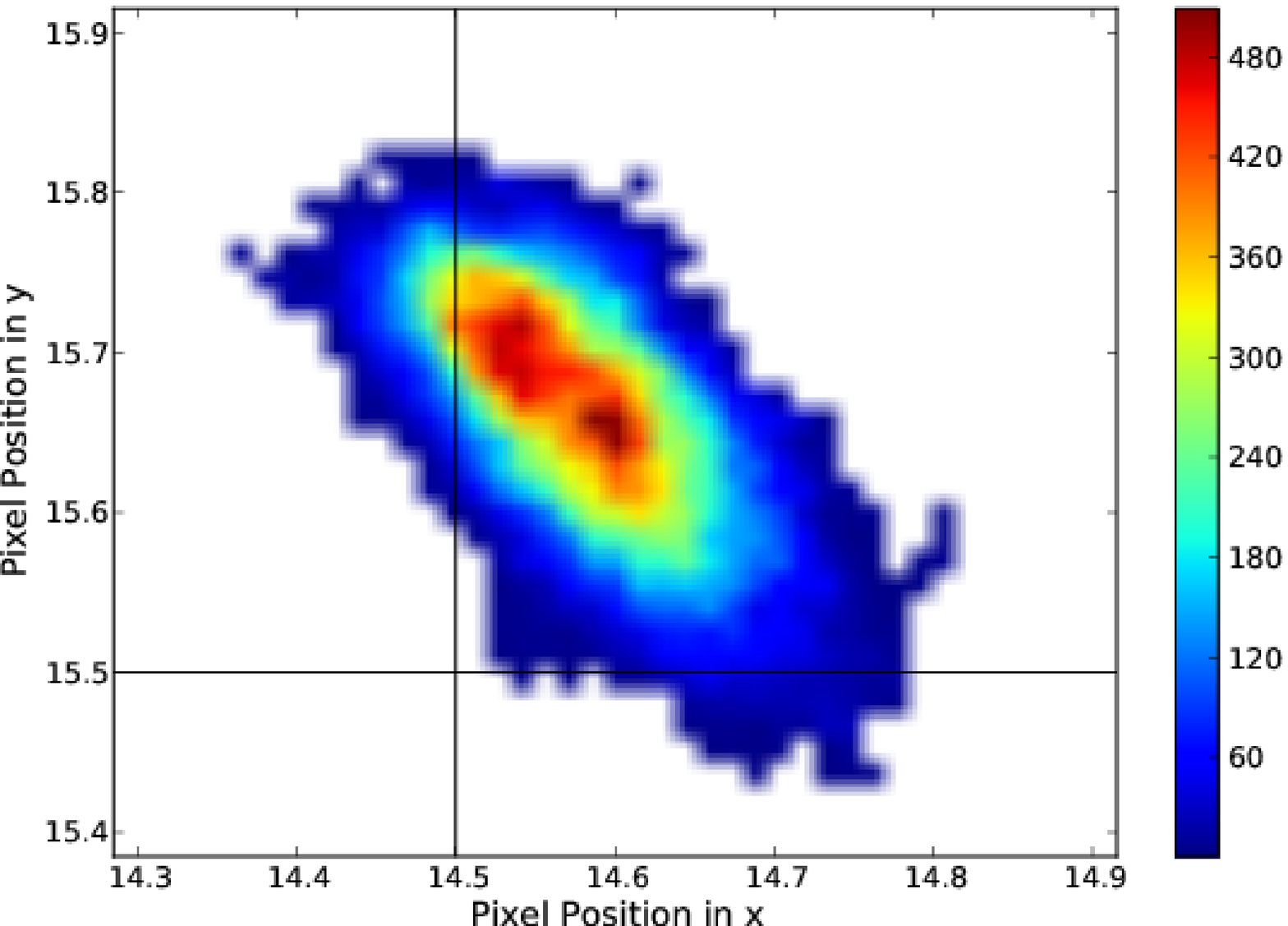}
\fi
\caption{
BLISS map and pointing histogram of HD149bs11.
{\em Top:} Redder (bluer) colors indicate higher (lower) subpixel sensitivity.  The horizontal and vertical black lines depict pixel boundaries.
{\em Bottom:} Colors indicate the number of points in a given bin, which, in this case, is 0.015 pixels in length and width.
By recalculating the map at each step of the MCMC or minimizer, this technique substantially
improves on that of \citet{Ballard2010b}, and beats all tested functional fits.
}
\label{fig:BLISS}
\end{figure}

We do not treat the knots as MCMC jump parameters.  Rather, we step all other free parameters from Eq.\ \ref{eqn:full}, generate a new model using these new jump parameters, then divide the observed flux by the new model (\math{F\sb{\rm IP}(x,y) = F\sb{obs} / F\sb{s} E(t) R(t) V(\nu) P(p)}).  Hypothetically, the residuals of \math{F\sb{\rm IP}(x,y)} contain only position-dependent flux variations.  The flux value of a particular knot is the mean of \math{F\sb{\rm IP}(x,y)} for the points associated with that knot.  We also tried median and weighted average knot values but the results did not improve and these calculations are much slower.  Next, we generate the sensitivity map ($M[x,y]$, Figure \ref{fig:BLISS}) by interpolating the flux from the knots to all of the observed points using BLI and/or NNI.  We tested various weighted smoothing functions when generating the sensitivity maps but, again, there was no improvement in the results.  Finally, $M(x,y)$ enters Eq.\ \ref{eqn:full} for comparison with the observed flux to obtain an estimate of the goodness of fit and determine the MCMC acceptance probability.  This process repeats for each step of the MCMC routine or minimizer.

\if\submitms y
\clearpage
\fi
\if\submitms y
\clearpage
\fi

\subsection{Determining The Optimal Bin Size}
\label{sec:binSize}

The accuracy of the BLISS mapping technique depends critically on the bin size, or resolution in position space; however, there is a trade-off between bin size and speed.  Decreasing the bin size requires more knots and runs slower but may be necessary to adequately resolve sensitivity changes on the pixel surface.  There is, however, a practical limit to how small the bins can be.  A bin for every measurement will always produce a perfect fit, resulting in a negative number of degrees of freedom and leaving the eclipse parameters unconstrained.  Bin sizes must be small enough to resolve real, small-scale variations on the pixel surface but large enough to mix in- and out-of-eclipse points.  This mixture helps to minimize correlations between the eclipse parameters and the knots in the sensitivity map.


To establish the optimal bin size, we consider a range of bin sizes for both BLI and NNI using the 3.6 {\micron} data set (HD149bs11), which has the strongest position-dependent systematic. We draw several conclusions from the results in Table \ref{table:BinSize}.  First, the SDNR (our measure of goodness of fit) decreases with decreasing bin size, indicating a better fit.  Unfortunately, the SDNR decreases indefinitely, so it cannot constrain the minimum bin size.  Second, BLI fits the data better than NNI for bin sizes greater than 0.015 pixels. The opposite is true for smaller bin sizes, which is counterintuitive because BLI should always outperform NNI, assuming no uncertainty in the position.  Thus, the bin size at which NNI outperforms BLI is indicative of the centering precision for a particular data set.  We estimate the precision for our analysis of HD149bs11 to be 0.009 pixels in $x$ and 0.007 pixels in $y$ by calculating the root-mean-squared (RMS) frame-to-frame position difference.  This agrees well with Figure \ref{fig:yPos} and is consistent with the cross-over bin size of 0.015 pixels, where NNI outperforms BLI.  Last, we place an upper limit on the bin size by noting that the eclipse depths become inconsistent with each other for pixel sizes \math{\geq 0.050} using BLI and \math{\geq 0.020} using NNI.  

We conclude that, whenever possible, BLI should be used with a bin size that is independent of the eclipse depth and has a lower SDNR than NNI.  We have found cases where BLI is never better than NNI.  In those instances, the position dependence is so weak that the intrapixel model component is unnecessary.  A better fit with BLI, compared to NNI, is thus a good indicator that a position-dependence systematic is present in a given data set.  For the test cases shown in Table \ref{table:BinSize}, using BLI with a bin size between 0.015 and 0.020 pixels is recommended based on our criteria.

Precise centering is important for this method because imprecision limits the smallest meaningful bin size.  Our preliminary work (see the SI of \citealp{Stevenson2010}) indicates that the Gaussian and least-asymmetry centering methods are better than the center of light; additional work is in preparation. 

\comment{
}

\if\submitms y
\clearpage
\fi
\begin{table}[t]
\caption{\label{table:BinSize} 
BLISS Map Test - Variable Bin Size}
\atabon\strut\hfill\begin{tabular}{cccc}
    \hline
    \hline
    Model       & Bin Size  & SDNR      & Eclipse Depth     \\
                & [Pixels]  &           & [\%]               \\
    \hline
    BLI         & 0.100     & 0.0028736 & 0.063 \pm\ 0.003 \\
    BLI         & 0.050     & 0.0028222 & 0.043 \pm\ 0.003 \\
    BLI         & 0.020     & 0.0028076 & 0.040 \pm\ 0.003 \\
    BLI         & 0.015     & 0.0028031 & 0.040 \pm\ 0.003 \\
    BLI         & 0.010     & 0.0027917 & 0.040 \pm\ 0.003 \\
    BLI         & 0.005     & 0.0027403 & 0.040 \pm\ 0.003 \\
    BLI         & 0.002     & 0.0024796 & 0.039 \pm\ 0.003 \\
    \hline
    NNI         & 0.100     & 0.0029435 & 0.071 \pm\ 0.003 \\
    NNI         & 0.050     & 0.0028527 & 0.054 \pm\ 0.003 \\
    NNI         & 0.020     & 0.0028116 & 0.044 \pm\ 0.003 \\
    NNI         & 0.015     & 0.0028024 & 0.041 \pm\ 0.003 \\
    NNI         & 0.010     & 0.0027865 & 0.041 \pm\ 0.003 \\
    NNI         & 0.005     & 0.0027109 & 0.040 \pm\ 0.003 \\
    NNI         & 0.002     & 0.0023773 & 0.039 \pm\ 0.003 \\
    \hline
\end{tabular}\hfill\strut\ataboff
\end{table}
\if\submitms y
\clearpage
\fi

\subsection{Comparing Intrapixel Models}

To compare the BLISS mapping technique with other intrapixel methods, we fit six different intrapixel models to the HD149bs11 data set.  These models are quadratic, cubic and sextic polynomials (including lower-order cross terms), B10's new weighted sensitivity function (fixing $\sigma\sb{x}$ to 0.021 and $\sigma\sb{y}$ to 0.0079), BLI, and NNI.  The eclipse depth in B10's model is slightly shallower than the other models, which are all well within 1\math{\sigma} of one another, but the uncertainties are essentially identical (see Table \ref{table:CompareModels}).  The BLISS models show significant improvement in SDNR compared to the others.  We cannot use the BIC to compare the BLISS model components with the polynomial model components for the reasons discussed in Appendix \ref{sec:tom}.

BLISS mapping represents a substantial improvements over polynomial model components because BLI and NNI can model real structure (such as pixelation) that cannot be modeled with low-order polynomials, they encounter fewer correlations between free parameters, and require fewer iterations to assess parameter uncertainties.

\if\submitms y
\clearpage
\fi
\begin{table}[t]
\caption{\label{table:CompareModels} 
BLISS Map Test - Comparing To Other Intrapixel Models}
\centering
\begin{tabular}{lccc}
    \hline
    \hline
    Model       & Bin Size      & SDNR      & Eclipse Depth      \\
                &               &           & [\%]               \\
    \hline
    Ballard     & 2.4 seconds\tablenotemark{a}   
                                & 0.0028230 & 0.034 \pm\ 0.003 \\
    Cubic       & -             & 0.0028180 & 0.039 \pm\ 0.003 \\
    Sextic      & -             & 0.0028157 & 0.040 \pm\ 0.003 \\
    Quadratic   & -             & 0.0028186 & 0.041 \pm\ 0.003 \\
    BLI         & 0.015 pixels  & 0.0028031 & 0.040 \pm\ 0.003 \\
    NNI         & 0.015 pixels  & 0.0028024 & 0.041 \pm\ 0.003 \\
    \hline
\end{tabular}
\begin{minipage}[t]{\linewidth}
\tablenotetext{1}{\ Longer bin sizes were considered but produced worse results.  Shorter bin sizes are prohibitively expensive to compute and are below the limit of detectable motion.}
\end{minipage}
\end{table}
\if\submitms y
\clearpage
\fi

\section{PRIMARY TRANSIT FITS AND RESULTS}
\label{sec:tranresults}

\begin{figure}[h]
\if\submitms y
  \setcounter{fignum}{\value{figure}}
  \addtocounter{fignum}{1}
  \newcommand\fignam{f\arabic{fignum}.ps}
\else
  \newcommand\fignam{./figs/hd149bp-rms.ps}
\fi
\strut\hfill
\includegraphics[width=\linewidth, clip]{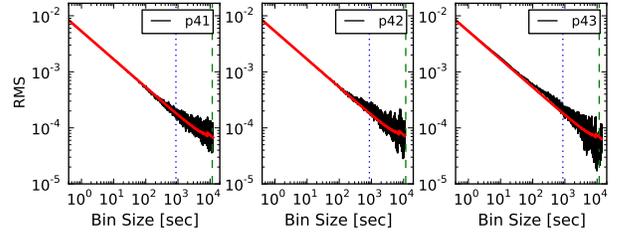}
\hfill\strut
\figcaption{\label{fig:trrms}
RMS residual flux \vs bin size for three HD 149026b transits.  Black vertical lines at each bin size depict 1$\sigma$ uncertainties on the RMS residuals ($RMS / \sqrt{2N}$, where $N$ is the number of bins).  The red line shows the predicted standard error for Gaussian noise, the dotted vertical blue line indicates the ingress/egress timescale, and the dashed vertical green line indicates the transit duration timescale.  Any RMS residuals that are several $\sigma$ above the red line would indicate correlated noise at that bin size.  When considering the effects of correlations on transit depth, the bin size of interest is the transit duration and not the ingress/egress time.
The shorthanded legend labels correspond to the last three characters in each event's label (e.g., p41 = HD149bp41).
}
\end{figure}

\begin{figure*}[htb]
\if\submitms y
  \setcounter{fignum}{\value{figure}}
  \addtocounter{fignum}{1}
  \newcommand\fignama{f\arabic{fignum}a.ps}
  \newcommand\fignamb{f\arabic{fignum}b.ps}
  \newcommand\fignamc{f\arabic{fignum}c.ps}
\else
  \newcommand\fignama{./figs/hd149bp4-raw.ps}
  \newcommand\fignamb{./figs/hd149bp4-bin.ps}
  \newcommand\fignamc{./figs/hd149bp4-norm.ps}
\fi
\strut\hfill
\includegraphics[width=0.32\linewidth, clip]{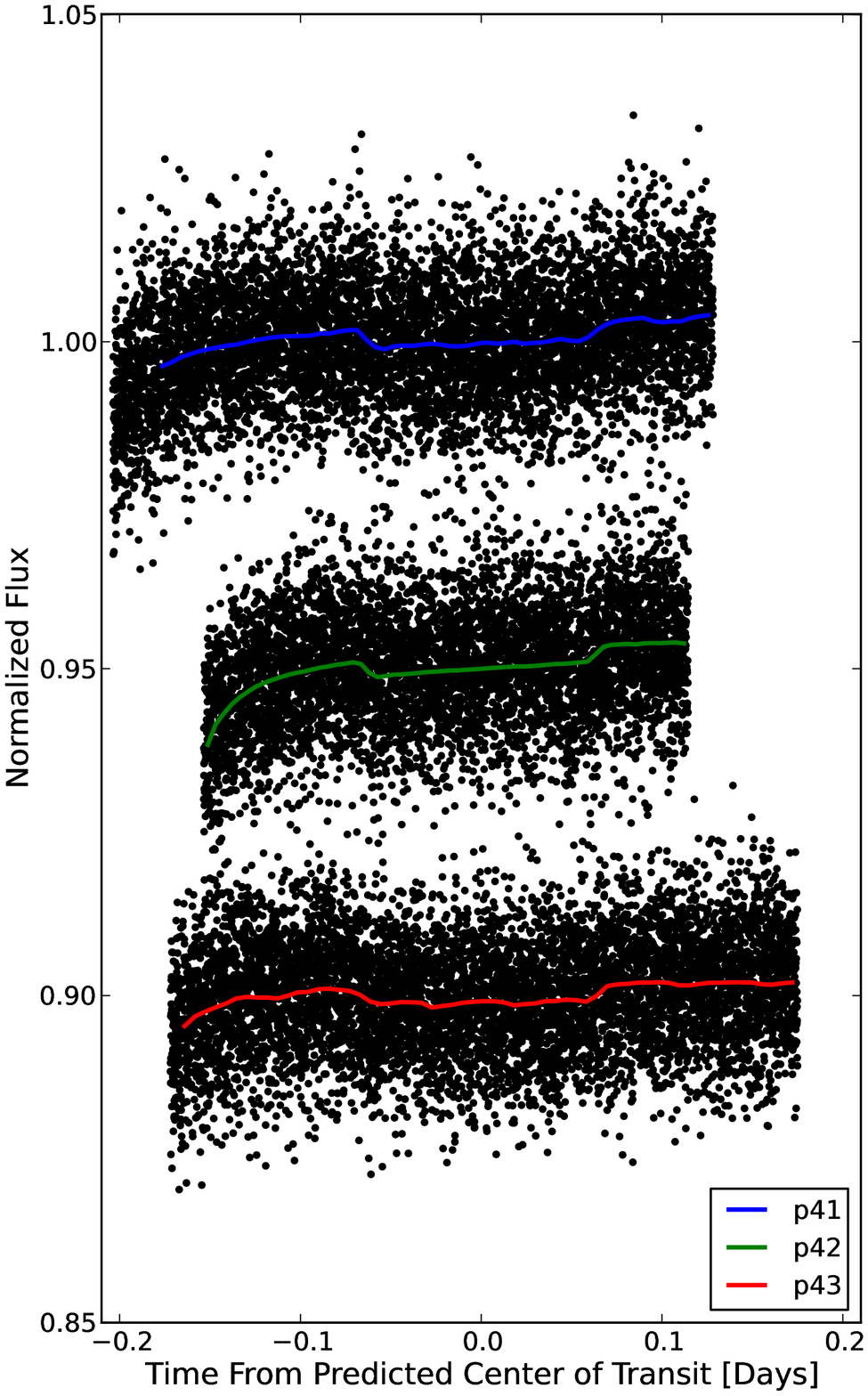}
\includegraphics[width=0.32\linewidth, clip]{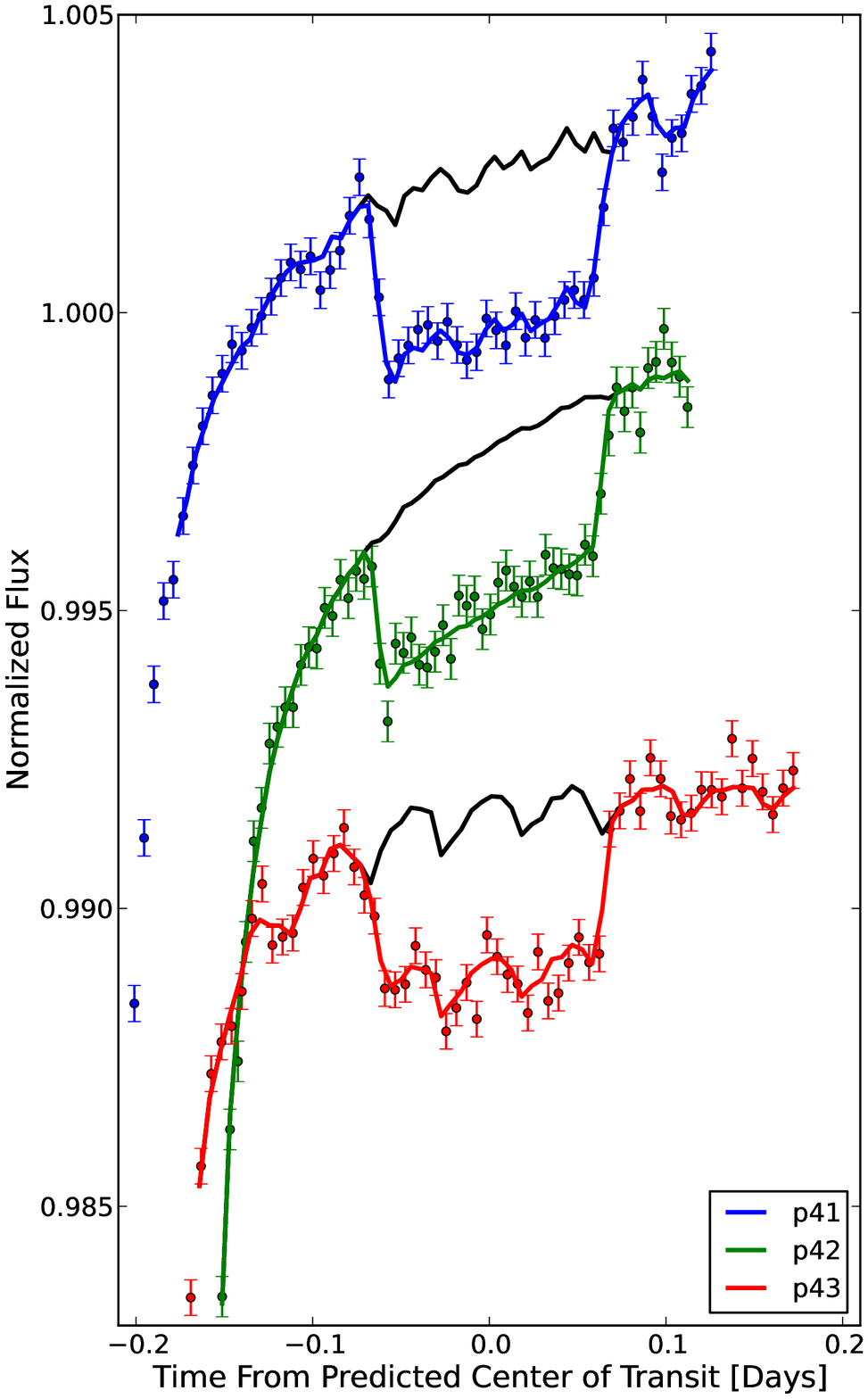}
\includegraphics[width=0.32\linewidth, clip]{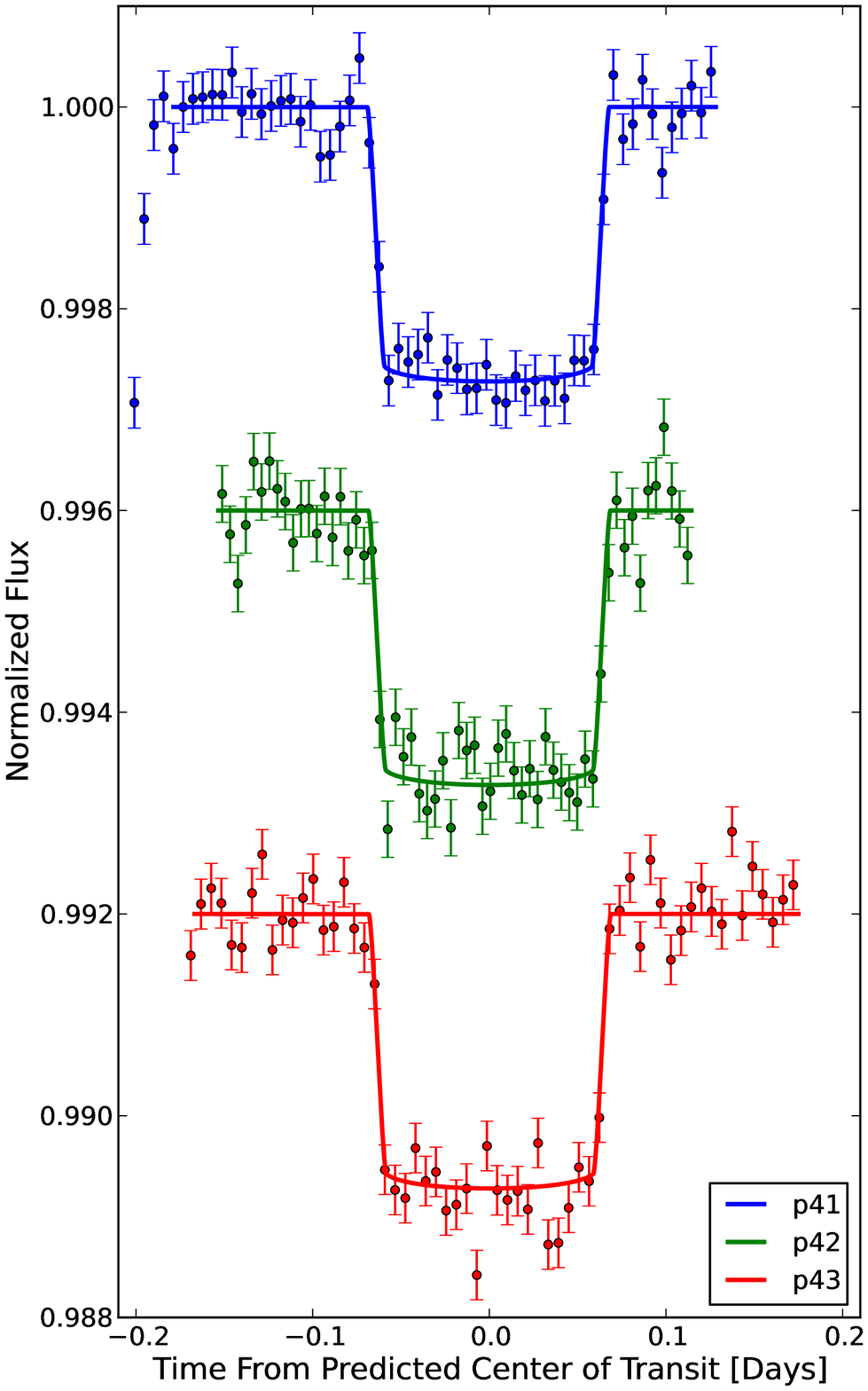}
\hfill\strut
\figcaption{\label{fig:tlc}
Raw (left), binned (center) and systematics-corrected (right) primary-transit light curves of HD 149026b at 8.0 {\microns}. The results are normalized to the system flux and shifted vertically for ease of comparison.  The colored lines are best-fit models, the black curves omit their transit model components, and the error bars are 1\math{\sigma} uncertainties.  The shorthanded legend labels correspond to the last three characters in each event's label (e.g., p41 = HD149bp41).  The pixelation effect (see Section \ref{subsec:pixelation}) is most prevalent in HD149bp43.
}
\end{figure*}

We present the scaling of the RMS model residuals \vs bin size (a test of correlation in time) in Figure \ref{fig:trrms}, the best-fit transit light curves in Figure \ref{fig:tlc}, a comparison between two fits in Table \ref{table:tranfits}, and the full set of best-fit transit parameters in Appendix \ref{sec:app}.  The electronic supplement contains light-curve files and supplementary figures.
Below, we discuss each observation to explain how we arrived at the final results.

\if\submitms y
\clearpage
\fi
\if\submitms y
\clearpage
\fi

\if\submitms y
\clearpage
\fi
\if\submitms y
\clearpage
\fi

\subsection{Three Fits at 8.0 \math{\mu}m}

At each 0.25-pixel increment in photometry aperture size, we model the time-dependent systematic with the ramps listed in Eqs.\ \ref{eqnse} - \ref{eqnquad}.  An aperture size of 3.5 pixels produces the lowest SDNR values for all ramps and for all transit events.  We estimate the background flux using an annulus from 7 to 15 pixels centered on the star.  We follow the method described in Section \ref{sec:binSize} when determining the optimal bin sizes of the BLISS maps.

\subsubsection{HD149bp41}

There are 26 consecutive frames (19494 to 19519) shifted horizontally by exactly one pixel; we flag these frames as bad.  After clipping the first 5,000 data points ($\sim$33.3 minutes, $q$ = 5,000), many of the ramps fit the time-dependent systematic equally well, but the fits exhibit a large range of radius ratios ($R\sb{p}/R\sb{\star}$ = 0.0494 to 0.0517).  Of the top three models shown in Table \ref{table:tranfits}, Eq.\ \ref{eqnsel} has the lowest BIC value and favors a moderate radius ratio of 0.0502 {\pm} 0.0011.  For comparison, \citet{Nutzman2009} and \citet{Carter2009} report values of 0.05158 \pm\ 0.00077 and 0.5188 \pm\ 0.00085, respectively.  To model the ramp, both adopt a quadratic function in $\ln(t)$ (Eq.\ \ref{eqnlog}) with $t\sb{0}$ fixed to a time a few minutes prior to the first observation.  Fixing parameters can cause a MCMC run to underestimate uncertainties of the remaining free parameters (see Section \ref{subsec:ortho}).  Instead, we orthogonalize the correlated parameters (system flux and all three ramp parameters in Eq.\ \ref{eqnsel}) to provide a coordinate system in which our MCMC routine can sample efficiently.  There is a weak position dependence (see Figure \ref{fig:tlc}) in both $x$ and $y$ directions that we model with the BLISS mapping technique using 0.030-pixel bins.

\subsubsection{HD149bp42}

For all ramps with reasonable fits, we find that the planet-to-star radius ratios range from 0.0444 to 0.513.  The three best models appear in Table \ref{table:tranfits} with a range of uncertainties in $R\sb{p}/R\sb{\star}$.  The ramp parameters from Eq.\ \ref{eqnse2}\sb{--} correlate most strongly with the radius ratio, resulting in an uncertainty that is twice that of Eq.\ \ref{eqnlog}.  Fixing ramp parameters in Eq.\ \ref{eqnse2}\sb{--} erroneously improves the radius ratio uncertainty to 0.008\%.  Equation \ref{eqnlog} has the lowest BIC value, so we use it to fit the full data set ($q$ = 0).  The data exhibit only a minor position dependence in the $x$ direction; however, the significant improvement in SDNR 
indicates that we should include the BLISS map during the joint model fit.

\subsubsection{HD149bp43}

Unlike the previous two data sets, HD149bp43 was preflashed to mitigate the ramp effect (see K09 for details).  Thus, we do not need to clip a significant initial portion of the data set or use a double-rising exponential.  Instead, we clip only the first 1000 data points ($\sim$6.7 minutes) and use Eq.\ \ref{eqnse}\sb{--} to model the time-dependent systematic.  As seen in Table \ref{table:tranfits}, the HD149bp43 transit depth is consistently deeper than the other two data sets.  The $R\sb{p}/R\sb{\star}$ parameter is independent of our choice of $q$ but is dependent on the choice of ramp model components, ranging from 0.0516 to 0.0536.  BIC favors the deepest transit depth, resulting in a radius ratio that is larger than other best-fit ratios by $\sim 4\sigma$.  For comparison, K09 use Eq.\ \ref{eqnlog} with a fixed $t\sb{0}$ parameter and report a radius ratio of 0.05253 \pm\ 0.00076.  We achieve the same underestimated uncertainty using a similarly constrained $t\sb{0}$ parameter.
A relatively strong position dependence is evident in Figure \ref{fig:tlc} and is modeled with a BLISS map using 0.050-pixel bins.

\if\submitms y
\clearpage
\fi
\begin{table}[ht]
\centering
\caption{\label{table:tranfits} 
Individual Transit Model Fits}
\begin{tabular}{cccccc}
    \hline
    \hline
    Label       & $R(t)$                & $M(x,y)$  & SDNR      & ${\Delta}$BIC & $R\sb{p}/R\sb{\star}$ \\
    \hline
    HD149bp41   & \ref{eqnsel}\sb{--}   & BLI       & 0.0083449 &      0.0      & 0.0502 \pm\ 0.0011    \\
    HD149bp41   & \ref{eqnlog}          & BLI       & 0.0083444 &      0.4      & 0.0517 \pm\ 0.0009    \\
    HD149bp41   & \ref{eqnll}           & BLI       & 0.0083444 &      0.4      & 0.0517 \pm\ 0.0010    \\
    \hline
    HD149bp42   & \ref{eqnlog}          & BLI       & 0.0083565 &      0.0      & 0.0503 \pm\ 0.0008    \\
    HD149bp42   & \ref{eqnll}           & BLI       & 0.0083564 &      4.0      & 0.0513 \pm\ 0.0009    \\
    HD149bp42   & \ref{eqnse2}\sb{--}   & BLI       & 0.0083562 &      5.6      & 0.0502 \pm\ 0.0016    \\
    \hline
    HD149bp43   & \ref{eqnse}\sb{--}    & BLI       & 0.0083681 &      0.0      & 0.0536 \pm\ 0.0008    \\
    HD149bp43   & \ref{eqnlog}          & BLI       & 0.0083682 &      6.5      & 0.0525 \pm\ 0.0010    \\
    HD149bp43   & \ref{eqnsel}\sb{--}   & BLI       & 0.0083685 &      7.0      & 0.0516 \pm\ 0.0011    \\
    HD149bp43   & \ref{eqnll}           & BLI       & 0.0083682 &      7.2      & 0.0527 \pm\ 0.0010    \\
    \hline
\end{tabular}
\end{table}
\if\submitms y
\clearpage
\fi

\subsection{Joint Fit}

We perform two joint-model fits, each requiring less than $2\times10^6$ iterations to estimate uncertainties.  The first considers only the three transits analyzed here while the second also considers the more precise NICMOS data from \citet{Carter2009} by placing priors on $i$ and $a/R\sb{\star}$.  
Both fits in Table \ref{table:transit} are consistent with previous results from \citet[$R\sb{p}/R\sb{\star} = 0.0522 \pm\ 0.0008$]{Knutson2010} and \citet[$R\sb{p}/R\sb{\star} = 0.0519 \pm\ 0.0008$]{Carter2009} and have improved estimates of the radius ratio.  The uncertainties in the duration and ingress/egress times for the independent fit are significantly larger than those from the fit with \citet{Carter2009} priors.

\if\submitms y
\clearpage
\fi
\begin{table}[ht]
\centering
\caption{\label{table:transit} 
Joint Transit Model Fits}
\begin{tabular}{@{}rcc@{}}
    \hline
    \hline
    Parameters              & Independent Fit               & \citet{Carter2009} Priors\tablenotemark{a} \\
    \hline
    $R\sb{p}/R\sb{\star}$   & 0.0514 \pm\ 0.0006            & 0.0518 \pm\ 0.0006            \\
    $i$ [$\sp{\circ}$]      & 87.2$^{+1.6}_{-2.1}$          & 84.6 \pm\ 0.5                 \\
    $a/R\sb{\star}$         & 6.8$^{+0.3}_{-0.7}$           & 5.98 \pm\ 0.17                \\
    Impact Parameter        & 0.33$^{+0.21}_{-0.19}$        & 0.57 \pm\ 0.04                \\
    Transit depth [\%]      & 0.264 \pm\ 0.006              & 0.268 \pm\ 0.006              \\
    Duration [t\sb{1}-t\sb{4}, hr] & 3.23$^{+0.04}_{-0.02}$ & 3.286 \pm\ 0.019              \\
    Ingress/Egress [hr]     & 0.178$^{+0.043}_{-0.015}$     & 0.234 \pm\ 0.012              \\
    HD149bp41   \\
    Midpoint (MJD\sb{UTC})\tablenotemark{b}
                            & 4327.3719 \pm\ 0.0005         & 4327.3720 \pm\ 0.0005         \\
    Midpoint (MJD\sb{TDB})\tablenotemark{b}
                            & 4327.3726 \pm\ 0.0005         & 4327.3727 \pm\ 0.0005         \\
    O-C (minutes)\tablenotemark{c}           
                            & -1.0 \pm\ 0.7                 & -0.9 \pm\ 0.8                 \\
    SDNR                    & 0.0083440                     & 0.0083440                     \\
    HD149bp42   \\
    Midpoint (MJD\sb{UTC})\tablenotemark{b}
                            & 4356.1316 \pm\ 0.0005         & 4356.1316 \pm\ 0.0005         \\
    Midpoint (MJD\sb{TDB})\tablenotemark{b}
                            & 4356.1323 \pm\ 0.0005         & 4356.1323 \pm\ 0.0005         \\
    O-C (minutes)\tablenotemark{c}           
                            & 0.2 \pm\ 0.7                  & 0.0 \pm\ 0.8                  \\
    SDNR                    & 0.0083550                     & 0.0083556                     \\
    HD149bp43   \\
    Midpoint (MJD\sb{UTC})\tablenotemark{b}
                            & 4597.7070 \pm\ 0.0004         & 4597.7068 \pm\ 0.0005         \\
    Midpoint (MJD\sb{TDB})\tablenotemark{b}
                            & 4597.7077 \pm\ 0.0004         & 4597.7075 \pm\ 0.0005         \\
    O-C (minutes)\tablenotemark{c}           
                            & 0.8 \pm\ 0.7                  & 0.6 \pm\ 0.6                  \\
    SDNR                    & 0.0083690                     & 0.0083691                     \\
    \hline
\end{tabular}
\begin{minipage}[t]{1.0\linewidth}
\tablenotetext{1}{{We place priors on $i$ and $a/R\sb{\star}$ using values from \citet{Carter2009}.}}
\tablenotetext{2}{{MJD = BJD - 2,450,000.}}
\tablenotetext{3}{{Computed using the period and ephemeris from \citet[$p$ = 2.8758925 \pm\ 0.0000023 days, $t\sb{0}$ = 2454597.70645 \pm\ 0.00018 BJD\sb{UTC}]{Knutson2009b}.}}
\end{minipage}
\end{table}
\if\submitms y
\clearpage
\fi

\section{SECONDARY-ECLIPSE FITS AND RESULTS}
\label{sec:eclresults}

There were 11 secondary-eclipse observations.  We considered whether the eclipse duration is consistent with the more-precise transit duration in these fits, which is likely if the orbit is nearly circular.  
In a joint fit of all secondary-eclipse events, the strong signal from HD149bs11 dominates the shared eclipse duration.  
The best-fit eclipse duration was 4.5 \pm\ 3.3 minutes longer than, but still consistent with, the transit duration from \citet{Carter2009}, and the mid-eclipse phases were in all but one case within 1.5\math{\sigma} of 0.5, together indicating circularity.  Since the transit and eclipse durations are consistent, we apply priors to the eclipse duration and ingress/egress times using the values given in Table \ref{table:transit}.
Unless otherwise stated, we estimated the background flux using an annulus from 7 to 15 pixels that was centered on the star.  We present the RMS model residuals in Figure \ref{fig:eclrms}, the best-fit light curves in Figure \ref{fig:lc}, and the best-fit parameters in Appendix \ref{sec:app}.  The electronic supplement contains light-curve files and supplementary figures.
Below, we discuss each observation in detail.

\if\submitms y
\clearpage
\fi
\begin{figure}[ht]
\if\submitms y
  \setcounter{fignum}{\value{figure}}
  \addtocounter{fignum}{1}
  \newcommand\fignam{f\arabic{fignum}.ps}
  \strut\hfill
  \includegraphics[width=0.7\linewidth, clip]{\fignam}
  \hfill\strut
\else
  \newcommand\fignam{./figs/hd149bs-rms2.ps}
  \strut\hfill
  \includegraphics[width=\linewidth, clip]{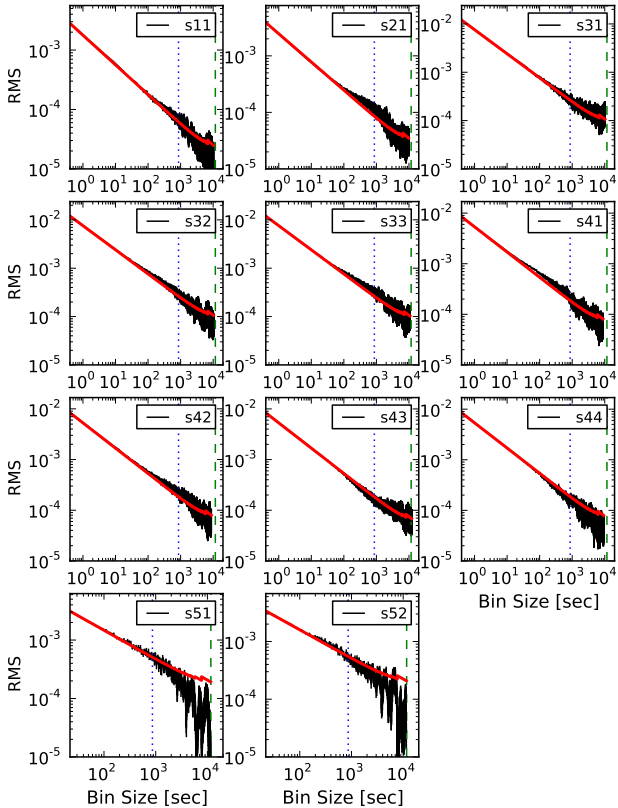}
  \hfill\strut
\fi
\figcaption{\label{fig:eclrms}
Same as Figure \ref{fig:trrms} except for eleven HD 149026b eclipses.
The shorthanded legend labels correspond to the last three characters in each event's label (e.g., s11 = HD149bs11).
}
\end{figure}
\if\submitms y
\clearpage
\fi

\if\submitms y
\clearpage
\fi
\begin{figure*}[t]
\if\submitms y
  \setcounter{fignum}{\value{figure}}
  \addtocounter{fignum}{1}
  \newcommand\fignama{f\arabic{fignum}a.ps}
  \newcommand\fignamb{f\arabic{fignum}b.ps}
  \newcommand\fignamc{f\arabic{fignum}c.ps}
  \strut\hfill
  \includegraphics[width=0.3\linewidth, clip]{\fignama}
  \includegraphics[width=0.3\linewidth, clip]{\fignamb}
  \includegraphics[width=0.3\linewidth, clip]{\fignamc}
  \hfill\strut
\else
  \newcommand\fignama{./figs/hd149b-bin.ps}
  \newcommand\fignamb{./figs/hd149b-norm1.ps}
  \newcommand\fignamc{./figs/hd149b-norm2.ps}
  \strut\hfill
  \includegraphics[height=11.5cm, clip]{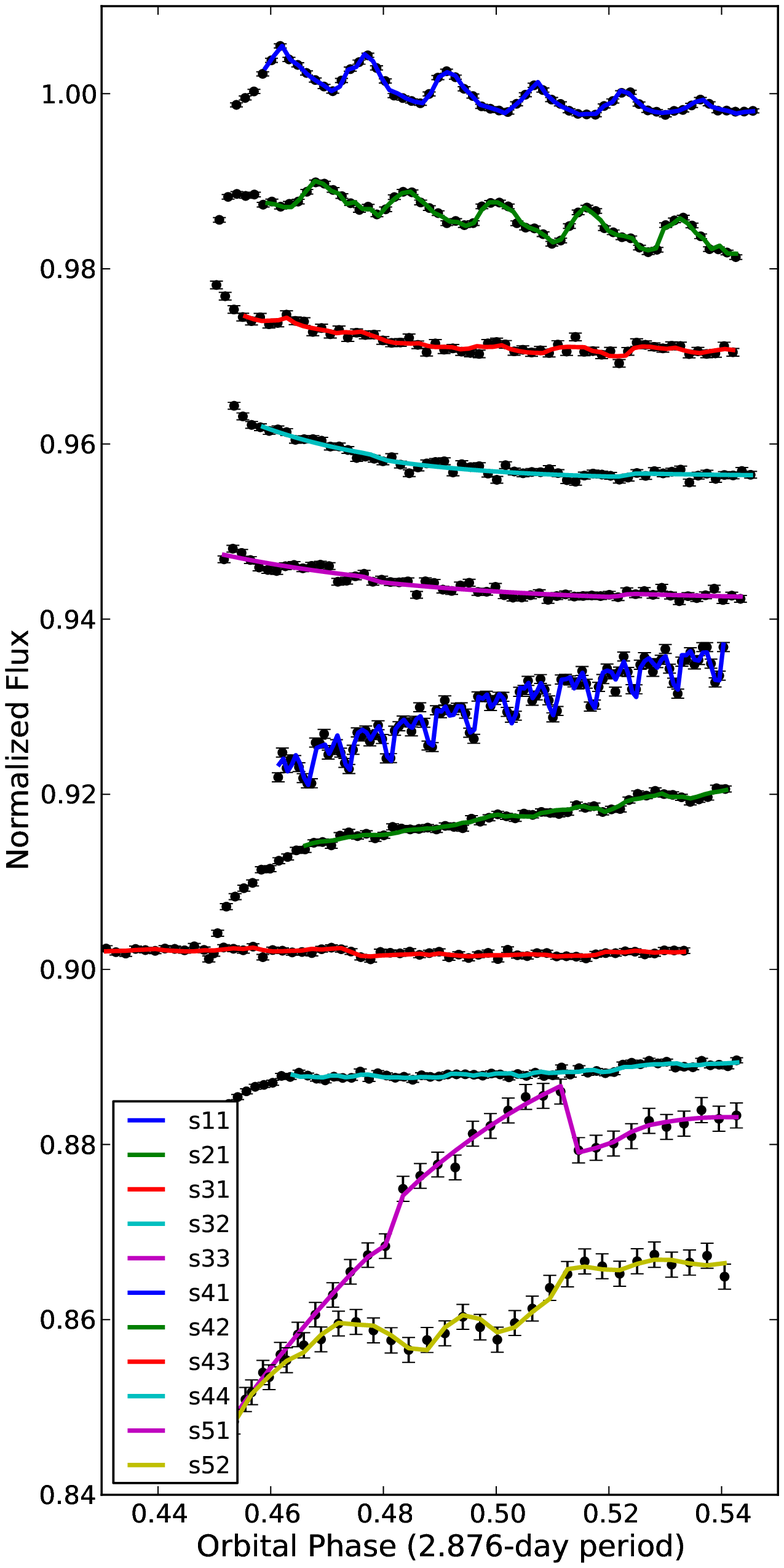}
  \includegraphics[height=11.5cm, clip]{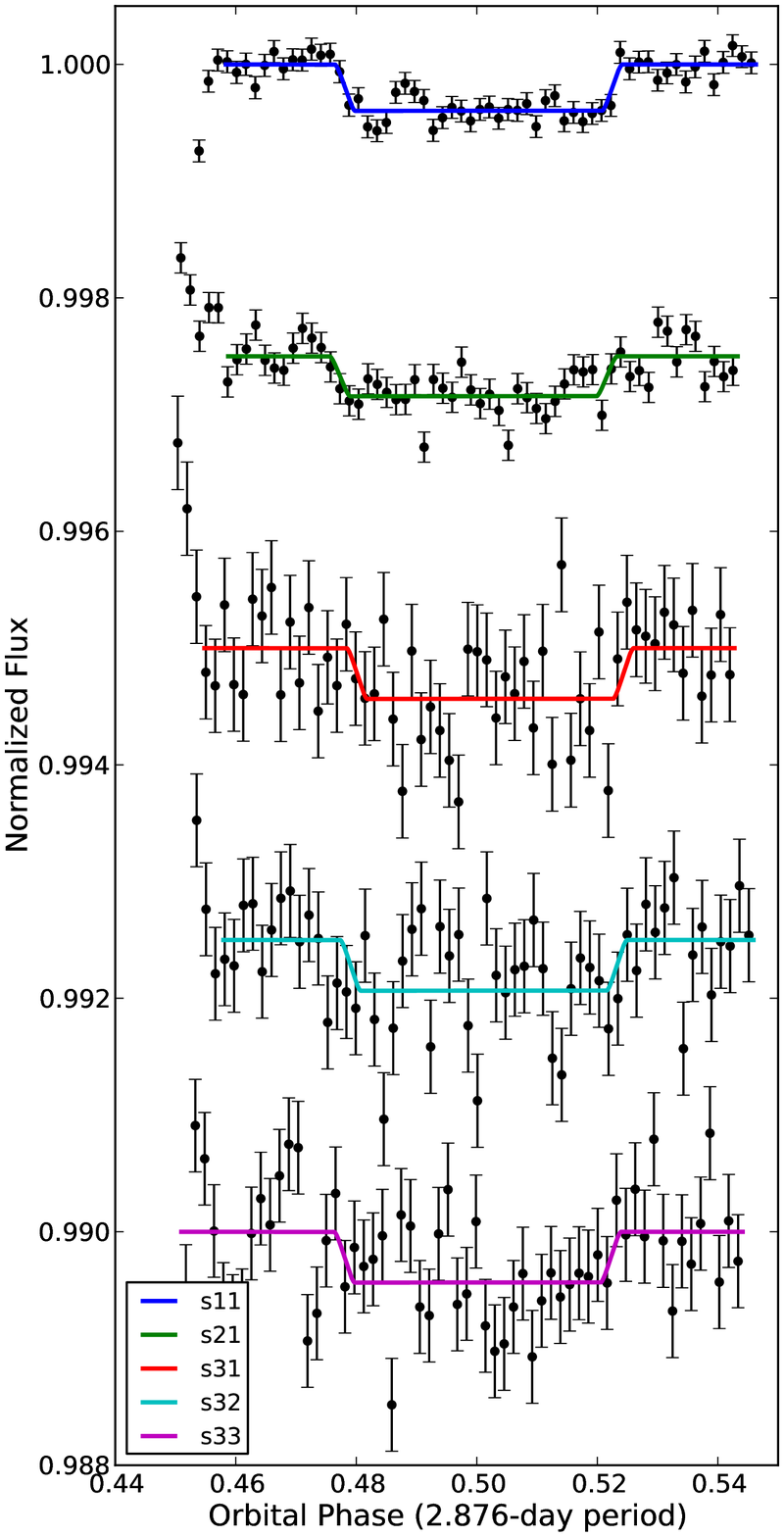}
  \includegraphics[height=11.5cm, clip]{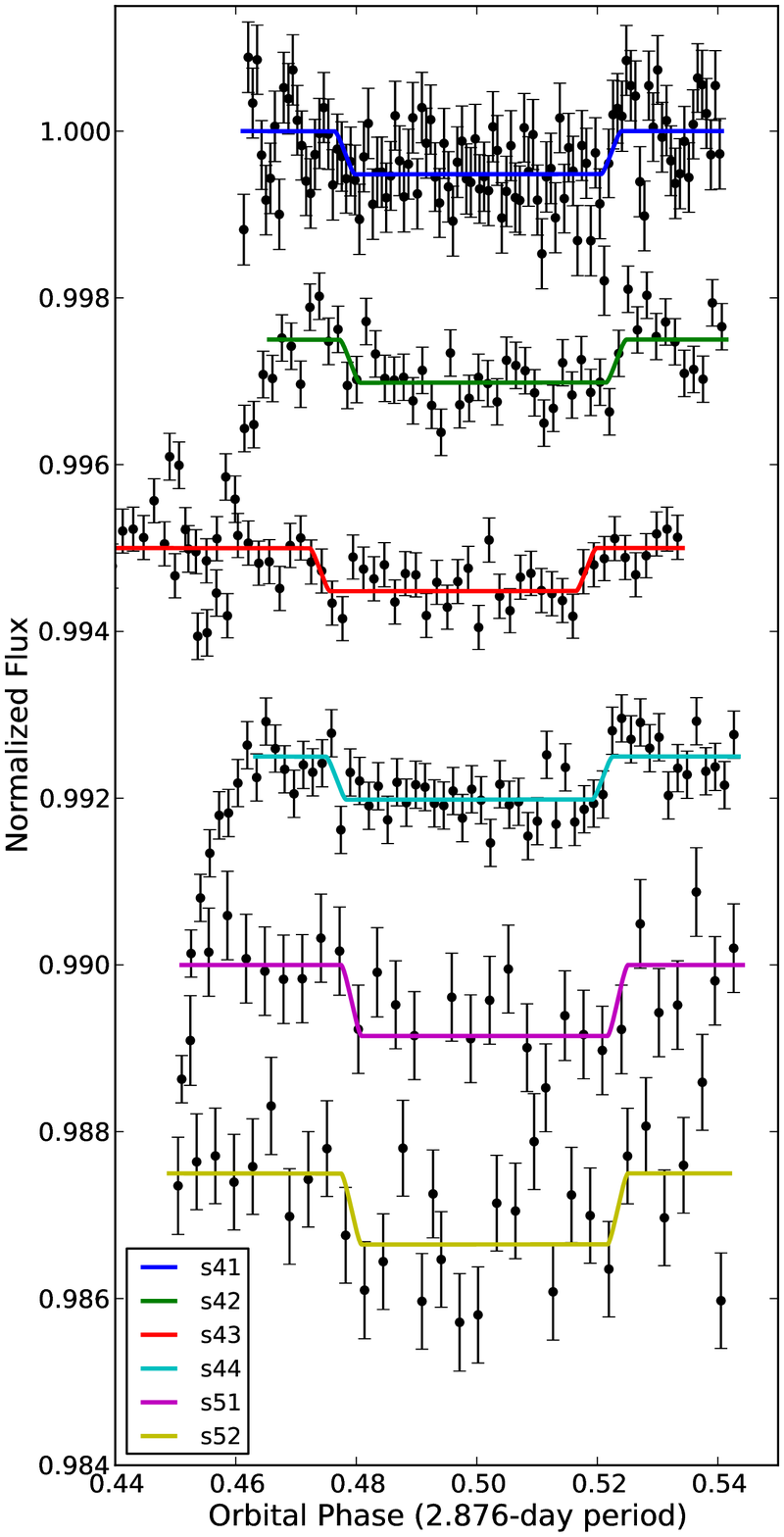}
  \hfill\strut
\fi
\figcaption{\label{fig:lc}
Binned (left) and systematics-corrected (center \& right) secondary-eclipse light curves of HD 149026b in five {\em Spitzer} channels. The results are normalized to the system flux and shifted vertically for ease of comparison.  The colored lines are best-fit models and the error bars are 1\math{\sigma} uncertainties.  The shorthanded legend labels correspond to the last three characters in each event's label (e.g., s11 = HD149bs11).
}
\end{figure*}
\if\submitms y
\clearpage
\fi

\subsection{Fit at 3.6 \math{\mu}m - HD149bs11}

\begin{figure}[ht]
\if\submitms y
  \setcounter{fignum}{\value{figure}}
  \addtocounter{fignum}{1}
  \newcommand\fignam{f\arabic{fignum}.ps}
\else
  \newcommand\fignam{./figs/hd149bs11-hist.ps}
\fi
\centering
\includegraphics[width=\linewidth, clip]{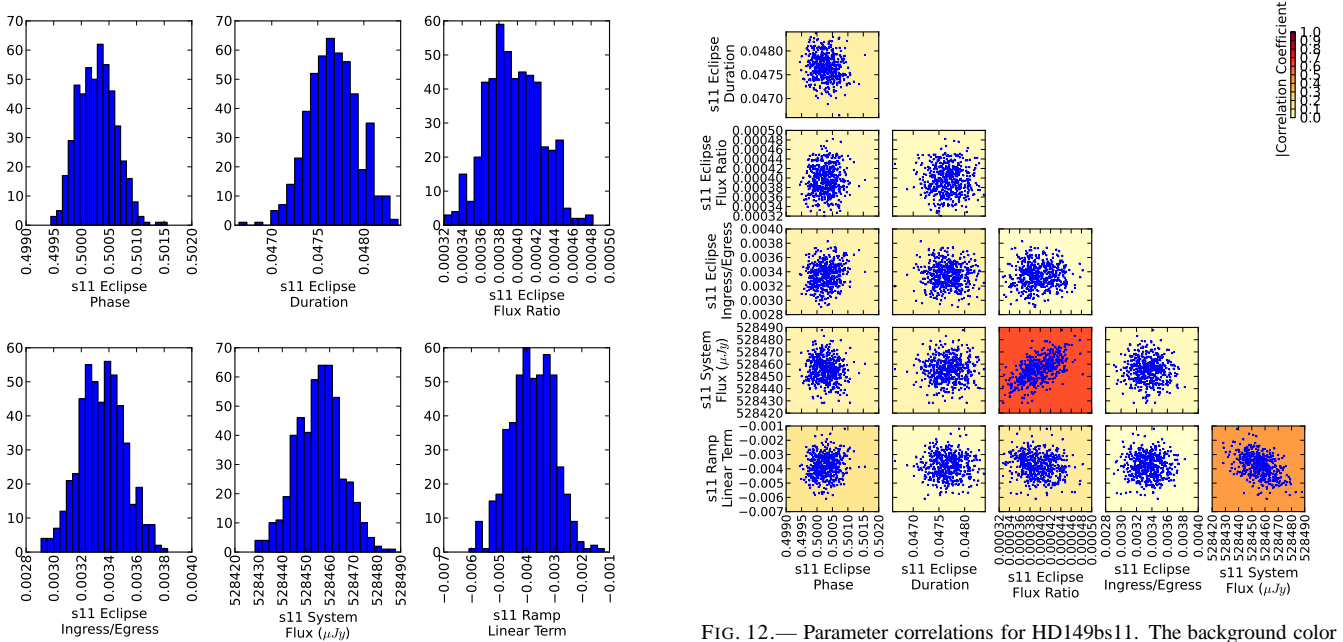}
\caption{Parameter histograms for HD149bs11.  We plot every 200\sp{th} step in the MCMC chain to decorrelate parameter values.  The BLISS map knots are similarly distributed.  Additional histograms are part of the electronic supplement.
\\
}
\label{fig:hd149bs11-hist}
\end{figure}

\begin{figure}[ht]
\if\submitms y
  \setcounter{fignum}{\value{figure}}
  \addtocounter{fignum}{1}
  \newcommand\fignam{f\arabic{fignum}.ps}
\else
  \newcommand\fignam{./figs/hd149bs11-corr.ps}
\fi
\centering
\includegraphics[width=\linewidth, clip]{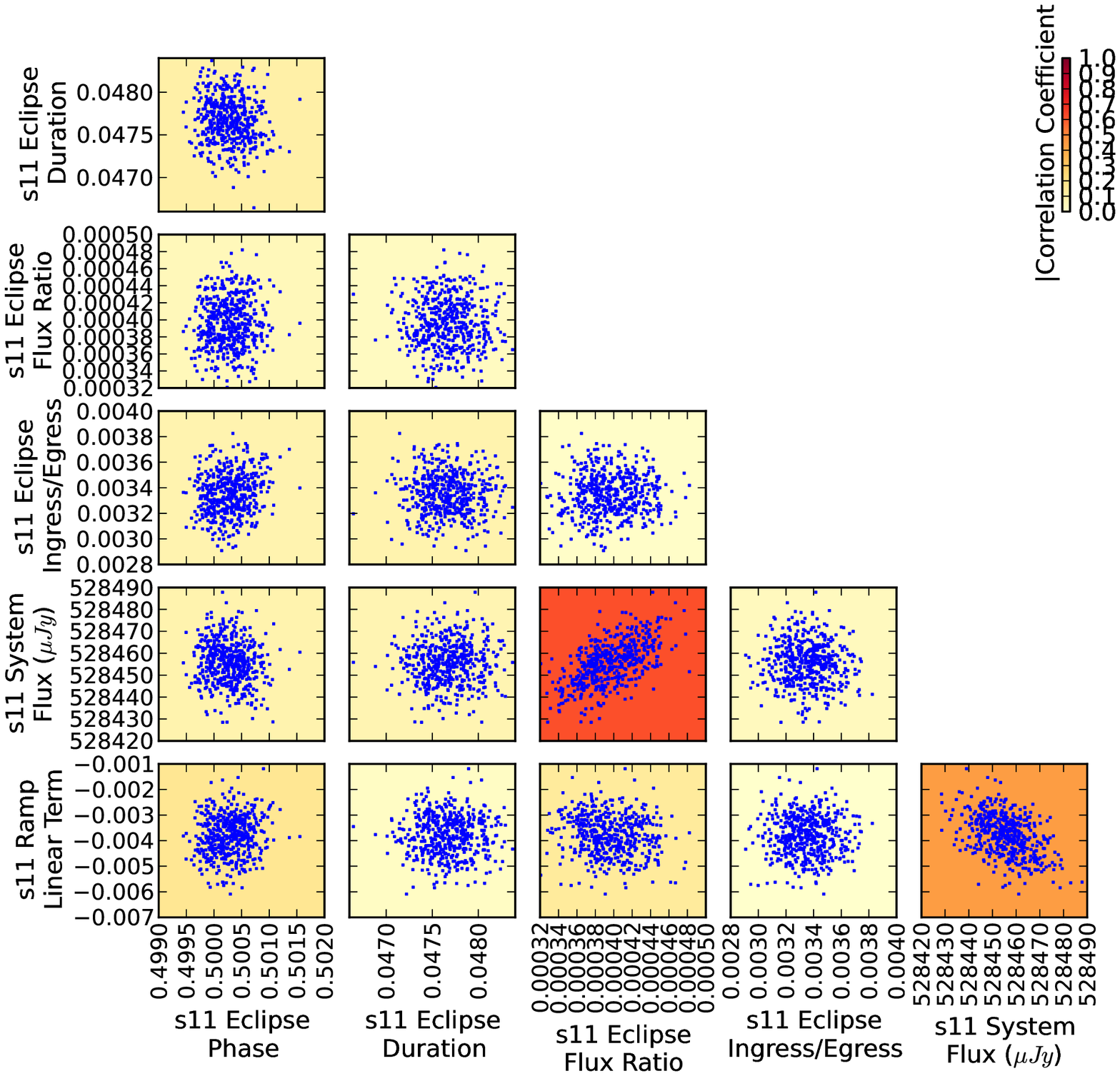}
\caption{Parameter correlations for HD149bs11.  The background color depicts the absolute value of the correlation coefficient.  The uncertainties produced from the MCMC method fully account for correlations between free parameters (e.g., eclipse flux ratio and system flux).  We plot every 200\sp{th} step in the MCMC chain to decorrelate parameter values.  Additional correlation plots are part of the electronic supplement.
}
\label{fig:hd149bs11-corrcoeffs}
\end{figure}

\begin{figure}[ht]
\if\submitms y
  \setcounter{fignum}{\value{figure}}
  \addtocounter{fignum}{1}
  \newcommand\fignam{f\arabic{fignum}.ps}
\else
  \newcommand\fignam{./figs/hd149bs11-fig701.ps}
\fi
\centering
\includegraphics[width=\linewidth, clip]{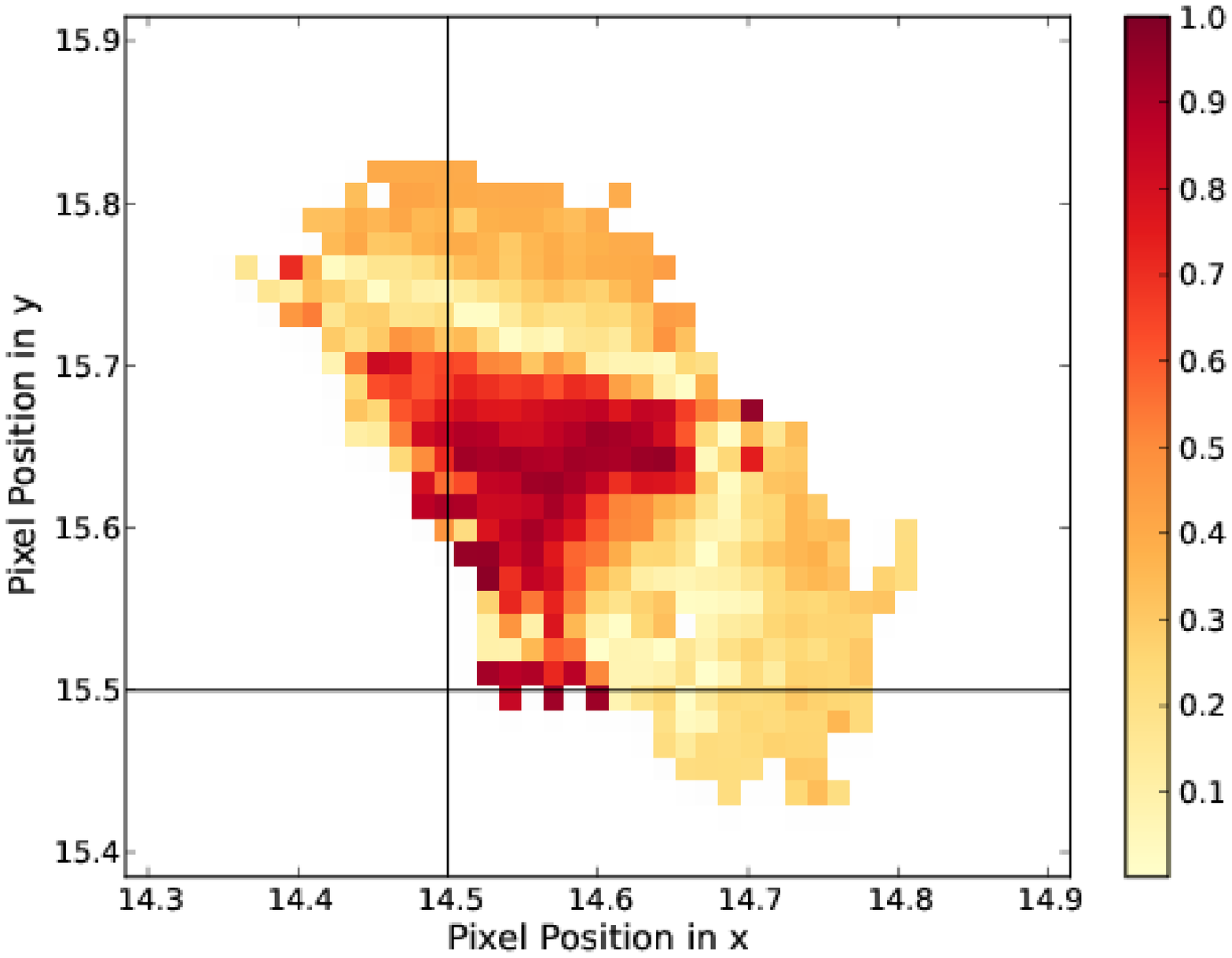}
\caption{Correlation coefficients between eclipse depth and computed BLISS map knots for HD149bs11.  The presence of relatively strong correlation regions (in red) indicates that computing the BLISS map at each step of an MCMC routine is necessary to assess the uncertainty on the eclipse depth correctly, as opposed to fixing the map as is done by B10.  In this case, fixing the BLISS map to its post-minimizer values leads to an erroneous 13\% decrease in the eclipse-depth error estimate.
}
\label{fig:hd149bs11-BLISScorrcoeffs}
\end{figure}

For this data set, we find that BLI outperforms NNI down to a bin size of 0.015 pixels when we exclude bins with less than 4 measurements.  Bins with fewer data points are insufficiently sampled to compute a reliable mean flux for the knot value.  The linear ramp (Eq.\ \ref{eqnlin}) fits best.  The posterior parameter distributions (histograms) and parameter correlation plots (including knot values in the BLISS map) are in Figures \ref{fig:hd149bs11-hist} - \ref{fig:hd149bs11-BLISScorrcoeffs}.  Similar plots at other wavelengths are included in the electronic supplement. 

\if\submitms y
\clearpage
\fi
\if\submitms y
\clearpage
\fi

\if\submitms y
\clearpage
\fi
\if\submitms y
\clearpage
\fi

\if\submitms y
\clearpage
\fi
\if\submitms y
\clearpage
\fi

\subsection{Fit at 4.5 \math{\mu}m - HD149bs21}

Using the strategy described in Section \ref{sec:binSize}, BLI achieves a better fit than NNI with bin sizes of 0.012 pixels along $x$ and 0.006 pixels along $y$.  Additionally, we fit the intrapixel sensitivity with three different polynomial models ranging between second and sixth order.  After clipping the first 5,000 points ($q$ = 5,000), the eclipse depths using various ramp and intrapixel model components are consistent (see Table \ref{table:cs21Ramps}); however, the SDNR clearly favors BLI and the BIC favors Eq.\ \ref{eqnse}\sb{+} to model the systematics.  To minimize the convergence time in our MCMC chains, we orthogonalize the eclipse depth, system flux, and both ramp parameters ($r\sb{0}$ and $r\sb{1}$).  All of the model fits exhibit, to various degrees, bimodal distributions in the eclipse-midpoint histograms.  The lesser peak occurs at a phase of 0.497 and is likely a result of the model trying to fit the eclipse egress to the points from phases of 0.514 to 0.520, which are consistently above the secondary eclipse by 1 - 2\math{\sigma} (see Figure \ref{fig:lc}, center panel).

\if\submitms y
\clearpage
\fi
\begin{table}[ht]
\centering
\caption{\label{table:cs21Ramps} 
HD149bs21 - Comparing Model Fits}
\begin{tabular}{cccccc}
    \hline
    \hline
    $R(t)$                              & $M(x,y)$      & SDNR      & \math{\Delta}BIC  & Ecl. Depth        \\
                                        &               &           &                   & [\%]              \\
    \hline
    \ref{eqnse}\sb{+}                   & BLI           & 0.0038800 &  0.0              & 0.034 \pm\ 0.006  \\
    \ref{eqnquad}                       & BLI           & 0.0038800 & 1.4               & 0.033 \pm\ 0.007  \\
    \ref{eqnsel}\sb{+}                  & BLI           & 0.0038800 & 10.8              & 0.033 \pm\ 0.006  \\
    \ref{eqnse}\sb{+}                   & Quad. Poly.   & 0.0038961 &  --               & 0.032 \pm\ 0.006  \\
    \ref{eqnse}\sb{+}                   & Cubic Poly.   & 0.0038954 &  --               & 0.033 \pm\ 0.006  \\
    \ref{eqnse}\sb{+}                   & Sextic Poly.  & 0.0038941 &  --               & 0.034 \pm\ 0.006  \\
    \hline
\end{tabular}
\end{table}
\if\submitms y
\clearpage
\fi

\subsection{Three Fits at 5.8 \math{\mu}m}

Using the processes described below, we choose the best-fit model for each event, then perform a joint fit with a single eclipse-depth parameter.

\subsubsection{HD149bs31}

Initially reported by \citet{Stevenson2010}, we find clear evidence of pixelation at 5.8 {\microns} (see Figure \ref{fig:hd149bs31-ip}).  A relatively large bin size of 0.04 pixels is appropriate for HD149bs31 using BLI in combination with Eq.\ \ref{eqnse}\sb{+} to express the time-dependent flux variation.  We orthogonalize the system flux and both ramp parameters when computing uncertainties.  The eclipse-midpoint histogram peaks at a phase of 0.502 and has a broad uncertainty of 0.005.  The best-fit eclipse depth is 0.044 {\pm} 0.016\%.

\subsubsection{HD149bs32}

The pixelation effect in HD149bs32 is weak because stellar centers fall predominantly near the middle of an interpolated subpixel (i.e., away from the blue peaks in the right panel of Figure \ref{fig:gj436bo41-BLISS}).  As such, an intrapixel model component is unnecessary and we model its systematics solely by Eq.\ \ref{eqnse}\sb{+}.  We orthogonolize the same parameters as with HD149bs31.  The eclipse-midpoint histogram shows a strong bimodal distribution, with peak phase values of 0.496 and 0.501.  The latter is favored with an eclipse depth of 0.038 {\pm} 0.017\%.
This data set points to the illegitimacy of fixing one of the ramp parameters (see Section \ref{subsec:ortho}).  The eclipse depth is correlated with the ramp parameters, so fixing one of them erroneously improves the eclipse depth uncertainty to 0.012\%.

\subsubsection{HD149bs33}

For the same reasons as with HD149bs32, no intrapixel model component is needed with HD149bs33.  We model the declining flux using Eq.\ \ref{eqnse}\sb{+}, orthogonolize the same three parameters as above, and find an eclipse depth of 0.047 \pm\ 0.016\%.  The histogram of eclipse midpoints is clearly bimodal but favors the best-fit value of 0.500$^{+0.003}_{-0.005}$.

\subsubsection{5.8 \math{\mu}m Joint Fit}

\begin{figure}[ht]
\if\submitms y
  \setcounter{fignum}{\value{figure}}
  \addtocounter{fignum}{1}
  \newcommand\fignam{f\arabic{fignum}.ps}
\else
  \newcommand\fignam{./figs/hd149b3b4-comb.ps}
\fi
\centering
\includegraphics[width=\linewidth, clip]{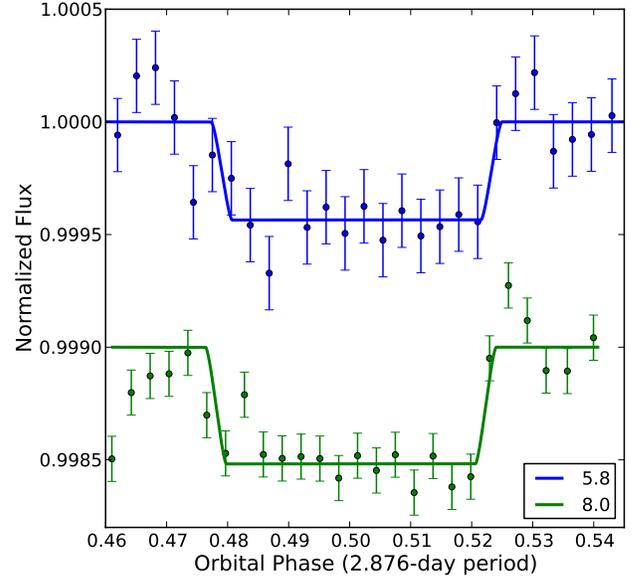}
\caption{Combined and binned eclipse light curves at 5.8 and 8.0 {\microns}.  Note the improved S/N achieved with combined modeling, compared to Figure \ref{fig:lc}.
The best joint-fit HD149bs32 and HD149bs41 models are plotted for comparison.}
\label{fig:hd149b3b4-comb}
\end{figure}

To improve S/N on the eclipse depth, we share this parameter in a joint fit of all three data sets.  We retain individual eclipse-midpoint times for the subsequent orbital analysis.  The MCMC chain converged after $6\times10^5$ iterations.  The combined light curve in Figure \ref{fig:hd149b3b4-comb} illustrates the improvement when compared to the three 5.8 {\micron} light curves in Figure \ref{fig:lc}.  The best simultaneous fit favors an eclipse depth of 0.044 \pm\ 0.010\%.

\subsection{Four Fits at 8.0 \math{\mu}m}

Similarly to the fits at 5.8 {\microns}, we fit models to each data set individually then use the best models in an $1.1\times10^6$ iteration joint fit that shares a common eclipse depth.

\subsubsection{HD149bs41}

The significant improvements in our pipeline since the original analysis by H07 warrant a new analysis of HD149bs41.  We follow all of our current techniques described in Section \ref{sec:obs} and test all of the listed ramp model components.  As with H07 and K09, we find that Eq.\ \ref{eqnse}\sb{--} best describes the overall ramp.  The smaller ramps associated with each telescope movement are best described by Eq.\ \ref{eqnvs}, according to BIC; however we also present H07's 12-point spline for comparison (see Table \ref{table:cs41Models}).  Each model employs a constant-flux offset at each of the nine nod positions, of which eight are free parameters as described in Section \ref{sec:obs}.

\begin{figure}[ht]
\if\submitms y
  \setcounter{fignum}{\value{figure}}
  \addtocounter{fignum}{1}
  \newcommand\fignam{f\arabic{fignum}.ps}
\else
  \newcommand\fignam{./figs/hd149bs41-PointingHist-60.ps}
\fi
\centering
\includegraphics[width=\linewidth, clip]{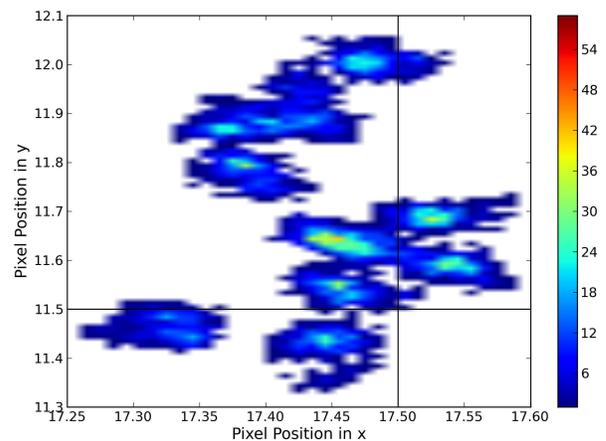}
\caption{Pointing histogram for one of nine nod positions of HD149bs41.  The small, 0.01-pixel bin size clearly shows that the positions of the 12 visits have very little overlap, resulting in a time-dependent position sensitivity and making the data impossible to model accurately using a BLISS map.  The small footprint size demonstrates the difficulty of making a definitive IRAC intrapixel map using all the stellar staring data in
each channel.  The horizontal and vertical black lines represent pixel boundaries.}
\label{fig:cs41pos}
\end{figure}

Due to the nodding motion with this particular data set, BLI and NNI are inappropriate models to use.  We can see in Figure \ref{fig:cs41pos}, which illustrates one of the nine nod positions, that the pixel position is slightly different for each of the twelve visits to this position.  This behavior introduces a strong time dependence in the position sensitivity correction that cannot be disentangled.  Our best attempt to correct for the position sensitivity uses a linear correction in two dimensions for each of the nine nod positions (\math{9\times}Linear).  Table \ref{table:cs41Models} compares the four best model combinations.  Compared to K07, our flux offsets are multiplicative rather than additive (see Eq.\ \ref{eqn:full}) and our final model does not fix either of the ramp parameters.  As a result, our eclipse-depth uncertainty is larger (0.049 \pm\ 0.016\%).

\if\submitms y
\clearpage
\fi
\if\submitms y
\clearpage
\fi

\if\submitms y
\clearpage
\fi
\begin{table}[ht]
\caption{\label{table:cs41Models} 
HD149bs41 - Comparing Model Fits}
\atabon\strut\hfill\begin{tabular}{@{\ }cccccc@{\ }}  
    \hline
    \hline
    R(t)                & M(x,y)                & Visit Sensitivity & SDNR      & \math{\Delta}BIC  & Ecl. Depth \\
                        &                       &                   &           &                   & [\%]   \\
    \hline
    \ref{eqnse}\sb{--}   & -                     & Eq.\ \ref{eqnvs} & 0.0084575 & {\bf 0}           & {\bf 0.049}   \\  
    \ref{eqnse}\sb{--}   & -                     & 12-pt Spline     & 0.0084559 & 59                & 0.049  \\
    \ref{eqnse}\sb{--}   & \math{9\times}Linear  & Eq.\ \ref{eqnvs} & 0.0084510 & 126               & 0.050  \\
    \ref{eqnse}\sb{--}   & \math{9\times}Linear  & 12-pt Spline     & 0.0084494 & 184               & 0.050  \\
    \hline
\end{tabular}\hfill\strut\ataboff
\end{table}
\if\submitms y
\clearpage
\fi

\subsubsection{HD149bs42} 

We use a 0.04-pixel bin size and only consider bins with at least eight points to ignore an outlier near subpixel location (14.36, 15.24).  
Table \ref{table:cs42Ramps} contains $\Delta$BIC values and best-fit eclipse depths from our least-squares minimizer for three different values of the clipping parameter: $q=$ 0, 5,000, and 10,000.  This parameter ignores the given number of data points from the beginning of the observation and is a common procedure \citep{Knutson2011} when trying to find the best-fitting ramp.  The table indicates that all but one of the eclipse depths are consistent for \math{q = 10,000} and that Eqs.\ \ref{eqnseq}\sb{--} and \ref{eqnse2}\sb{--} are consistent at all three $q$ values.  Our MCMC routine finds strong non-linear correlations in all ramp model components except Eq.\ \ref{eqnquad}, which exhibits linear correlations that are easily handled by orthogonalizing the system flux and both ramp parameters.  We use Eq.\ \ref{eqnquad} with $q$ = 10,000 in the joint model fit.  The eclipse depth for the competing solution (Eqn.\ \ref{eqnse}\sb{--}) differs by less than 1$\sigma$.



\if\submitms y
\clearpage
\fi
\begin{table}[ht]
\caption{\label{table:cs42Ramps} 
HD149bs42 - Comparing Model Fits}
\atabon\strut\hfill\begin{tabular}{@{\ }ccccccc@{\ }}
    \hline
    \hline
    q:                  & 0                 & 0     & 5000                  & 5000     & 10000              & 10000     \\
    R(t)                & Ecl. Depth        & \math{\Delta}BIC 
                                                    & Ecl. Depth            & \math{\Delta}BIC 
                                                                                        & Ecl. Depth        & \math{\Delta}BIC  \\
                        & [\%]              &       & [\%]                  &           & [\%]              &           \\
    \hline
    \ref{eqnse}\sb{--}  & 0.149             & 59    & 0.111                 &  1        & 0.068             &  0        \\
    \ref{eqnsel}\sb{--} & 0.030             &  4    & 0.056                 &  2        & 0.068             & 11        \\
    \ref{eqnseq}\sb{--} & 0.065             &  7    & 0.065                 &  9        & 0.063             & 19        \\
    \ref{eqnse2}\sb{--} & 0.065             &  8    & 0.062                 &  9        & 0.061             & 20        \\
    \ref{eqnll}         & 0.132             & 47    & 0.099                 &  6        & 0.067             & 10        \\
    \ref{eqnlq}         & 0.096             & 37    & 0.086                 & 15        & 0.065             & 20        \\
    \ref{eqnlog}        & 0.087             &  0    & 0.071                 &  0        & 0.068             & 10        \\
    \ref{eqnl4q}        & 0.062             & 16    & 0.052                 & 17        & 0.049             & 28        \\
    \ref{eqnquad}       & 0.197             &205    & 0.128                 & 31        & {\bf 0.064}       &  {\bf 0}        \\
    \hline
\end{tabular}\hfill\strut\ataboff
\end{table}
\if\submitms y
\clearpage
\fi

\subsubsection{HD149bs43} 

We choose a relatively large bin size of 0.03 pixels because the intrapixel effect is minimal (position dependence is flat) and we do not want to overfit the edges of position space where there are few data points.  Smaller bin sizes result in similar eclipse depths.  Without BLI, we find an eclipse depth of 0.040 \pm\ 0.008\%, nearly identical to K09; however, this fit does not have the lowest BIC value (\math{\Delta}BIC = 112).  There is a small but clear position dependence that, once accounted for, results in a marginally deeper eclipse of 0.044 \pm\ 0.008\%.  We confirm that the eclipse midpoint is noticeably earlier (by 4$\sigma$) than the expected phase value of 0.5, but do not claim a detection of eclipse timing variation because HD149bs21 measures the preceding eclipse with a much stronger S/N and is consistent with a circular orbit.  Fixing the phase of mid-eclipse to 0.5 results in a marginally shallower eclipse depth (0.039 \pm\ 0.008\%) and a larger SDNR value.

\subsubsection{HD149bs44}

Relative to the other 8.0 {\micron} events, HD149bs44 exhibits significantly larger SDNR and uncertainty scaling factor values (see Table \ref{tab:fits2} in Appendix \ref{sec:app}).  The uncertainty scaling factor renormalizes the error bars such that \math{\chi_{\nu}^{2}=1}, so a smaller scaling factor indicates a better fit.  For this noisy data set, the models achieve relatively poor fits compared to other 8.0 {\micron} data sets.

The ramp models that produce the most consistent eclipse depths in Table \ref{table:cs44Ramps} are Eqs.\ \ref{eqnseq}\sb{--} and \ref{eqnquad} for $q=$ 5,000, 7,500, and 10,000.  Smaller values of $q$ produce inconsistent eclipse depths for all ramp models; larger $q$ values do not provide sufficient out-of-eclipse baselines.  For $q=$ 7,500, most of the ramps find eclipse depths that are in agreement with the consistent values given by Eqs.\ \ref{eqnseq}\sb{--} and \ref{eqnquad}.  Of these ramps, Eqs.\ \ref{eqnse}\sb{--} and \ref{eqnquad} share the lowest BIC value; however, the quadratic parameter in Eq.\ \ref{eqnquad} correlates strongly with the eclipse depth, resulting in a larger uncertainty (0.017 \vs 0.013), so we select Eq.\ \ref{eqnse}\sb{--} with $q$ = 7,500 for our final joint model.
This data set also applies a BLISS map with 0.025-pixel bin sizes and at least ten points per bin.


\if\submitms y
\clearpage
\fi
\begin{table}[ht]
\caption{\label{table:cs44Ramps} 
HD149bs44 - Comparing Model Fits}
\atabon\strut\hfill\begin{tabular}{@{\ }c@{\ }c@{\ }c@{\ }c@{\ }c@{\ }c@{\ }c@{\ }}
    \hline
    \hline
    q:                  & 5000      & 5000  & 7500                  & 7500      & 10000     & 10000     \\
    R(t)                & Ecl. Depth & \math{\Delta}BIC 
                                            & Ecl. Depth & \math{\Delta}BIC 
                                                                                & Ecl. Depth  & \math{\Delta}BIC \\
                        & [\%]      &       & [\%]                  &           & [\%]        &         \\
    \hline
    \ref{eqnse}\sb{--}  & 0.072     &  0    & {\bf0.066} 
                                                                    & {\bf 0}   & 0.085     &  0        \\
    \ref{eqnsel}\sb{--} & 0.075     & 11    & 0.069                 & 11        & 0.085     & 11        \\
    \ref{eqnseq}\sb{--} & 0.073     & 23    & 0.068                 & 22        & 0.069     & 24        \\
    \ref{eqnse2}\sb{--} & 0.072     & 22    & 0.066                 & 22        & 0.085     & 21        \\
    \ref{eqnll}         & 0.074     & 11    & 0.069                 & 11        & 0.082     & 11        \\
    \ref{eqnlq}         & 0.089     & 17    & 0.096                 & 10        & 0.086     & 22        \\
    \ref{eqnl4q}        & 0.073     & 33    & 0.067                 & 22        & 0.081     & 32        \\
    \ref{eqnquad}       & 0.073     &  1    & 0.069 
                                                                    &  0        & 0.069     &  3        \\
    \hline
\end{tabular}\hfill\strut\ataboff
\end{table}
\if\submitms y
\clearpage
\fi

\comment{
\begin{table}[ht]
\caption{\label{table:cs44Ramps} 
HD149bs44 - Comparing Model Fits}
\atabon\strut\hfill\begin{tabular}{cccccc}
    \hline
    \hline
    R(t)                & SDNR      & \math{\Delta}BIC   & Ecl. Depth [\%]  \\
    \hline
    \ref{eqnse}\sb{--}\tablenotemark{1}   & 0.0085244 & 0             & 0.068 \pm\ 0.010      \\
    \ref{eqnse}\sb{--}   & 0.0085244 & 11            & 0.068 \pm\ 0.010      \\
    \ref{eqnsel}\sb{--}  & 0.0085243 & 21            & 0.070 \pm\ 0.011      \\
    \ref{eqnseq}\sb{--}  & 0.0085244 & 32            & 0.067 \pm\ 0.011      \\
    \ref{eqnse2}\sb{--}  & 0.0085244 & 32            & 0.068 \pm\ 0.011      \\
    \ref{eqnll}         & 0.0085243 & 21            & 0.069 \pm\ 0.011      \\
    \ref{eqnlq}         & 0.0085219 & 20            & 0.095 \pm\ 0.014      \\
    \ref{eqnl4q}        & 0.0085244 & 43            & 0.066 \pm\ 0.015      \\
    \ref{eqnquad}       & 0.0085244 & 11            & 0.067 \pm\ 0.010      \\
    \hline
\end{tabular}\hfill\strut\ataboff
\tablenotetext{1}{$t\sb{0}$ parameter is fixed to its best-fit value.}
\end{table}
}

\subsection{Two Fits at 16 \math{\mu}m}

Each event consists of 1050 exposures, divided serially into three 350-image sequences at non-overlapping stellar centers on the detector.  The second sequence ($p$=1) is completely within the secondary eclipse and, because of the free position offset parameter, has no impact on the eclipse depth (but still helps to model the systematics).  Unfortunately, this means that the eclipse depth is completely determined by the small number of points after ingress in the first sequence and before egress in the last sequence.

To avoid residual effects from potentially bright previous targets, neither of the two 16 {\micron} observations was positioned at the center of the array.  As a result, the outer radius of the sky annulus is limited to 11 and 12 pixels for HD149bs51 and HD149bs52, respectively, to avoid the flux falloff at the edges of the blue peak-up array.  We apply both aperture and optimal photometry \citep{Horne1986, Deming2005}; the former produces cleaner results for both data sets. 



\begin{figure}[htb]
\if\submitms y
  \setcounter{fignum}{\value{figure}}
  \addtocounter{fignum}{1}
  \newcommand\fignama{f\arabic{fignum}a.ps}
  \newcommand\fignamb{f\arabic{fignum}b.ps}
  \centering
  \includegraphics[width=0.7\columnwidth, clip]{\fignama}
  \includegraphics[width=0.7\columnwidth, clip]{\fignamb}
\else
  \newcommand\fignama{./figs/hd149bs51_AllPhotap.ps}
  \newcommand\fignamb{./figs/hd149bs52_AllPhotap.ps}
  \centering
  \includegraphics[width=\columnwidth, clip]{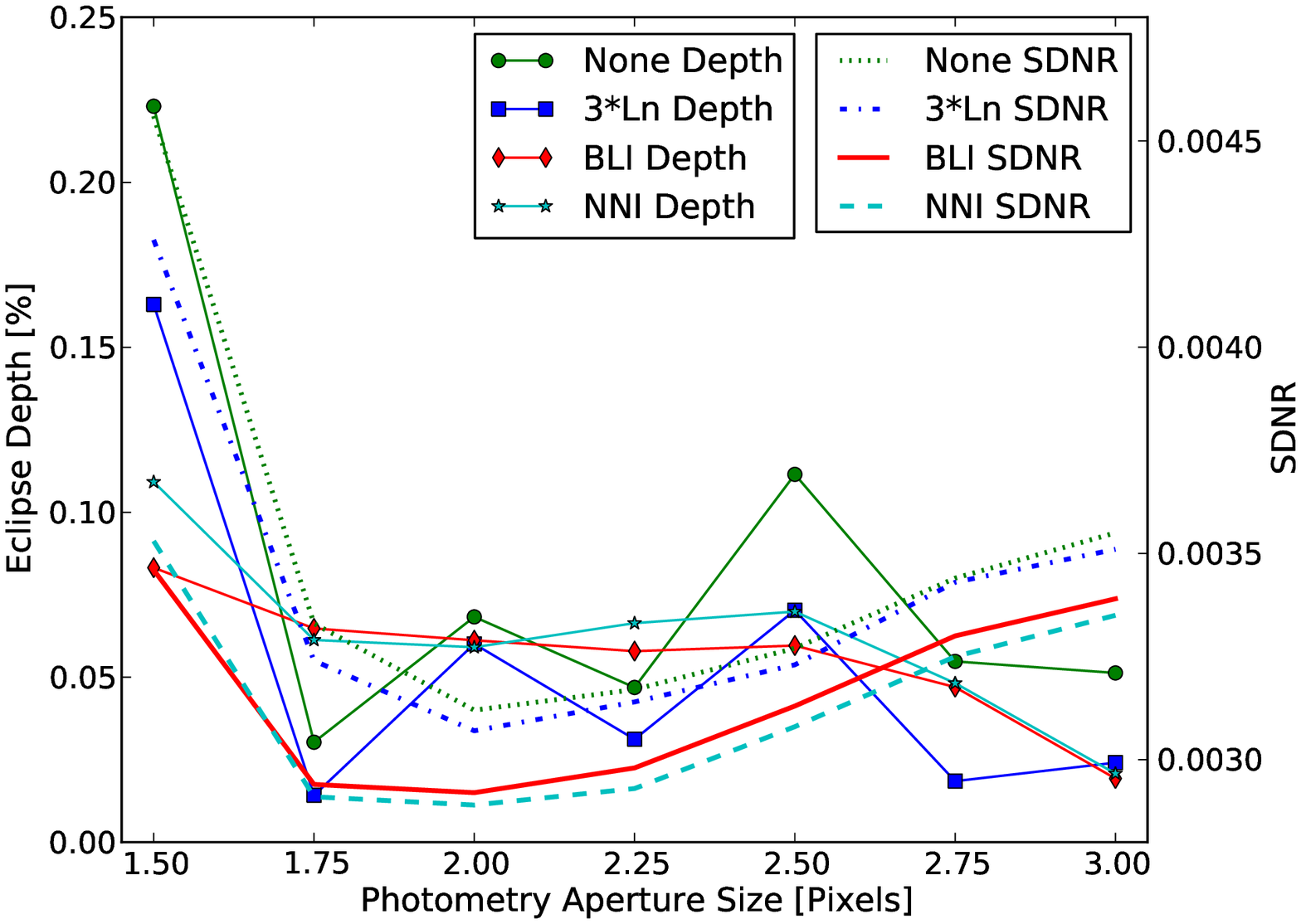}
  \includegraphics[width=\columnwidth, clip]{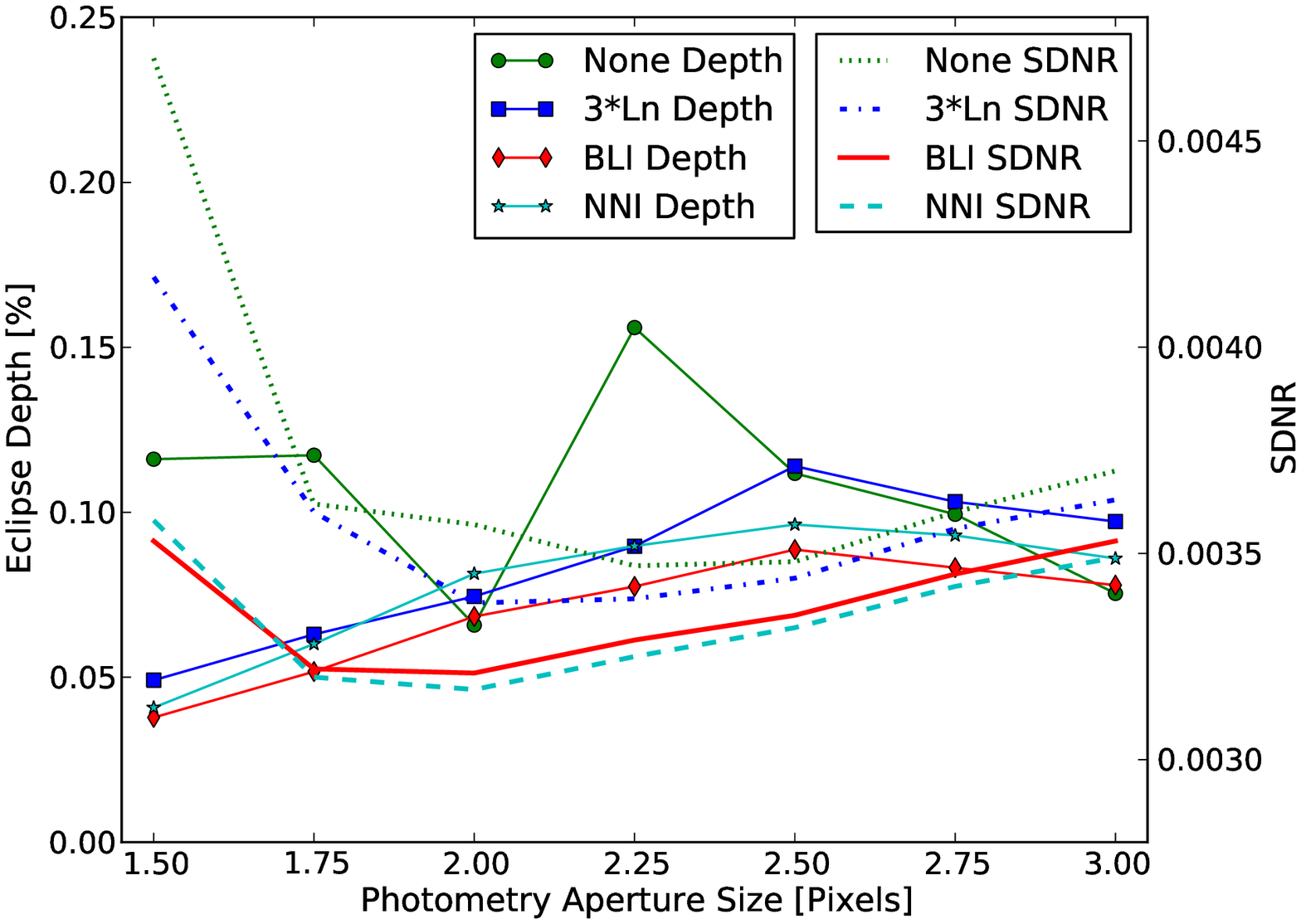}
\fi
\caption{Best-fit eclipse depths and corresponding SDNR values for HD149bs51 and HD149bs52.
Both are plotted as a function of photometry aperture size for the HD149bs51 (upper panel) and HD149bs52 (lower panel) data sets.  The labels refer to the type of intrapixel model component used.  Here, `None' uses no model, `3*Ln' uses a linear function in $x$ and $y$ at each of the three positions, and `BLI' and `NNI' both use our new BLISS mapping technique with 0.02-pixel bins.  In all but one case, the best aperture size is 2.0 pixels, according to SDNR.  Due to the weak eclipse signal, the phase of mid-eclipse is fixed to 0.5.  A typical \math{1\sigma} eclipse-depth uncertainty is 0.037\%.}
\label{fig:hd149bs51apsizes}
\end{figure}

Figure \ref{fig:hd149bs51apsizes} displays the best-fit eclipse depths and corresponding SDNR values versus aperture size for four competing models in each data set.  The first applies no intrapixel correction, the second uses three linear components (one at each position), the third uses BLI, and the final model uses NNI.  The first two models also apply a position offset to account for the differences in pixel sensitivity.  This offset is redundant for BLI and NNI.

Although these models find similar (\math{< 1\sigma}) eclipse depths at the best aperture size of 2.0 pixels, this is not the case for other aperture sizes.  Models using BLI or NNI produce the most consistent eclipse depths and result in lower SDNR values (see Tables \ref{table:cs51Models} and \ref{table:cs52Models}).  We test a range of bin sizes for BLISS mapping but find that NNI consistently outperforms BLI.  We take this as evidence that there are insufficient data points in most bins to model the weak position dependence accurately.  As such, we consider only the last two models and select no intrapixel model for HD149bs51 and the \math{3\times}linear model for HD149bs52 to perform the joint fit.  For improved convergence in our MCMC chains, we orthogonalize the system fluxes and both ramp terms in Eqs.\ \ref{eqnquad} and \ref{eqnse}\sb{--} for HD149bs51 and HD149bs52, respectively.  Due to the weak eclipse signal, the midpoint is fixed to a phase of 0.5 for the individual fits; the joint fit achieves a sufficient S/N to share the eclipse midpoint as a free parameter.  The final model fit uses $1.6\times10^6$ iterations and is in Table \ref{tab:fits2} of Appendix \ref{sec:app}.

\if\submitms y
\clearpage
\fi
\begin{table}[ht]
\caption{\label{table:cs51Models} 
HD149bs51 - Comparing Model Fits}
\atabon\strut\hfill\begin{tabular}{ccccc}
    \hline
    \hline
    R(t)            & M(x,y)                & SDNR      & \math{\Delta}BIC  & Ecl. Depth [\%]   \\
    \hline
    \ref{eqnquad}   & NNI                   & 0.002890  & 0                 & 0.059 \pm\ 0.038  \\
    \ref{eqnquad}   & BLI                   & 0.002924  & 21                & 0.061 \pm\ 0.035  \\
    \ref{eqnquad}   & -                     & 0.003119  & 169               & 0.068 \pm\ 0.038  \\
    \ref{eqnquad}   & \math{3\times}Linear  & 0.003066  & 175               & 0.060 \pm\ 0.037  \\
    \hline
\end{tabular}\hfill\strut\ataboff
\end{table}
\if\submitms y
\clearpage
\fi


\if\submitms y
\clearpage
\fi
\begin{table}[ht]
\caption{\label{table:cs52Models} 
HD149bs52 - Comparing Model Fits}
\atabon\strut\hfill\begin{tabular}{ccccc}
    \hline
    \hline
    R(t)                & M(x,y)                & SDNR      & \math{\Delta}BIC  & Ecl. Depth [\%]   \\
    \hline
    \ref{eqnse}\sb{--}   & NNI                   & 0.003166  & 0                 & 0.081 \pm\ 0.036  \\
    \ref{eqnse}\sb{--}   & BLI                   & 0.003211  & 23                & 0.068 \pm\ 0.035  \\
    \ref{eqnse}\sb{--}   & \math{3\times}Linear  & 0.003378  & 175               & 0.074 \pm\ 0.038  \\
    \ref{eqnse}\sb{--}   & -                     & 0.003574  & 243               & 0.066 \pm\ 0.034  \\
    \hline
\end{tabular}\hfill\strut\ataboff
\end{table}
\if\submitms y
\clearpage
\fi


\if\submitms y
\clearpage
\fi
\if\submitms y
\clearpage
\fi

We find that the eclipse depths for both data sets vary with the choice of eclipse midpoint.  Using the midpoint from our joint model (0.5015, see Table \ref{tab:fits2}), we find that the best-fit eclipse depth differs by up to $1\sigma$ from the values listed in Tables \ref{table:cs51Models} and \ref{table:cs52Models}, where the midpoint is fixed to 0.5.
We also test joint fits without the data at $p=1$ and find a comparable joint-fit eclipse depth of 0.099 \pm\ 0.035\%.  The small change is likely the result of weak correlations with the ramp parameters.  We include all positions in the final fit as this results in a lower SDNR and a small improvement in the eclipse depth uncertainty.

\section{ATMOSPHERE}
\label{sec:atm}
 
In order to understand the atmosphere of the planet better, we compare 
our measured flux ratios to those generated from a model atmosphere.  We 
simulate the atmosphere of HD 149026b using the model presented by 
\citet{Fortney05,Fortney2006,Fortney08}.  See \citet{Fortney08} for a 
description of the heritage of the model, which includes solar system 
planets and brown dwarfs in addition to exoplanets.  The chemical 
mixing ratios used assume chemical equilibrium, following 
\citet{Lodders02}, at both solar metallicity (``$1\times$ solar'') as 
well as $30\times$ solar, using the abundances of \citet{Lodders03}. 
The opacity database is described by \citet{Freedman08}, with an update 
to include CO$_2$ opacity.
 
We have generated chemistry/opacity grids with and without the opacity 
of gaseous TiO/VO.  These gases, which are strong absorbers of optical 
flux, may be responsible for the temperature inversions diagnosed in the 
atmosphere of some planets \citep[e.g.,][]{Hubeny03,Fortney2006,Fortney08}, but see \citet{Spiegel2009} for important caveats.  The mid-infrared flux ratios observed for hot Jupiters are quite 
diverse, but high flux ratios (and corresponding large brightness 
temperatures) in the mid-infrared, together with small 3.6-to-4.5 
{\micron} ratios, have been found in models with temperature inversions 
\citep{Fortney2006,Burrows07,Burrows08,Madhu2009,Spiegel2010}, and these models have had some 
success in comparisons with data \citep[see][for a review]{Seager2010}. 
  For HD 149026b, we find a 3.6-to-4.5 {\micron} ratio $>1$, similar to 
HD 189733b, which indicates no temperature inversion.
 
\begin{figure}[htb]
\if\submitms y
  \setcounter{fignum}{\value{figure}}
  \addtocounter{fignum}{1}
  \newcommand\fignam{f\arabic{fignum}.eps}
\else
  \newcommand\fignam{./figs/HD149_ratios_2011.eps}
\fi
\centering
\includegraphics[width=\linewidth, clip]{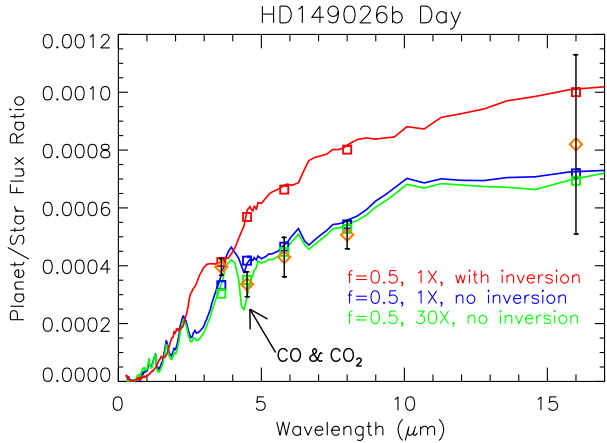}
\caption{Atmospheric models of the dayside of HD 149026b.  The red line 
depicts a model with a temperature inversion, $f=0.50$, solar 
metallicity, and an effective day-side temperature, $T$\sb{eff}, of 2067 K.  
The blue model has no temperature inversion, $f=0.50$, 
solar metallicity, and $T$\sb{eff} = 2056 K.  
The green model is similar to the blue model but 
with enhanced metallicity ($30\times$ solar) and $T$\sb{eff} = 2081 K.  
Model band-averaged ratios are shown as squares while the data points are orange diamonds. 
Models that lack temperature inversions and include a high atmospheric 
metallicity best match the observed data.}
\label{fig:atm}
\end{figure}

We compare the measured flux ratios to three different models in Figure 
\ref{fig:atm}.  All assume redistribution of absorbed stellar flux over 
the dayside of the planet only \citep[$f=1/2$, as described by][]{Fortney07b}.  We show a $1\times$ solar model with TiO/VO opacity (which 
yields an inversion) and the same model with TiO/VO opacity neglected. 
We also plot a similar no-inversion model at $30\times$ solar 
metallicity.  This last model leads to mixing ratios $\sim$30 and 
$\sim$900 times larger for CO and CO$_2$, respectively, compared to the 
solar metallicity case \citep{Zahnle09}.  The dramatic increases in CO 
and CO$_2$ lead to the most notable spectral difference, the much deeper 
absorption due to the overlapping 4.5 {\micron} band of CO and 4.2 
{\micron} band of CO$_2$.  Such a high metallicity for this atmosphere 
may well be realistic, as Saturn is $\sim10\times$ solar in carbon 
\citep{Flasar05}, while Uranus and Neptune are $\sim30-60\times$ enriched.
 
\if\submitms y
\clearpage
\fi
\if\submitms y
\clearpage
\fi

Models with a temperature inversion similar to the red line shown in 
Figure \ref{fig:atm} are clearly disfavored, given the 3.6-to-4.5 
{\micron} ratio $>1$ and the large flux ratios at redder wavelengths. 
Our best fit for the three models is the $30\times$ enhanced, 
no-inversion model.  Only the 3.6 {\micron} point is outside our 
1$\sigma$ error bar.  Even with the strong CO/CO$_2$ absorption, we 
still cannot quite match the observed 3.6-to-4.5 {\micron} ratio.  At 
face value, this would imply even larger CO and/or CO$_2$ mixing ratios; 
however, this may not necessarily be the case, as some other modelers 
\citep[e.g.][]{Burrows08} generally show a deeper absorption feature 
at 4.5 {\microns} than we obtain with our models.  This is likely due to 
differences in the temperature gradient as a function
of height in the different models, as 
this helps to control the depth of absorption features.  A steeper 
gradient leads to deeper absorption features.
 
Less-efficient redistribution of absorbed flux leads to a hotter 
dayside, and still would yield a satisfactory fit to the observations, 
albeit near the top of the 1$\sigma$ error bars.  Given the 8.0 
{\micron} phase curve of K09, which showed modest 
day/night phase variation, such a hot dayside (which would leave little 
energy for the night side) is not favored.  We recommend that a coupled 
3D dynamics/radiative transfer model be run for the system, to 
understand if the implied day- and night-side temperatures can be 
matched.  \citet{Showman09} had good success in matching the phase curve 
and dayside photometry of HD 189733b.   These models tend to show better 
day-night homogenization of temperature contrasts than one would assume 
from the fact that our best-fit 1D model assumes all absorbed flux is 
re-radiated on the planet's day side.  Additionally, near-infrared 
fluxes from the JHK-band \citep[e.g.,][]{Croll2010}, where this planet is 
brightest, would help to understand the dayside luminosity; however, given the small planet-to-star radius ratio, this may have to wait for the James Webb Space Telescope.
 
\citet{Knutson2010} suggest that the absence of a temperature inversion within an exoplanet atmosphere correlates with higher levels of chromospheric activity from the host star.  The lack of a temperature inversion in HD 149026b does not agree with HD 149026's relatively low activity level, but this may be due to the exoplanet's high density \citep{Knutson2010}.
 
\begin{figure*}[tb]
\if\submitms y
  \setcounter{fignum}{\value{figure}}
  \addtocounter{fignum}{1}
  \newcommand\fignam{f\arabic{fignum}.eps}
\else
  \newcommand\fignam{./figs/contrib.PT.149.2011.eps}
\fi
\centering
\includegraphics[width=15.0cm, clip]{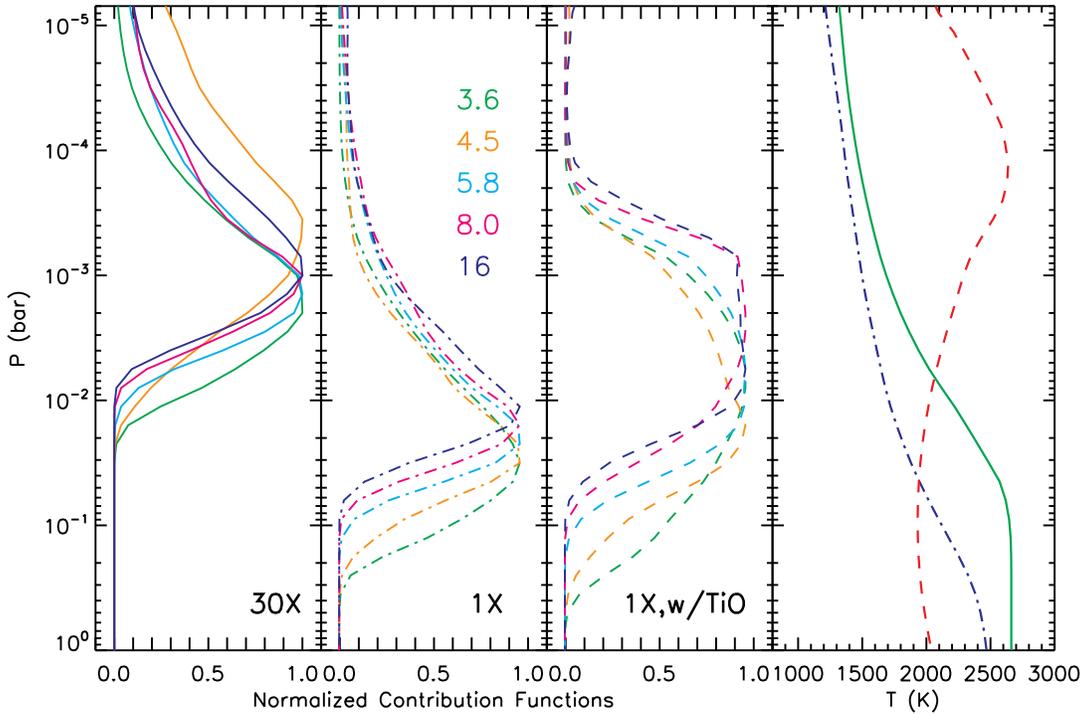}
\caption{Contribution functions and atmospheric pressure-temperature profiles.
{\em Left 3 panels}:  The contribution functions in the five Spitzer bandpasses are for the
three models shown in Figure \ref{fig:atm}.  Emission generally comes from higher in
the atmosphere for the metal-enriched model (left) and, to a lesser extent, the model that features
a temperature inversion (right).  {\em Right-most panel}:  The atmospheric pressure-temperature profiles are for these same models, colored to match Figure \ref{fig:atm}.  The 30$\times$ solar no-inversion model is everywhere warmer than the 1$\times$ solar no-inversion model, but they have very similar $T_{\rm eff}$ values, since the emission from the 30$\times$ model comes from much higher in the atmosphere.}
\label{fig:pt}
\end{figure*}

The pressure-temperature profiles for the three atmospheric models are 
shown in Figure \ref{fig:pt}.  Also plotted are the contribution 
functions \citep[e.g.,][]{ChambHunt} for thermal flux in each of the five 
{\em Spitzer} bandpasses.  Contribution functions trend towards lower 
pressure with enhanced metallicity for all bandpasses, but move most 
dramatically at 4.5 {\microns} due to CO and CO$_2$.  At constant metallicity, a temperature inversion 
tends to smear the contributions over a wider pressure range, favoring 
lower pressures due to the hot upper atmosphere, but to a lesser extent 
compared to models with increasing metallicity.

\if\submitms y
\clearpage
\fi
\if\submitms y
\clearpage
\fi

\section{ORBIT}
\label{sec:orbit}

We have collected a total of eleven individual {\em Spitzer} secondary-eclipse observations with useful timing of HD 149026b over a 3.5 year baseline.  These times constrain \math{e \cos \omega}, where $e$ is the eccentricity and $\omega$ is the argument of periapsis, and can be used to establish eccentricity limits on the planet's orbit.  We use BJD\sb{TDB} given in Tables \ref{tab:fits1} and \ref{tab:fits2} and correct for the eclipse-transit light-time (42 seconds).  The mean eclipse phase, using the K09 ephemeris, is \math{0.49997 \pm 0.00028}, suggesting that \math{e \cos \omega =  -0.00003 \pm 0.00044}.  The data are consistent with a circular orbit (\math{e < 0.0013}).  The times of secondary eclipse do not show any significant trends and do not have a period that differs significantly from the period determined from transit and radial velocity data. Such a difference would indicate apsidal motion or other secular effects \citep{Gimenez1995, Heyl2007}.  An MCMC ephemeris fit to our secondary-eclipse times gives a period of \math{2.875884 \pm 0.000006} days, and a fit of all the available transit times gives a period of \math{2.8758922 \pm 0.0000015} days.  The difference between the two periods is not significant (1.4$\sigma$).  If the measured period difference is due to apsidal motion, then it would indicate that \math{\dot{\omega}e\sin \omega = (9 \pm 7) \times 10^{-5}} \degree /day \citep{Gimenez1995}, where $\dot{\omega}$ is the rate of apsidal precession.  Further secondary-eclipse observations will refine the secondary-eclipse period.

We use our primary-transit and secondary-eclipse data to perform a fit, as described by \citet{Campo2011}, that also incorporates other available transit data \citep{Carter2009, Winn2008, Charbonneau2006} and radial velocity data \citep{Sato2005, Butler2006}.  When we fit an eccentric orbit to the available data (Table \ref{tab:orbit}), we determine that \math{e = 0.154 \pm 0.016}.  Although this is a 10$\sigma$ eccentricity, it is almost completely dominated by the \math{e \sin \omega} component, leading us to believe that the eccentricity may be an overestimate \citep{Laughlin2005}.  This is possible when the peaks of the radial velocity curve, where the waveform is most sensitive to changes in \math{e \sin \omega}, are undersampled.  The eccentricity affects the symmetry of the RV curve, so when both peaks are not well sampled, the best-fit solution may misrepresent the actual eccentricity of the planet's orbit.  Indeed, 16 of the 23 usable RV data points were taken at a transit phase greater than 0.5 and there are no data points between phase values of 0.1 and 0.3.  The dearth of RV measurements near 0.25 signifies that only one of the two peaks is adequately constrained.  To best refine the value of \math{ e \sin \omega}, we require additional RV measurements between phases 0 and 0.5, particularly near 0.25.

In a comparison fit assuming a circular orbit (see Table \ref{tab:orbit2}), where the RV curve is perfectly sinusoidal and symmetric, BIC is worse than for the eccentric fit.  Despite this, the undersampled radial velocity data and the high degree of consistency of the eclipse phases with 0.5 make it unlikely that the orbit of this planet has an eccentricity greater than the maximum value of \math{|e \cos \omega|}. A near-perfect alignment of the system's semi-major axis with our line-of-sight ($\omega \sim 90\degree$) would be necessary, but the agreement between the transit and eclipse durations (see Section \ref{sec:eclresults}) argues against this scenario.
Acknowledging that our secondary-eclipse timing measurements yield little information about \math{e \sin \omega}, we present both solutions without judgment.  Although an eccentric orbit is unlikely, it cannot be ruled out with the data currently available.

\if\submitms y
\clearpage
\fi
\begin{table}[ht]
\centering
\caption{\label{tab:orbit} Eccentric Orbital Model}
\begin{tabular}{rr@{\,{\pm}\,}lr@{\,}}
\hline
\hline
Parameter                                       & \mctc{Value}              \\
\hline
\math{e \sin \omega}              	            & 0.154         & 0.016     \\
\math{e \cos \omega}              		        & -0.00037      & 0.00044   \\
\math{e}                          		        & 0.154         & 0.016     \\
\math{\omega} (\degree)           		        & 90.14         & 0.16       \\
\math{P} (days)                   		        & 2.8758919     & 0.0000014 \\
\math{T\sb{0}} (MJD\sb{TDB})\tablenotemark{a}   & 4597.70716    & 0.00016   \\
\math{K} (ms\sp{-1})                            &  47.4         & 1.1       \\
\math{\gamma} (ms\sp{-1})                       &  -4.3         & 0.6       \\
BIC                                             &  \mctc{123}              \\    
\hline
\end{tabular}
\begin{minipage}[t]{0.63\linewidth}
\tablenotetext{1}{MJD\sb{TDB} = BJD\sb{TDB} - 2,450,000}
\end{minipage}
\end{table}
\if\submitms y
\clearpage
\fi

\if\submitms y
\clearpage
\fi
\begin{table}[ht]
\centering
\caption{\label{tab:orbit2} Circular Orbital Model}
\begin{tabular}{rr@{\,{\pm}\,}lr@{\,}}
\hline
\hline
Parameter                                       & \mctc{Value}              \\
\hline
\math{P} (days)                   		        & 2.8758916     & 0.0000014 \\
\math{T\sb{0}} (MJD\sb{TDB})\tablenotemark{a}   & 4597.70713    & 0.00016   \\
\math{K} (ms\sp{-1})                            &  42.6         & 0.9       \\
\math{\gamma} (ms\sp{-1})                       &  -1.6         & 0.6       \\
BIC                                             &  \mctc{179}              \\    
\hline
\end{tabular}
\begin{minipage}[t]{0.63\linewidth}
\tablenotetext{1}{MJD\sb{TDB} = BJD\sb{TDB} - 2,450,000}
\end{minipage}
\end{table}
\if\submitms y
\clearpage
\fi

\section{CONCLUSIONS}
\label{sec:concl}

Over 3.5 years, {\em Spitzer} observed three primary transits and eleven secondary eclipses of HD 149026b in five infrared wavelengths.  We utilize multiple observations for channels with the weakest eclipse depths to improve S/N estimates and better constrain the dayside atmospheric composition.  
The addition of a third transit event at 8.0 {\microns} confirms previous results \citep{Nutzman2009, Carter2009, Knutson2009b} and offers an improved constraint on the planet-to-star radius ratio ($R\sb{p}/R\sb{\star}$).
A new eclipse analysis of HD149bs41 confirms the findings from \citet{Knutson2009b}, namely that an eclipse depth of $\sim0.05\%$ fits best at 8.0 {\microns}.  However, we find a larger uncertainty due to correlations between the eclipse depth and ramp parameters that were not fully explored because one of the ramp parameters was previously fixed.

The atmosphere is explained well by a 1D chemical-equilibrium model.  A temperature inversion is no longer favored when fitting the observed planet-to-star flux ratios.  The best-fit model includes large amounts of CO and CO\sb{2}, moderate heat redistribution ($f=0.5$), and strongly enhanced metallicity ($30\times$ solar).
Using the times from our secondary-eclipse observations, we find no deviations from a circular orbit at the \math{1\sigma} level.  However, given the available RV data, we cannot completely rule out an eccentric orbit with an unlikely orbital alignment.

We present a new technique, called BiLinearly-Interpolated Subpixel Sensitivity (BLISS) mapping, to model {\em Spitzer's} position-dependent systematics (intrapixel variability and pixelation).  In all cases tested to date, BLISS mapping outperforms previous methods in both speed and goodness of fit.
We also apply an orthogonalization technique for linearly-correlated parameters that accelerates the convergence of Markov chains that employ the Metropolis random walk sampler.

\acknowledgments

We thank N. Lust, contributors to SciPy, Matplotlib, and the Python Programming
Language, the free and open-source community, the
NASA Astrophysics Data System, and the JPL Solar System Dynamics group
for software and services.  This work is based on
observations made with the {\em Spitzer Space Telescope}, which is operated
by the Jet Propulsion Laboratory, California Institute of Technology
under a contract with NASA.  Support for this work was provided by
NASA through an award issued by JPL/Caltech.
\\

\bibliography{ms}

\newpage
\begin{appendices}

\section{Model comparison with complex models}
\label{sec:tom}

\newcommand{\like}{\mathcal{L}}  
\newcommand{\mlike}{\mathcal{M}}  

In Section \ref{sec:bilinint}, we show that BLISS mapping improves the SDNR of models that include an intrapixel sensitivity term.
Elsewhere in this paper and in our
earlier work \citep{Campo2011} we use the Bayesian Information Criterion (BIC) to
compare models of other systematic error components (e.g., rival parametric
ramp models), but we did not use the BIC to compare intrapixel sensitivity
models.  In this Appendix we discuss the approximations underlying the BIC
to elucidate when it is useful, and in particular, to explain why we do not
use it (or similar statistical criteria) for comparison of intrapixel
sensitivity models.

The BIC provides an asymptotic approximation to quantities that may be used
for Bayesian quantification of model uncertainty.  The most simple use of
the BIC is to approximate Bayesian explanatory model selection.
A basic distinction among model selection criteria is between explanatory
criteria (that seek the model that best describes the processes that
produced the available data) and predictive criteria (that seek the model
that will make the most accurate predictions of future data based on the
available data).  For example, the well-known Akaike Information Criterion 
(AIC = $\chi^2 + 2k$)
is a predictive criterion.  Although different from the BIC in rationale, its
derivation invokes similar asymptotic approximations to those we describe
here for the BIC, and we consider it to be similarly hampered for comparing
large models.
To illuminate the nature of the approximations underlying the BIC, we
sketch its derivation here.

Consider a set of models $\{M_i\}$ for observed data $D$.  Let $\theta_i$
denote the parameters of model $i$, with $k_i$ dimensions.  The (Bayesian)
posterior probability for model $i$ is proportional to the product of a prior
probability for the model, and the model's {\em marginal likelihood},
\begin{equation}
\mlike_i = \int d\theta_i\; \pi_i(\theta_i) \like_i(\theta_i),
\label{mlike-def}
\end{equation}
where $\pi_i(\theta_i) \equiv p(\theta_i|M_i)$ is the prior probability density
function (PDF) for the model's parameters, and
$\like_i(\theta_i) \equiv p(D|\theta_i, M_i)$ is the likelihood function,
the probability for the data presuming the model holds with parameters
$\theta_i$ (the likelihood function is proportional to
$\exp[-\chi^2(\theta_i)/2]$ for our models).  As its name suggests, the
role of the marginal likelihood in quantifying model uncertainty is
completely analogous to the role of the more familiar likelihood function in
quantifying parameter uncertainty (within a particular model).
Note that $\mlike_i$ is an {\em average} of the likelihood function for
model $i$, not a {\em maximum}:  in Bayesian inference, the weight of the
evidence for a model is given by the typical value of the likelihood
function for its parameters, not the optimum (largest) value.

For nonlinear models with more than a few dimensions, calculation
of the integral in Eq.\ \ref{mlike-def} is not feasible using
standard quadrature methods.  The development of algorithms for accurate 
calculation of marginal likelihoods is an active research area, and existing
algorithms are typically problem-specific and computationally expensive
(see Clyde et al.\ 2007 for a recent review targeting astronomers).

\begin{figure}
\centering
\includegraphics[width=0.8\linewidth, clip]{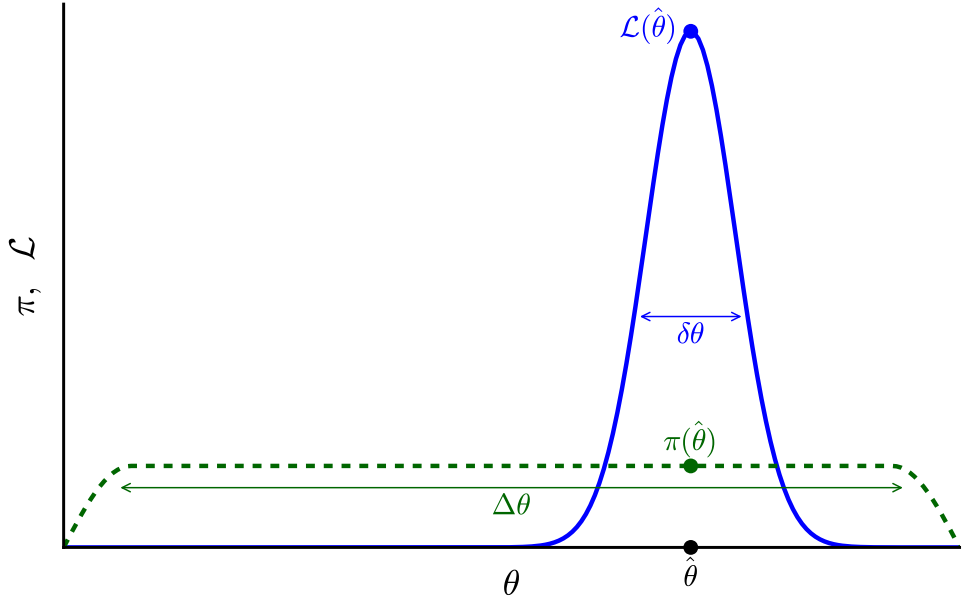}
\caption{Ingredients for the derivation of the BIC approximation to the
marginal likelihood. Curves show the likelihood function (solid blue) and
prior PDF (dashed green), with characteristic widths $\delta\theta$ and
$\Delta\theta$.  Points show the maximum likelihood parameter estimate,
$\hat\theta$, and the values of the likelihood and prior at $\hat\theta$.}
\label{fig:MargLike}
\end{figure}

The BIC approximates $-2\ln\mlike$ for straightforward models in the limit
of voluminous data, i.e., asymptotically (we drop the model index, $i$, when
referring to a generic model, to simplify notation).  In this limit, the
likelihood function $\like(\theta)$ will be strongly peaked at the
maximum likelihood estimate, $\hat\theta$, and it will be much more
strongly concentrated than the prior; Figure~\ref{fig:MargLike} depicts the
situation.  As a first step in approximating the integral in
equation~\ref{mlike-def}, we  evaluate the prior at $\hat\theta$ and pull
it out of the integral, so
\begin{equation}
\mlike 
  \approx \pi(\hat\theta) \int d\theta\;  \like(\theta).
\label{mlike-pi-hat}
\end{equation}
The prior PDF has dimensions of $[1/\theta]$, and we can express
$\pi(\hat\theta)$ as the inverse of a local prior scale, $\Delta\theta$ (for
a normalized flat prior spanning a range $\Delta\theta$, we have
$\pi(\hat\theta) = 1/\Delta\theta$ exactly).  We next approximate the
integral of the likelihood function in Eq.\ \ref{mlike-pi-hat} as the
product of its height (the maximum likelihood value, $\like(\hat\theta)$)
and a characteristic width, $\delta\theta$ (describing the posterior
uncertainty in $\theta$).  This gives
\begin{equation}
\mlike
  \approx \like(\hat\theta) \frac{\delta\theta}{\Delta\theta}.
\label{mlike-ockham}
\end{equation}
With suitable definitions of the prior and posterior scales, we can
make this equation exact 
(e.g., for a one-dimensional model
with a Gaussian likelihood function of standard deviation $\sigma$, and a
flat prior with range $\Delta\theta\gg \sigma$, as long as $\hat\theta$ is
not near a boundary of the prior range, then setting $\delta\theta =
\sigma\sqrt{2\pi}$ $\approx 1.06$ times the full width at half
maximum, makes the equation accurate).
The factor multiplying the maximum
likelihood is sometimes called the {\em Ockham factor} and will be $\le 1$;
it quantifies how much of the parameter space of the model is wasted, in the
sense of including parameter values that are ruled out by the data.
Note that for dimension $k>1$, these scales are {\em volumes}, i.e.,
products of scales in each dimension.

Asymptotically, for a simple model of fixed dimension, we expect the
uncertainty for each parameter to scale like $1/\sqrt{N}$ for sample
size $N$.  So for a model of dimension $k$, we expect $\delta\theta$
to eventually decrease proportional to $N^{-k/2}$.  Let $N_a$ be the
sample size where the asymptotic behavior kicks in, and $\delta\theta_a$
be the typical scale of the uncertainties at that sample size.
Then we expect
\begin{equation}
\mlike
  \approx \like(\hat\theta) \frac{\delta\theta_a}{\Delta\theta}
  \left(\frac{N_a}{N}\right)^{k/2}.
\label{mlike-asymp}
\end{equation}
In the simple case of estimating linear parameters for data with additive
Gaussian noise of fixed noise variance, we expect asymptotic behavior
right away, so $N_a = 1$.
Now take the logarithm and group terms according to their dependence on $N$:
\begin{equation}
\ln\mlike
  \approx \ln\like(\hat\theta) - \frac{k}{2}\ln N
  + \ln \left(\frac{N_a^{k/2} \delta\theta_a}{\Delta\theta}\right).
\label{log-mlike}
\end{equation}
The first term may have a nontrivial $N$ dependence; the last term is
constant with respect to $N$.  Schwarz (1978) derived a more precise
expression like this, explicitly calculating the first term in the case of
linear models with sampling distributions in the exponential family (which
includes, e.g., normal, Poisson, and multinomial distributions).  In that
case the maximum log-likelihood term is expected to grow increasingly
negative, roughly proportionally to $N$.

The BIC keeps the $N$-dependent terms in Eq.\ \ref{log-mlike} and multiplies by $-2$;
\begin{equation}
\mbox{BIC} = -2\ln\like(\hat\theta) + k\ln N = \chi^2 + k\ln N.
\label{BIC-def}
\end{equation}
It is attributed to Schwarz (and sometimes called the Schwarz criterion),
but notably, Schwarz did not drop the $N$-independent term in his
approximation to $\ln\mlike$, although he termed it a ``residual'' with
respect to the $N$-dependent terms.

If the models under consideration are considered equally probable a priori,
the most probable model is the one with the largest marginal likelihood.
BIC-based model selection uses the BIC to approximate the logarithm of the
marginal likelihood, choosing the model with the smallest value of the BIC.
The derivation sketched above provides some insight regarding when we might
expect this procedure to identify the highest likelihood model.
There are two main considerations.

First, the BIC is an asymptotic criterion.  Its accuracy requires sample
sizes large enough so that the parameter uncertainties are decreasing at the
$O(N^{-k/2})$ rate.  For complex models with many parameters, this is not
simply a matter of sample sizes being ``large enough.''  For some
models---e.g., nonparametric models---the number of parameters may grow with
sample size (explicitly or effectively); for others, some parameters may be
sensitive to only a subset of the data.  For example, the BLISS model uses a
piecewise linear intrapixel sensitivity map, so a particular coefficient is
determined by only a subset of the image pixels.  In these cases, a BIC-like
criterion may be valid, but with the $k\ln N$ term replaced with a term that
more accurately describes the asymptotic behavior of parameter
uncertainties. Determining the form of such a term can be subtle (see Kass
and Raftery 1995, \S~4.2).  In these settings, it may take very large sample
sizes to reach asymptotic behavior.

Second, the BIC drops a constant (in $N$) term from the logarithm of the
marginal likelihood---Schwarz's ``residual.''  That is, it drops a
{\em multiplicative factor} from the estimated marginal likelihood.  This
factor depends on the prior volume, $\Delta\theta$.  For models with
many similar degrees of freedom, like the various intrapixel sensitivity
models, the prior volume is the product of the ranges of many
variables.  It can vary sensitively with the choice of a priori scale
per parameter, and if the competing models have different types of
parameters, the omitted residual terms may be very different from one
model to another.  The difference between the residual terms can be
large when the models are large.  As a result, the change in the BIC between
two large models cannot be relied upon for identifying the model with
the larger marginal likelihood.

Kass and Wasserman (1995; KW95) have examined the role of the residual term
in the asymptotic approximation of the log marginal likelihood,
arguing that for some problems there may be a reasonable argument
for it to be negligible.  Note that the last term in Eq.\ \ref{log-mlike}
will vanish if
\begin{equation}
\Delta\theta = N_a^{k/2} \delta\theta_a.
\label{unit-info}
\end{equation}
When the asymptotic sample size $N_a=1$ (e.g., for linear models, like
estimating the mean of a normal distribution with known variance), this
requirement corresponds to having the prior range equal to the width of the
likelihood for a single-sample dataset.  For $N_a>1$, the $N_a^{k/2}$ factor
scales the $\delta\theta_a$ likelihood volume to what the single-sample
volume would be if the model were asymptotic starting with $N=1$.  Thus KW95
dubbed a prior satisfying Eq.\ \ref{unit-info} a {\em unit information
prior}, i.e., a prior that is as informative as a single datum.\footnote{The
KW95 derivation is of course more careful than that sketched here.  They
account for correlations between parameters, replacing
Eq.\ \ref{unit-info} with a relationship between Hessian matrices of the
prior and likelihood.}
To the extent that one could consider the uncertainty scale associated
with a single measurement to reflect prior uncertainty, the
BIC may be an accurate approximation to $\ln\mlike$.

However, even when a unit information prior scale appears reasonable, for
large models, even a small variation from this scale for the prior range of
each parameter could produce a large net residual term.  We thus do not
consider the unit information prior argument to provide a sound
justification for using the BIC to compare large models.

For these reasons, in our work we limit use of the BIC to comparing
small models, or large models that are nested, so that rival models
share the vast majority of parameters.  For example, we rely on the
BIC to compare different ramp models that share a common BLISS map
model, but we do not consider the BIC to be valid for comparing, say,
a model using a BLISS map to a model using B10's intrapixel
variability correction, or to polynomial intrapixel models.

The residual issue, in part, reflects an inherent weakness of marginal-likelihood 
based model comparisons with large models.  Small changes
in the per-parameter prior scale for such models can lead to large changes
in marginal likelihoods, even with an accurate numerical calculation of
the marginal likelihoods.  This motivates developing a way to allow
the scale to adapt to the data.  In a Bayesian framework, this can
be accomplished using a hierarchical model to implement regularization.
One considers the prior range to be an adjustable parameter itself
that is learned from the data.  This can be done in a manner that
essentially lets the data determine the effective number of free
parameters (the total number of knots) required for the intrapixel map.  
We would then compare our best fit using BLISS mapping to those obtained using other intrapixel models and select the appropriate model component.
Such calculations are
beyond the scope of this paper, but would be a productive avenue for future research.



\clearpage
\section{Best-Fit Parameters}
\label{sec:app}

\begin{table*}[ht]
\centering
\caption{\label{tab:trfits} Best-Fit Joint Transit Light-Curve Parameters}
\if\submitms y
{\footnotesize
\fi
\begin{tabular}{rr@{\,{\pm}\,}lr@{\,{\pm}\,}lr@{\,{\pm}\,}l}
\hline
\hline
Parameter                                                        &   \mctc{HD149bp41}            &   \mctc{HD149bp42}            &   \mctc{HD149bp43}      \\
\hline
Wavelength ({\microns})                                          &   \mctc{   8.0   }            &   \mctc{   8.0   }            &   \mctc{   8.0   }               \\
Array Position (\math{\bar{x}}, pix)                             &   \mctc{      14.93       }   &   \mctc{      14.96       }   &   \mctc{      15.14       }  \\
Array Position (\math{\bar{y}}, pix)                             &   \mctc{      15.14       }   &   \mctc{      14.56       }   &   \mctc{      14.47       }  \\
Position Consistency\tablenotemark{1} (\math{\delta\sb{x}}, pix) &   \mctc{      0.013       }   &   \mctc{      0.011       }   &   \mctc{      0.011       }   \\
Position Consistency\tablenotemark{1} (\math{\delta\sb{y}}, pix) &   \mctc{      0.012       }   &   \mctc{      0.013       }   &   \mctc{      0.013       }   \\
Aperture Size (pix)                                              &   \mctc{      3.5         }   &   \mctc{      3.5         }   &   \mctc{      3.5           }     \\
Sky Annulus Inner Radius (pix)                                   &   \mctc{      7.0         }   &   \mctc{      7.0         }   &   \mctc{      7.0           }      \\
Sky Annulus Outer Radius (pix)                                   &   \mctc{     15.0         }   &   \mctc{      15.0        }   &   \mctc{     15.0           }      \\
System Flux \math{F\sb{s}} (\micro Jy)                          &         117229 & 14           &         117460 & 60           &        117356 & 13           \\
Transit Midpoint\tablenotemark{2} (MJD\sb{UTC})                  &      4327.3720 & 0.0005       &      4356.1315 & 0.0005       &     4597.7067 & 0.0005         \\
Transit Midpoint\tablenotemark{2} (MJD\sb{TDB})                  &      4327.3727 & 0.0005       &      4356.1323 & 0.0005       &     4597.7075 & 0.0005        \\
\math{R\sb{p}/R\sb{\star}}                                       &         0.0518 & 0.0006       &         0.0518 & 0.0006       &        0.0518 & 0.0006            \\
cos i                                                            &          0.095 & 0.009        &          0.095 & 0.009        &         0.095 & 0.009             \\
\math{a/R\sb{\star}}                                             &           5.98 & 0.17         &           5.98 & 0.17         &          5.98 & 0.17              \\
Ramp Equation (\math{R(t)})                                      &  \mctc{\ref{eqnsel}\sb{--}}   & \mctc{\ref{eqnlog}      }     &   \mctc{  \ref{eqnse}\sb{--}      }      \\
Ramp, \math{r\sb{0}}                                             &         24    & 4             &      \mctc{   0      }        &          18    & 2            \\
Ramp, \math{r\sb{1}}                                             &     -0.6      & 0.7           &      \mctc{   0      }        &         4.1    & 1.4           \\
Ramp, \math{r\sb{2}}                                             &      0.0110  & 0.0015         &      \mctc{   0      }        &     \mctc{    0     }         \\
Ramp, \math{r\sb{6}}                                             &     \mctc{    0      }        &      0.0009   & 0.0005        &     \mctc{    0     }         \\
Ramp, \math{r\sb{7}}                                             &     \mctc{    0      }        &     -0.00048  & 0.00011       &     \mctc{    0     }         \\
Ramp, \math{t\sb{0}}                                             &     \mctc{    0      }        &     -0.0248   & 0.0011        &     \mctc{    0     }         \\
BLISS Map (\math{M(x,y)})                                        &   \mctc{       Yes        }   &   \mctc{       Yes        }   &   \mctc{       Yes        }    \\
Min. Number of Points Per Bin                                    &   \mctc{        4         }   &   \mctc{        4         }   &   \mctc{        4         }   \\
Total Frames                                                     &   \mctc{      67008       }   &   \mctc{      54080       }   &   \mctc{      70000       }     \\
Rejected Frames (\%)                                             &   \mctc{     0.714840      }   &   \mctc{     0.377219     }   &   \mctc{     0.504286     }   \\
Frames Used\tablenotemark{3}                                     &   \mctc{      61520       }   &   \mctc{      53865       }   &   \mctc{      68646       }    \\
Free Parameters                                                  &   \mctc{        8         }   &   \mctc{        5         }   &   \mctc{        4         }    \\
AIC Value                                                        &   \mctc{     184044       }   &   \mctc{     184044       }   &   \mctc{     184044       }    \\
BIC Value                                                        &   \mctc{     184216       }   &   \mctc{     184216       }   &   \mctc{     184216       }   \\
SDNR                                                             &   \mctc{    0.0083440     }   &   \mctc{    0.0083556     }   &   \mctc{    0.0083691     }   \\
Uncertainty Scaling Factor                                       &   \mctc{     0.818677     }   &   \mctc{     0.818525     }   &   \mctc{     0.821463     }  \\
Photon-Limited S/N (\%)                                          &   \mctc{       83.1       }   &   \mctc{       83.1       }   &   \mctc{       82.6       }  \\
\hline
\end{tabular}
\if\submitms y
}
\fi
\begin{minipage}[t]{0.67\linewidth}
\tablenotetext{1}{RMS frame-to-frame position difference.}
\tablenotetext{2}{MJD = BJD - 2,450,000.}
\tablenotetext{3}{We exclude frames during instrument/telescope settling, for insufficient points at a given knot, and for bad pixels in the photometry aperture.}
\end{minipage}
\end{table*}
\if\submitms y
\clearpage
\fi

\begin{table*}[ht]
\centering
\caption{\label{tab:fits1} Best-Fit Joint Eclipse Light-Curve Parameters}
\if\submitms y
{\tiny
\fi
\begin{tabular}{rr@{\,{\pm}\,}lr@{\,{\pm}\,}lr@{\,{\pm}\,}lr@{\,{\pm}\,}lr@{\,{\pm}\,}l}
\hline
\hline
Parameter                                                        &   \mctc{HD149bs11}            &   \mctc{HD149bs21}            &   \mctc{HD149bs31}            &   \mctc{HD149bs32}            &   \mctc{HD149bs33} \\
\hline
Wavelength ({\microns})                                          &   \mctc{   3.6   }            &   \mctc{   4.5   }            &   \mctc{   5.8   }            &   \mctc{   5.8   }            &   \mctc{   5.8   }           \\
Array Position (\math{\bar{x}}, pix)                             &   \mctc{      14.58       }   &   \mctc{      14.57       }   &   \mctc{      14.50       }   &   \mctc{      14.42       }   &   \mctc{      14.78       }   \\
Array Position (\math{\bar{y}}, pix)                             &   \mctc{      15.66       }   &   \mctc{      14.99       }   &   \mctc{      14.37       }   &   \mctc{      14.17       }   &   \mctc{      14.67       }   \\
Position Consistency\tablenotemark{1} (\math{\delta\sb{x}}, pix) &   \mctc{      0.009       }   &   \mctc{      0.009       }   &   \mctc{      0.020       }   &   \mctc{      0.020       }   &   \mctc{      0.015       }   \\
Position Consistency\tablenotemark{1} (\math{\delta\sb{y}}, pix) &   \mctc{      0.007       }   &   \mctc{      0.004       }   &   \mctc{      0.018       }   &   \mctc{      0.013       }   &   \mctc{      0.017       }   \\
Aperture Size (pix)                                              &   \mctc{       3.75       }   &   \mctc{       2.75       }   &   \mctc{       2.75       }   &   \mctc{       2.75       }   &   \mctc{       2.75       }   \\
Sky Annulus Inner Radius (pix)                                   &   \mctc{       7.0        }   &   \mctc{       7.0        }   &   \mctc{       7.0        }   &   \mctc{       7.0        }   &   \mctc{       7.0        }   \\
Sky Annulus Outer Radius (pix)                                   &   \mctc{       15.0       }   &   \mctc{       15.0       }   &   \mctc{       15.0       }   &   \mctc{       15.0       }   &   \mctc{       15.0       }   \\
System Flux\tablenotemark{2} \math{F\sb{s}} (\micro Jy)          &        528456   & 10          &        334037   & 40          &        216010   & 70          &        214102   & 40          &        212621   & 80          \\
Eclipse Depth (\%)                                               &        0.040 & 0.003          &         0.034 & 0.006         &         0.044 & 0.010         &         0.044 & 0.010         &         0.044 & 0.010         \\
Brightness Temperature (K)                                       &         2000   & 60           &         1650   & 120          &        1600   & 200           &        1600   & 200           &        1600   & 200           \\
Eclipse Midpoint\tablenotemark{3} (orbits)                       &        0.5003 & 0.0004        &         0.4994 & 0.0007       &         0.502 & 0.004         &\mctc{0.501$^{+0.002}_{-0.005}$}&\mctc{0.500$^{+0.003}_{-0.005}$}\\
Eclipse Midpoint\tablenotemark{4} (MJD\sb{UTC})                  &      4535.8756 & 0.0010       &      4596.2669 & 0.0019       &       4325.941 & 0.013        &      4633.658 & 0.010       &      4903.989 & 0.012       \\
Eclipse Midpoint\tablenotemark{4} (MJD\sb{TDB})                  &      4535.8764 & 0.0010       &      4596.2676 & 0.0019       &       4325.942 & 0.013        &      4633.659 & 0.010       &      4903.990 & 0.012       \\
Eclipse Duration (\math{t\sb{\rm 4-1}}, hrs)                     &           3.29 & 0.02         &           3.29 & 0.02         &           3.29 & 0.02         &          3.29 & 0.02          &           3.29 & 0.02         \\
Ingress/Egress Time (\math{t\sb{\rm 2-1}}, hrs)                  &          0.230 & 0.011        &          0.234 & 0.012        &          0.234 & 0.012        &          0.234 & 0.012        &          0.234 & 0.012        \\
Ramp Equation (\math{R(t)})                                      &   \mctc{  \ref{eqnlin}    }   &   \mctc{\ref{eqnse}\sb{+} }   &   \mctc{\ref{eqnse}\sb{+} }   &   \mctc{\ref{eqnse}\sb{+} }   &   \mctc{\ref{eqnse}\sb{+} }   \\
Ramp, \math{r\sb{0}}                                             &   \mctc{      0           }   &            29 & 9             &            30 & 6             &            42 & 5             &            26 & 5             \\
Ramp, \math{r\sb{1}}                                             &   \mctc{      0           }   &             7 & 4             &             8 & 3             &            14 & 2             &           6.4 & 2.0           \\
Ramp, \math{r\sb{2}}                                             &       -0.0038 & 0.0008        &   \mctc{      0           }   &   \mctc{      0           }   &   \mctc{      0           }   &   \mctc{      0           }   \\
BLISS Map (\math{M(x,y)})                                        &   \mctc{       Yes        }   &   \mctc{       Yes        }   &   \mctc{       Yes        }   &   \mctc{        No        }   &   \mctc{        No        }  \\
Min. Number of Points Per Bin                                    &   \mctc{        4         }   &   \mctc{        5         }   &   \mctc{        5         }   &   \mctc{        -         }   &   \mctc{        -         }   \\
Total Frames                                                     &   \mctc{      54080       }   &   \mctc{      54080       }   &   \mctc{      54080       }   &   \mctc{      54080       }   &   \mctc{      54080       }   \\
Rejected Frames (\%)                                             &   \mctc{     0.308802     }   &   \mctc{     0.160873     }   &   \mctc{     0.268121     }   &   \mctc{     0.277367     }   &   \mctc{     0.312500     }   \\
Frames Used\tablenotemark{5}                                     &   \mctc{      50769       }   &   \mctc{      48769       }   &   \mctc{      50919       }   &   \mctc{      50930       }   &   \mctc{      53911       }   \\
Free Parameters                                                  &   \mctc{        6         }   &   \mctc{        7         }   &   \mctc{        7         }   &   \mctc{        4         }   &   \mctc{        4         }   \\
AIC Value                                                        &   \mctc{     50775        }   &   \mctc{     48776        }   &   \mctc{     155775       }   &   \mctc{     155775       }   &   \mctc{     155775       }   \\
BIC Value                                                        &   \mctc{     50828        }   &   \mctc{     48837        }   &   \mctc{     155924       }   &   \mctc{     155924       }   &   \mctc{     155924       }   \\
SDNR                                                             &   \mctc{    0.00280365    }   &   \mctc{    0.00388070    }   &   \mctc{    0.0120115     }   &   \mctc{    0.0119551     }   &   \mctc{    0.0119188     }   \\
Uncertainty Scaling Factor                                       &   \mctc{     0.401020     }   &   \mctc{     0.489317     }   &   \mctc{     0.914694     }   &   \mctc{     1.02508      }   &   \mctc{     1.02476      }   \\
Photon-Limited S/N (\%)                                          &   \mctc{       93.9       }   &   \mctc{       89.9       }   &   \mctc{       73.6       }   &   \mctc{       73.7       }   &   \mctc{       74.3       }   \\
\hline
\end{tabular}
\if\submitms y
}
\fi
\begin{minipage}[t]{0.91\linewidth}
\tablenotetext{1}{RMS frame-to-frame position difference.}
\tablenotetext{2}{We multiply the HD149bs32 and HD149bs33 measured system fluxes by 0.968 to correct for an IRAC flux conversion issue in the S18.18 pipeline.}
\tablenotetext{3}{Based on the period and ephemeris time given by \citet{Knutson2009b}.}
\tablenotetext{4}{MJD = BJD - 2,450,000.}
\tablenotetext{5}{We exclude frames during instrument/telescope settling, for insufficient points at a given knot, and for bad pixels in the photometry aperture.}
\end{minipage}
\end{table*}
\if\submitms y
\clearpage
\fi

\begin{table*}[ht]
\centering
\caption{\label{tab:fits2} Best-Fit Joint Eclipse Light-Curve Parameters}
\if\submitms y
{\tiny
\fi
\begin{tabular}{@{}r@{\ }r@{\,{\pm}\,}l@{\ }r@{\,{\pm}\,}l@{\ }r@{\,{\pm}\ }l@{\ }r@{\,{\pm}\,}l@{\ }r@{\,{\pm}\,}l@{\ }r@{\,{\pm}\,}l@{}}
    \hline
\hline
Parameter                                                        &   \mctc{HD149bs41}            &   \mctc{HD149bs42}            &   \mctc{HD149bs43}            &   \mctc{HD149bs44}            &   \mctc{HD149bs51}            &   \mctc{HD149bs52}   \\
\hline
Wavelength ({\microns})                                          &   \mctc{   8.0   }            &   \mctc{   8.0   }            &   \mctc{   8.0   }            &   \mctc{   8.0   }            &   \mctc{    16   }            &   \mctc{    16   }   \\
Array Position (\math{\bar{x}}, pix)                             &   \mctc{      15.06       }   &   \mctc{      14.40       }   &   \mctc{      15.04       }   &   \mctc{      14.64       }   &   \mctc{      26.07       }   &   \mctc{      23.95       }   \\
Array Position (\math{\bar{y}}, pix)                             &   \mctc{      14.45       }   &   \mctc{      15.09       }   &   \mctc{      14.13       }   &   \mctc{      15.08       }   &   \mctc{      22.16       }   &   \mctc{      20.65       }  \\
Position Consistency\tablenotemark{1} (\math{\delta\sb{x}}, pix) &   \mctc{      0.091       }   &   \mctc{      0.013       }   &   \mctc{      0.011       }   &   \mctc{      0.012       }   &   \mctc{      0.016       }   &   \mctc{      0.018       }   \\
Position Consistency\tablenotemark{1} (\math{\delta\sb{y}}, pix) &   \mctc{      0.077       }   &   \mctc{      0.011       }   &   \mctc{      0.012       }   &   \mctc{      0.011       }   &   \mctc{      0.021       }   &   \mctc{      0.018       }   \\
Aperture Size (pix)                                              &   \mctc{       4.0        }   &   \mctc{       3.5        }   &   \mctc{       3.75       }   &   \mctc{       3.5        }   &   \mctc{       2.0        }   &   \mctc{       2.0        }   \\
Sky Annulus Inner Radius (pix)                                   &   \mctc{       7.0        }   &   \mctc{       7.0        }   &   \mctc{       7.0        }   &   \mctc{       7.0        }   &   \mctc{       6.0        }   &   \mctc{       6.0        }   \\
Sky Annulus Outer Radius (pix)                                   &   \mctc{       15.0       }   &   \mctc{       15.0       }   &   \mctc{       15.0       }   &   \mctc{       15.0       }   &   \mctc{       11.0       }   &   \mctc{       12.0       }   \\
System Flux\tablenotemark{2} \math{F\sb{s}} (\micro Jy)          &        117190   & 100         &        116960   & 10          &        117818   & 6           &        118365   & 70          &         18618   & 6           &         18805   & 11          \\
Eclipse Depth (\%)                                               &         0.052 & 0.006         &         0.052 & 0.006         &         0.052 & 0.006         &         0.052 & 0.006         &         0.085 & 0.032         &         0.085 & 0.032         \\
Brightness Temperature (K)                                       &         1650   & 110          &         1650   & 110          &         1650   & 110          &         1650   & 110          &        1800   & 600           &        1800   & 600           \\
Eclipse Midpoint\tablenotemark{3} (phase)                        &        0.5002 & 0.0007        &        0.5010 & 0.0009        &        0.4961 & 0.0012        &        0.4988 & 0.0006        &        0.5013 & 0.0014        &        0.5013 & 0.0014        \\
Eclipse Midpoint\tablenotemark{4} (MJD\sb{utc})                  &       3606.962  & 0.002       &      4567.513  & 0.003        &      4599.133  & 0.003        &      4912.613  & 0.002        &      4317.311  & 0.004        &      4343.194  & 0.004        \\
Eclipse Midpoint\tablenotemark{4} (MJD\sb{tdb})                  &       3606.963  & 0.002       &      4567.513  & 0.003        &      4599.134  & 0.003        &      4912.614  & 0.002        &      4317.311  & 0.004        &      4343.194  & 0.004        \\
Eclipse Duration (\math{t\sb{\rm 4-1}}, hrs)                     &           3.29 & 0.02         &           3.29 & 0.02         &           3.29 & 0.02         &           3.29 & 0.02         &           3.29 & 0.02         &           3.29 & 0.02         \\
Ingress/Egress Time (\math{t\sb{\rm 2-1}}, hrs)                  &          0.234 & 0.012        &          0.234 & 0.012        &          0.234 & 0.012        &          0.234 & 0.012        &          0.234 & 0.012        &          0.234 & 0.012        \\
Ramp Equation (\math{R(t)})                                      &   \mctc{ \ref{eqnse}\sb{--}}  &   \mctc{ \ref{eqnquad}}       &   \mctc{\ref{eqnlin}      }   &  \mctc{\ref{eqnse}\sb{--}}    &   \mctc{\ref{eqnquad}     }   &   \mctc{\ref{eqnse}\sb{--} }  \\
Ramp, \math{r\sb{0}}                                             &          15.4 & 1.2           &   \mctc{      0           }   &   \mctc{      0           }   &            14 & 8             &   \mctc{      0           }   &            62 & 10            \\
Ramp, \math{r\sb{1}}                                             &           3.1 & 0.5           &   \mctc{      0           }   &   \mctc{      0           }   &             0 & 4             &   \mctc{      0           }   &            24 & 4             \\
Ramp, \math{r\sb{2}}                                             &   \mctc{      0           }   &         0.083 & 0.002         &       -0.0014 & 0.0016        &   \mctc{      0           }   &         0.413 & 0.012         &   \mctc{      0           }   \\
Ramp, \math{r\sb{3}}                                             &   \mctc{      0           }   &         -0.67 & 0.11          &   \mctc{      0           }   &   \mctc{      0           }   &          -5.1 & 0.2           &   \mctc{      0           }   \\
BLISS Map (\math{M(x,y)})                                        &   \mctc{        No        }   &   \mctc{       Yes        }   &   \mctc{       Yes        }   &   \mctc{       Yes        }   &   \mctc{        No        }   &   \mctc{        No        }   \\
Min. Number of Points Per Bin                                    &   \mctc{        -         }   &   \mctc{        8         }   &   \mctc{        4         }   &   \mctc{        10        }   &   \mctc{        -         }   &   \mctc{        -         }   \\
Total Frames                                                     &   \mctc{      44352       }   &   \mctc{      54080       }   &   \mctc{      60500       }   &   \mctc{      54080       }   &   \mctc{       1050       }   &   \mctc{       1050       }   \\
Rejected Frames (\%)                                             &   \mctc{     0.47574      }   &   \mctc{     0.432692     }   &   \mctc{     0.634711     }   &   \mctc{     0.488166     }   &   \mctc{     0.285714     }   &   \mctc{     0.571429     }   \\
Frames Used\tablenotemark{5}                                     &   \mctc{      44041       }   &   \mctc{      48801       }   &   \mctc{      60100       }   &   \mctc{      46274       }   &   \mctc{       1047       }   &   \mctc{       1044       }   \\
Free Parameters                                                  &   \mctc{        15        }   &   \mctc{        4         }   &   \mctc{        3         }   &   \mctc{        4         }   &   \mctc{       10         }   &   \mctc{        12        }   \\
AIC Value                                                        &   \mctc{     194243       }   &   \mctc{     194243       }   &   \mctc{     194243       }   &   \mctc{     194245       }   &   \mctc{     2113         }   &   \mctc{     2113         }   \\
BIC Value                                                        &   \mctc{     194518       }   &   \mctc{     194518       }   &   \mctc{     194518       }   &   \mctc{     194540       }   &   \mctc{     2237         }   &   \mctc{     2237         }   \\
SDNR                                                             &   \mctc{    0.00845740    }   &   \mctc{    0.00833381    }   &   \mctc{    0.00837245    }   &   \mctc{    0.00847674    }   &   \mctc{    0.00311603    }   &   \mctc{    0.00337752    }   \\
Uncertainty Scaling Factor                                       &   \mctc{     0.816657     }   &   \mctc{     0.813220     }   &   \mctc{     0.819215     }   &   \mctc{     0.981709     }   &   \mctc{     0.374578     }   &   \mctc{     0.402919     }   \\
Photon-Limited S/N (\%)                                          &   \mctc{       82.3       }   &   \mctc{       83.1       }   &   \mctc{       82.4       }   &   \mctc{       81.6       }   &   \mctc{       27.2       }   &   \mctc{       24.9       }   \\
\hline
\end{tabular}
\if\submitms y
}
\fi
\begin{minipage}[t]{0.95\linewidth}
\tablenotetext{1}{RMS frame-to-frame position difference.}
\tablenotetext{2}{We multiply the HD149bs44 measured system flux by 0.973 to correct for an IRAC flux conversion issue in the S18.18 pipeline.}
\tablenotetext{3}{Based on the period and ephemeris time given by \citet{Knutson2009b}.}
\tablenotetext{4}{MJD = BJD - 2,450,000.}
\tablenotetext{5}{We exclude frames during instrument/telescope settling, for insufficient points at a given knot, and for bad pixels in the photometry aperture.}
\end{minipage}
\end{table*}
\if\submitms y
\clearpage
\fi

\end{appendices}

\end{document}